\documentclass[12pt]{article}

\usepackage{fullpage}
\usepackage{microtype} 

\usepackage[utf8]{inputenc} 
\usepackage[T1]{fontenc}
\usepackage{helvet} 
\usepackage{sectsty} 
 
\makeatletter\renewcommand{\@dotsep}{1000} 

\usepackage{dsfont}
\usepackage{mathtools,amssymb,amsmath,bbold,mathrsfs}
\usepackage{mathalfa}
\usepackage{euscript}
\usepackage[bbgreekl]{mathbbol}
\DeclareSymbolFontAlphabet{\mathbbm}{bbold}
\DeclareSymbolFontAlphabet{\mathbb}{AMSb}
\usepackage{bm} 
\usepackage{braket}
\usepackage{slashed}
\usepackage{multicol}
\usepackage{accents}
\usepackage{csquotes}
\usepackage{upgreek}

\DeclareMathAlphabet{\mpala}{T1}{ppl}{m}{it}

\DeclareMathAlphabet{\esstix}{U}{esstixcal}{m}{n}
\DeclareMathAlphabet{\bdx}{U}{BOONDOX-cal}{m}{n}
\DeclareMathAlphabet{\dcal}{U}{dutchcal}{m}{n}
\DeclareMathAlphabet{\bdcal}{U}{dutchcal}{b}{n}
\DeclareMathAlphabet{\bbdx}{U}{BOONDOX-cal}{b}{n}
\DeclareMathAlphabet{\mathpzc}{OT1}{pzc}{b}{n}

\makeatletter
\g@addto@macro\bfseries{\boldmath}
\makeatother

\usepackage[table,dvipsnames]{xcolor}
\usepackage{tcolorbox}
\usepackage{enumitem}
\usepackage{cite}
\usepackage[colorlinks=true,linkcolor=purple,citecolor=purple,linktocpage=true, urlcolor=NavyBlue]{hyperref}
\usepackage{accents}
\usepackage{harmony}
\usepackage[font=small,labelfont=bf]{caption}
\usepackage{graphicx} 
\usepackage{subcaption} 
\usepackage{adjustbox}
\usepackage[normalem]{ulem}

\newcommand{\be}{\begin{equation}}
\newcommand{\ee}{\end{equation}}
\newcommand{\bea}{\begin{eqnarray}}
\newcommand{\eea}{\end{eqnarray}}
\newcommand{\bee}{\begin{equation} \begin{aligned}}
\newcommand{\eee}{\end{aligned} \end{equation}}
\newcommand{\lb}{\label}

\newcommand{\sss}{\scriptscriptstyle}

\newcommand{\cL}{\mathcal{L}}

\newcommand{\M}{\mathcal{M}} 
\newcommand{\NP}{\mathcal{N}} 
\renewcommand{\H}{\mathcal{H}} 
\renewcommand{\S}{\mathcal{S}} 
\newcommand{\Time}{\mathcal{T}} 

\newcommand{\form}{}
\renewcommand{\vec}{\uline}
\newcommand{\rd}{\mathrm{d}} 
\newcommand{\Lie}{\cL} 
\newcommand{\exd}{\form{\rd}} 
\newcommand{\DN}{\text{D}} 

\newcommand{\kv}{K} 
\renewcommand{\u}{u} 
\renewcommand{\k}{k}
\newcommand{\n}{n}

\newcommand{\uu}{\bdx{u}} 
\newcommand{\kk}{\bdx{k}}
\newcommand{\nn}{\bdx{n}}

\newcommand{\uAH}{\mathsf{u}} 
\newcommand{\kAH}{\mathsf{k}}
\newcommand{\nAH}{\mathsf{n}}

\newcommand{\bvol}{\form{\epsilon}}

\newcommand{\volN}{\form{\epsilon}_{\sss \NP}}
\newcommand{\volM}{\form{\epsilon}_{\sss \M}}
\newcommand{\volSR}{\mathring{\form{\epsilon}}_{\sss {\S}}}
\newcommand{\volS}{\form{\epsilon}_{\sss \S}}

\newcommand{\e}{\mathrm{e}} 

\newcommand{\pa}{\partial}

\newcommand{\mr}{\mathring}
\newcommand{\Newton}{\mathrm{G}_{\sss \mathrm{N}}}

\newcommand{\vtheta}{\vartheta}
\newcommand{\ba}{\bar{\alpha}}
\newcommand{\bkappa}{\overline{\kappa}}
\newcommand{\vkappa}{\varkappa}
\newcommand{\bvkappa}{\overline{\varkappa}}
\newcommand{\vrho}{\varrho}
\newcommand{\A}{\esstix{A}}
\renewcommand{\P}{\esstix{P}}
\newcommand{\bP}{\overline{\P}}
\newcommand{\m}{\esstix{M}}
\newcommand{\E}{\esstix{E}}
\newcommand{\Ent}{\esstix{S}}

\newcommand{\dd}{\! \cdot \!}

\newcommand{\AH}{{\sss \H}}
\newcommand{\PAH}{\mathsf{P}}
\newcommand{\bPAH}{\overline{\mathsf{P}}}
\newcommand{\KAH}{\mathsf{K}}

\newcommand{\kappaAH}{\upkappa}
\newcommand{\bkappaAH}{\overline{\upkappa}}
\newcommand{\rhoAH}{\uprho}

\newcommand{\emp}[1]{{\color{NavyBlue}\emph{#1}}}

\newcommand{\step}[1]{\color{purple!70}#1\color{black}}

\definecolor{Sepia}{RGB}{244, 232, 201}

\newcommand{\bbox}{\begin{tcolorbox}[
  colback=NavyBlue!5,
  colframe=white,    
  boxrule=0pt,       
  arc=0mm, 
  left=5pt, right=5pt, top=5pt, bottom=5pt 
]}
\newcommand{\ebox}{\end{tcolorbox}}

\definecolor{paletan}{RGB}{245, 245, 220}
\definecolor{palegray}{RGB}{230,230,230}

\setlist[itemize]{left=0pt, labelsep=1em, labelindent=0pt, align=parleft, listparindent=0pt}

\begin{document}
\title{\Large{{\bfseries Revisiting Spherically Symmetric Spacetime I: \\ Geometro-Hydrodynamics\\
}}
}
\author{Puttarak Jai-akson$^1$\thanks{\href{mailto:puttarak.jaiakson@gmail.com}{\texttt{puttarak.jaiakson@gmail.com}}} \ and Yuki Yokokura$^{1,2}$\thanks{\href{mailto:yuki.yokokura@kek.jp}{\texttt{yuki.yokokura@kek.jp}}}
}
\date{\small{\textit{
$^1$ RIKEN iTHEMS, Wako, Saitama 351-0198, Japan\\
$^2$ Theory Center, KEK, Tsukuba, Ibaraki 305-0801, Japan \\
}}}

\maketitle

\begin{abstract}

This series of works revisits the geometry, dynamics, and covariant phase space of spherically symmetric spacetimes with the aim of exploring the thermodynamics of spacetime from their dynamical properties. In this first paper, we examine the geometry from the perspective of a foliation by spherical hypersurfaces. Using the rigging technique, we first define a local frame adapted to these slices and reconstruct the geometry and dynamics fully. We clarify the connection of the frame adapted to constant-radius slices, to the Kodama vector and Misner-Sharp energy. Through frame transformations, we then show that the gravitational dynamics in a general foliation-adapted frame can be interpreted as hydrodynamics, i.e., geometro-hydrodynamics: the Einstein equations exhibit the gravitational analogs of the Euler and Young-Laplace equations, and the spacetime can be viewed as the worldvolume of a concentric stack of "gravitational bubbles"---spherical collective modes with the Misner-Sharp energy density and a geometric pressure. We apply this framework to apparent horizons and study the dynamics. Finally, we demonstrate that a similar geometro-hydrodynamic picture holds in Lovelock gravity. These results provide a fresh perspective on this class of spacetimes and lay the foundation for understanding their thermodynamic properties.

\end{abstract}

\thispagestyle{empty}
\newpage
\setcounter{page}{1}


\hrule
\setcounter{tocdepth}{2}
\tableofcontents
\vspace{0.7cm}
\hrule

\newpage
\section{Introduction}
In this series of works, we will revisit the geometry and dynamics of spherically symmetric spacetimes, highlighting their resemblance to hydrodynamics and their connection to thermodynamics. Our ultimate goal is to gain a deeper understanding of the thermodynamics of spacetime based on its dynamics.

Gravity is universal in that any object carrying energy attracts others. This feature is most clearly manifested in spherically symmetric systems, where motion is effectively restricted to the radial direction, and in a strong-gravity limit, matter collapses to black holes. Despite their apparent simplicity, spherically symmetric spacetimes have long served as a cornerstone of general relativity, providing a powerful framework that captures essential aspects of the interaction between matter and gravity. They remain a testing ground for various approaches to quantum black holes and quantum gravity. 
For instance, they describe the classical black holes \cite{Poisson:2009pwt}, regular black-hole metrics \cite{1968qtr..conf...87B,Dymnikova:1992ux,Hayward:2005gi}, as well as exotic compact objects and quantum black holes\cite{Visser:1992qh,Mazur:2001fv,Barcelo:2009tpa,Kawai:2014afa,Cardoso:2019rvt}; they model quantum collapsing processes \cite{Frolov:1981mz,Kawai:2013mda,Haggard:2014rza,Kelly:2020lec};
they are also useful in studying thermodynamics of self-gravitating systems \cite{Sorkin:1981wd,Yokokura:2023wxp}; in the context of quantum gravity, spherically reduced models play a crucial role in canonical quantization approaches \cite{Thiemann:1992jj,Kastrup:1993br,Kuchar:1994zk,Lau:1995fr,Pullin2008,Gambini:2020nsf}, in mini-superspace formulations \cite{Ashtekar:2005qt,Bodendorfer:2019cyv,Geiller:2020xze}, as well as in an effective model coupled to quantum matters \cite{Livine:2025soz}; and they also appear in lower-dimensional theories \cite{TEITELBOIM198341,JACKIW1985343,Callan:1992rs}. Altogether, spherically symmetric models provide a framework in which the conceptual and technical challenges of gravity and spacetime can be explored in a simplified yet physically meaningful manner.


The thermodynamics of spacetime started with the pioneering works on black hole thermodynamics by Bekenstein \cite{Bekenstein1973} and Hawking \cite{Hawking:1975vcx}, which showed that black holes behave as thermodynamic objects with entropy proportional to the surface area of their event horizon. This perspective was further enriched to a general spacetime by Jacobson’s seminal result \cite{Jacobson:1995ab}, demonstrating that Einstein equations can be interpreted as an equation of state arising from thermodynamic principles. In these studies, it was discussed that spacetime possesses thermodynamic properties through the quantum nature of matter (the Compton wavelength of matter particles, particle creation in a time-dependent spacetime, and the Unruh effect in the vacuum of matter fields). 

However, if spacetime is composed of some (still unknown) microscopic constituents, a thermodynamic/hydrodynamic description of spacetime should manifest directly from their statistical behavior in a limit where the number of constituents becomes large (just as the behavior of water as a fluid emerges from the statistical behavior of numerous water molecules). Therefore, the Einstein equations, expected to emerge in this limit, should inherently possess the thermodynamic/hydrodynamic behavior. In this series, we utilize the properties of spherically symmetric spacetimes and explore the thermodynamics of spacetime more directly from the classical dynamics of spacetime itself. 



In thermodynamics and hydrodynamics, energy plays a crucial role \cite{Landau:1980mil, landau1987fluid}. However, in general relativity, it is non-trivial to identify a well-defined notion of energy because, due to the equivalence principle and diffeomorphism invariance, there is no local covariant energy or unique time coordinate. 
Several approaches have been developed to address this difficulty: for example, the Brown-York prescription \cite{Brown:1992br}, Hamiltonian formulations \cite{Arnowitt:1962hi,REGGE1974286}, and asymptotic charge constructions \cite{Bondi:1962px,Sachs:1962wk,Wald:1999wa,Barnich:2001jy}. Here, the special properties of spherically symmetric spacetime come into play. In spherically symmetric spacetimes, there exists the Misner-Sharp energy \cite{Misner:1973prb}, the locally conserved energy (even in dynamical cases) containing intrinsic mass and gravitational energy, which matches the ADM (Arnowitt-Deser-Misner) energy in the asymptotically flat limit \cite{Misner:1973prb,Hayward:1994bu}. Furthermore, spherically symmetric spacetimes have the Kodama vector \cite{Kodama:1979vn}, which provides a preferred notion of time evolution and leads to the Misner-Sharp energy naturally (see Sec.\ref{sec:Kodama} for a review). The Misner-Sharp energy and the Kodama vector play the basic roles of energy and time-evolution vector in spherically symmetric settings.

This first paper is devoted to constructing a geometric framework suitable for studying the thermodynamics of spherically symmetric spacetime. In particular, it seeks to recast the dynamics of gravity as hydrodynamics by using the Misner-Sharp energy fully, and identify a dictionary that translates geometric quantities (such as acceleration and curvatures) in terms of thermodynamic variables (such as energy and pressure). Importantly, this does not refer to ordinary matter fluids living on top of spacetime, but rather to spacetime itself behaving as a hydrodynamic system. To distinguish such a "gravitational fluid" from a standard matter fluid, this hydrodynamic description of spacetime geometry is referred to as \emp{geometro-hydrodynamics}\footnote{To the best of our knowledge, this term was first coined by Hu \cite{Hu:1996gk}, in analogy with Wheeler’s geometrodynamics.}.


The idea that gravitational dynamics can be reinterpreted in hydrodynamic terms is not new. 
A prominent example,  the black hole membrane paradigm \cite{Damour:1978cg,thorne1986black,Price:1986yy}, embodies this idea by connecting gravitational physics at the black hole horizon to hydrodynamics.
In this work, we aim to find a geometry-hydrodynamics dictionary that is different from that of the membrane paradigm and applicable to arbitrary spherical hypersurfaces (not necessarily horizons) in spherically symmetric spaces.

Thus, the objectives of this work are twofold: 
\begin{enumerate}
\item To demonstrate that the dynamics of spherically symmetric spacetime governed by the Einstein equations can be recast in the form of hydrodynamic equations, leading to a dictionary between the geometric and thermodynamic quantities. 
\item To develop a mathematical toolbox for studying the dynamics of general spherically symmetric spacetimes in terms of various frames associated with spherical hypersurfaces.
\end{enumerate}


\paragraph{Key ideas and main results:} \hfill

The main idea behind our geometric construction is to view spacetime as a foliation by codimension-1 spherical slices (hypersurfaces). Conventionally, a spherically symmetric spacetime $\M$ is regarded as the product  $\NP \times \S_r$ of a 2-dimensional Lorentzian plane $\NP$ and a round $(d-2)$-dimensional spherical shell $\S_r$ of areal radius $r: \NP \to \mathbb{R}$.  The plane $\NP$ provides the normal directions to the spherical shell. Alternatively, one may view the spacetime as $\M = \bigcup_\sigma \Sigma_\sigma$, a stack of slices $\Sigma_\sigma$ labeled by a function $\sigma(y)$ on the plane $\NP$. We will use the latter perspective, which offers several advantages.

\begin{figure}[h]
\centering
\includegraphics[width=\textwidth]{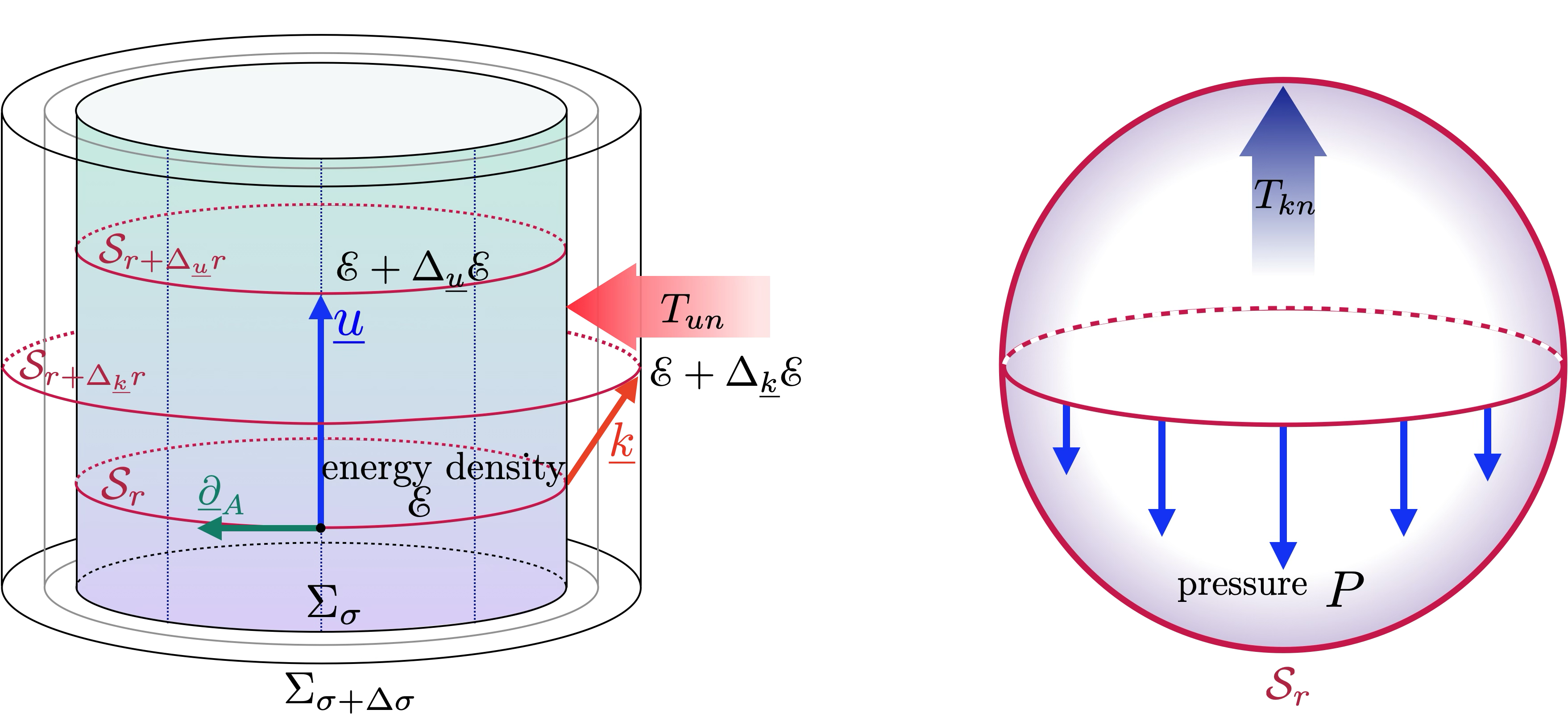} 
\caption{
Spacetime as the worldvolume of a concentric stack of gravitational bubbles.  
\textbf{(Left)} Spherically symmetric spacetime $\M$ is viewed as a stack of spherical slices $\Sigma_{\sigma}$. The tangent space $T\M$ is spanned by $(\vec{\u}, \vec{\k}, \vec{\pa}_A)$, where $\vec{\u}$ lies along $\Sigma_\sigma$, $\vec{\pa}_A$ are tangent to the gravitational bubble on $\S_r$, and $\vec{\k}$ is transverse, connecting one slice $\Sigma_{\sigma}$ to another $\Sigma_{\sigma + \Delta \sigma}$. The radius of $\S_r$ can vary in both directions $\vec{\u}$ and $\vec{\k}$. Each gravitational bubble carries the Misner-Sharp energy density $\E$. Its infinitesimal change along the temporal direction, $\Delta_{\vec{\u}}\E = (\Delta t)\Lie_{\u}\E$, over a time duration $\Delta t$, is governed by the Einstein equation and depends on the matter energy flux $T_{\u\n}$, analogous to the Euler equation for energy. The change $\Delta_{\vec{\k}}\E = (\Delta\sigma)\Lie_{\k}\E$ describes how $\E$ varies in the transverse direction and is likewise determined by the Einstein equation. \textbf{(Right)} On each gravitational bubble on $\S_r$, one can also define the notion of pressure (or surface tension), which equilibrates with the matter pressure $T_{\k\n}$ through the Einstein equation, interpreted as the Young-Laplace equation.
}
\label{fig:summary}
\end{figure}

First, a foliation naturally introduces a frame adapted to it. In this work, we construct such a frame using the rigging technique \cite{Mars:1993mj}. This technique provides a unified mathematical framework for describing the geometry of hypersurfaces of all causal types (spacelike, timelike, and null) and also accommodates situations in which the causal character of the hypersurface changes locally. The foliation-adapted frame consists of basis vectors $(e_{\mathscr{A}}, \vec{\pa}_A)$ and their dual 1-forms $(e^{\mathscr{A}}, \rd z^A)$, where $(\vec{\pa}_A, \rd z^A)$ are coordinate bases on the spherical shell $\S_r$. The frame $e_{\mathscr{A}} = (\vec{\u}, \vec{\k})$ spans the tangent space $T\NP$ to the normal plane $\NP$, while the coframe $e^{\mathscr{A}} = (\k, \n)$ spans its dual space $T^*\NP$. Their components in the coordinates $y^a$ on $\NP$ are $e_{\mathscr{A}}{}^a = (\u^a, \k^a)$ and $e_a{}^{\mathscr{A}} = (\k_a, \n_a)$, obeying the orthogonality conditions: $e_{\mathscr{A}}{}^a e_a{}^{\mathscr{B}} = \delta_{\mathscr{A}}^{\mathscr{B}}$ and $e_a{}^{\mathscr{A}}e_{\mathscr{A}}{}^b = \delta_a^b$. The pair $(\vec{\u}, \k)$ spans the vertical (or temporal, if the hypersurface is timelike) subspace of $\Sigma_\sigma$ and its dual, while the complementary pair $(\vec{\k}, \n)$ provides the transverse directions to $\Sigma_\sigma$. See the left of Fig.\ref{fig:summary}. The vector $\vec{\k}$ is called the rigging and is chosen here, for convenience, to be null. Geometrically, it is the generator of null geodesics passing through $\Sigma_\sigma$. The vertical vector $\vec{\u}$ serves as a generator of $\Sigma_\sigma$, so that $\Sigma_\sigma = \Time_\sigma \times \S_r$, where $\Time_\sigma \subset \NP$ is an integral curve of $\vec{\u}$. The geometry of both the slice $\Sigma_\sigma$ and the ambient spacetime $\M$ can then be constructed using these bases of the adapted frame.

The main advantage of decomposing spacetime into multiple slices, $\M = \bigcup_\sigma \Sigma_\sigma$, is that it naturally reveals its hydrodynamic aspects, as we will show. Each slice $\Sigma_\sigma$ can be regarded as the worldvolume of a gravitational fluid living on the space $\S_r$ and evolving in time along the direction of the vertical vector $\vec{\u}$. The Einstein equation governs this evolution of the fluid variables and additionally constrains how their values vary from one slice to another, which determines the distribution in the radial direction. 

In particular, our main focus is on radial slices $\Sigma_r$ defined by $\sigma(y)=r(y)$, hypersurfaces of constant areal radius of the spherical shell $\S_r$. There are several reasons why radial slices are preferable. First, the areal radius $r$, which is information inherent to spherically symmetric spacetimes, is not a background structure, unlike a time function typically introduced in the ADM formalism \cite{Kuchar:1994zk}, and the radial slice is a natural covariant structure. Second, in the radial slice, $r$ plays the role of an evolution parameter, moving from one $\Sigma_r$ to another, and the Einstein equations can be reformulated to propagate a consistently given boundary data radially, reconstructing the bulk spacetime \cite{Freidel:2008sh, Cianfrani:2013oja}. In our geometro-hydrodynamic context, an equation controls the variation of the hydrodynamic data between slices. Third, the Kodama vector \cite{Kodama:1979vn} provides a preferred temporal direction in general spherically symmetric spacetimes; the Kodama vector defines the Misner-Sharp energy, and the flow generated by the Kodama vector preserves the radius $r$ and sweeps out the constant-$r$ slice $\Sigma_r$. We will show that in the radial slice, the vertical frame $\vec{\u}$ is precisely the Kodama vector, thereby providing a connection between the rigging technique and the Kodama formalism. 

Now, we provide the main result. 
The Einstein equations, when evaluated in the radial frame $\bdx{e}_{\mathscr{A}} = (\vec{\uu}, \vec{\kk})$ adapted to the radial slice $\Sigma_r$, simplify considerably and reveal their hydrodynamic nature, leading to the dictionary between gravity and fluid. By comparing those equations with the hydrodynamic equations, we can find that the energy density and pressure of the gravitational fluid are given by, respectively, 
\begin{equation}
    \E := \frac{\m}{\A} \, , \qquad \text{and} \qquad \P := \frac{1}{8\pi\Newton}\vkappa -  \left(\frac{d-3}{d-2}\right) \E,
\end{equation}
where  $\m$ denotes the Misner-Sharp energy, $\A$ the area of the spherical shell $\S_r$, and $\vkappa$ the acceleration of the Kodama vector $\vec{\uu}$. In particular, the \emp{gravitational pressure} $\P$ 
appears as the surface pressure (tension) 
in an equation with the same form as the Young-Laplace equation \cite{landau1987fluid}.

Remarkably, this correspondence holds in a  general frame $e_{\mathscr{A}} = (\vec{\u},\vec{\k})$ adapted to an arbitrary slice $\Sigma_\sigma$. Indeed, by constructing the transformation rules $\bdx{e}_{\mathscr{A}} \mapsto e_{\mathscr{A}}$ from the radial frame $\bdx{e}_{\mathscr{A}}$ to a general one $ e_{\mathscr{A}}$ and applying them to the Einstein equations in the radial frame, we can derive the Einstein equations in any chosen frame. In particular, the normal components of the Einstein equation, $G_{ab} = 8\pi\Newton T_{ab}$, can be recast as 
\begin{align}\label{summary_eq}
 \left(\Lie_{\u} + \theta_{(\u)}\right)\E + P \theta_{(\u)}= \left( \sigma' \right)^{-1}T_{\u \n}  \, ,  
 \quad P \vtheta =T_{\k \n} \, ,
 \quad \text{and}
 \quad \left(\Lie_{\k} + \theta_{(\k)}\right)\E=-T_{\uu \k} \,,
\end{align}
where the gravitational pressure $P$ in a general frame is given by 
\begin{align*}
P := \P + \frac{\dot{\sigma}}{\sigma'} \bP ~~~{\rm with}~~~\bP := \frac{ \bvkappa }{8\pi\Newton}. 
\end{align*}
Here, $\bvkappa$ is the acceleration of the rigging vector $\vec{\kk}$ associated with the radial slice $\Sigma_r$; the expansion of the area (or the extrinsic curvature of the sphere) is denoted by $\vtheta = \frac{\rd}{\rd r} \ln \A$; $\theta_{(\u)} = \vtheta \u^a \pa_a r$ and $\theta_{(\k)} = \vtheta  \k^a \pa_a r$ represent the area expansions along $\vec{\u}$ and $\vec{\k}$, respectively; and the functions $\dot{\sigma} = \uu^a \pa_a \sigma$ and $\sigma' = \kk^a \pa_a \sigma$ serve as parameters for the frame transformation.

The equations \eqref{summary_eq} provide a geometro-hydrodynamic interpretation in a general foliation-adapted frame. 
See Fig.\ref{fig:summary}. 
The first equation takes exactly the same form as the Euler equation for energy \cite{landau1987fluid}, governing the evolution of the energy density $\E$ along the fluid vector $\vec{\u}$ under the external energy source $\left( \sigma' \right)^{-1}T_{\u \n}$. The second one is the Young-Laplace equation, 
where the product of the gravitational pressure $P$ (or surface tension $-P$) and the extrinsic curvature $\vartheta$ balances the radial matter pressure $T_{\k \n}$. The fact that $\E$ and $P$ appear consistently in both equations supports their interpretation strongly. 
Here, the Euler equation for velocity field does not appear because, due to spherical symmetry, the gravitational fluid does not flow in the tangential direction $\partial_A$. Therefore, this gravitational fluid behaves as a spherical collective mode (which we call \emp{"gravitational bubble"}\footnote{While the term "gravitational bubble" is also used in other contexts (e.g., \cite{Hawking:1979pi,Berezin:2014dma}), there is no direct connection to our idea.}) with $\E$ and $P$, merely contracting or expanding. The bulk space can be regarded as a concentric stack of gravitational bubbles. Here, the third equation comes into play; it connects gravitational bubbles by determining how the energy density $\E$ is distributed across different slices $\Sigma_\sigma$ along the null geodesic generated by $\vec{\k}$, although its counterpart does not exist in the standard hydrodynamics.\footnote{Note that the equation for the tangent direction, $G^A{}_A=8\pi\Newton T^A{}_A$, automatically holds via the Bianchi identity, $\nabla_\mu G^{\mu\nu}=0$, and the energy-momentum conservation, $\nabla_\mu T^{\mu\nu}=0$, if the other equations are satisfied.} 
Note that the non-trivial dynamics in spherically symmetric spacetime requires some deviation from the pure-gravity Einstein equations (Birkhoff's theorem and \cite{Livine:2025soz}), which should be reflected in the geometro-hydrodynamics.

Furthermore, to examine the robustness of this correspondence, particularly whether its validity depends on the details of gravitational dynamics, we perform a similar analysis for Lovelock gravity and find that a generalized geometro-hydrodynamics holds.

Thus, (a part of) the dynamics of spherically symmetric spacetime shares the same structure as hydrodynamic equations, providing an evidence for the geometro-hydrodynamics that describes the spacetime as the worldvolume of a stack of spherical collective modes (gravitational bubbles) of its (unknown) microscopic constituents possessing the energy density $\E$ and pressure $P$. 
This holds universally for any spherical slice in general spherically symmetric spacetime, any associated frame, and different types of gravitational dynamics. Our formulation provides a fundamental framework for exploring the thermodynamics of spacetime from a dynamics perspective.

This article is organized as follows. In Sec.\ref{sec:SSS}, we review the standard geometry of spherically symmetric spacetimes and introduce the Kodama vector, Misner-Sharp energy, and gravitational potential. In Sec.\ref{sec:slice}, we examine the geometry from the perspective of foliations by spherical slices using the rigging technique. We also discuss the radial slices and its connection with the Kodama formalism. In Sec.\ref{sec:Hydro}, we derive the geometro-hydrodynamic equations in an arbitrary frame by utilizing the frame transformations. In Sec.\ref{sec:AH}, we apply our framework to apparent horizons and derive their geometro-hydrodynamic equations and some interesting mathematical results. In Sec.\ref{sec:LL}, we extend the analysis to Lovelock gravity. Finally, we summarize and conclude in Sec.\ref{sec:conclusion}.

\paragraph{Notations:} The notations we will use here are listed below:
\begin{itemize}[itemsep=1ex, parsep=0pt, topsep=0pt]
\item Greek letters $\mu,\nu,\rho,...$ are spacetime indices and raised/lowered by a spacetime metric $g_{\mu \nu}$ and its inverse $g^{\mu \nu}$
\item Small Latin letters $a,b,c,...$ are indices on the normal plane $\NP$ and raised/lowered by a normal metric $h_{ab}$ and its inverse $h^{ab}$.
\item The capital Latin letters $A,B,C,...$ are indices on a $(d-2)$-dimensional sphere, $\S_r$, and raised/lowered by a sphere metric $q_{AB} = g_{AB}$ and its inverse $q^{AB}$.
\item The script capital Latin letters $\mathscr{A},\mathscr{B},\mathscr{C},...$ are indices for the frame on $T\NP$.
\item Vectors are presented with underlined letters such as $\vec{u}, \vec{k}, ....$
\item The Levi-Civita connections on $\M$ and $\NP$ are denoted by $\nabla_\mu$ and $\DN_a$, respectively. The Lie derivative along a vector $\vec{V}$ is denoted by $\Lie_{V}$. 
\item The frames adapted to different foliations are denoted by distinct letter styles. See Tab.\ref{tab:frames}.
\end{itemize}


\section{Spherically Symmetric Spacetimes} \lb{sec:SSS}

We begin with a brief review of spherically symmetric spacetimes and their properties, including the definitions of the Misner-Sharp energy and the Kodama vector, and we set up the notation that will be used throughout this article.

We consider a $d$-dimensional spherically symmetric spacetime $\M$ endowed with a Lorentzian metric $g = g_{\mu\nu} \exd x^\mu \otimes \exd x^\nu$ and a Levi-Civita connection $\nabla_\mu$. We work with coordinates $x^\mu = (y^a, z^A)$ in which the metric components exhibit a block-diagonal form, meaning the spacetime is a warped product manifold\footnote{A warped product manifold $\M = \NP \times_r \mr{\S}$, where $r: \NP \to \mathbb{R}$ is a warping function, has a fiber bundle structure, $\pi_\NP: \M \to \NP$ and $\pi_{\mr{\S}}: \M \to \mr{\S}$. Its metric is given by $g = \pi_\NP^* (h) + (r \circ \pi_\NP)^2 \pi_{\mr{S}}^*(\mr{q})$, where $h$ and $\mr{q}$ are metrics on $\NP$ and $\mr{\S}$, respectively.} \cite{o1983semi, Carot1999},
\begin{align}
g_{\mu \nu}\exd x^\mu \otimes \exd x^\nu = h_{ab} \exd y^a \otimes \exd y^b + r^2 \mr{q}_{AB} \exd z^A \otimes \exd z^B. \lb{metric}
\end{align}
Here, we denote by $\mr{q}_{AB}(z)$ the components of a metric on a unit round $(d-2)$-sphere $\mr{\S}$ in angular coordinates $(z^A)$. Attached uniformly to every point on $\mr{\S}$ is a normal 2-plane, also called the temporal-radial plane $\NP$, equipped with a Lorentzian metric $h = h_{ab} \exd y^a \otimes \exd y^b$ in general coordinates $(y^a)$ and a Levi-Civita connection $\DN_a$. A scalar field $r(y)$ plays the role of an \emp{areal radius} and labels a family of round spheres $\S_r$, also called \emp{spherical shells}, which are homothetic to $\mr{\S}$ and are equipped with an induced metric $q = q_{AB} \exd z^A \otimes \exd z^B$, with components given by $q_{AB} = r^2 \mr{q}_{AB}$. 

In this $(2+(d-2))$-decomposition of the metric \eqref{metric}, the spacetime volume element is given by $\sqrt{-g} = r^{d-2} \sqrt{- h} \sqrt{\mr{q}}$. The canonical volume form on the spacetime is $\volM = \volN \wedge \volS$. Here, $\volN$ is the volume form on the normal $\NP$, and $\volS = r^{d-2} \volSR$, where $\volSR$ denotes the volume form on the unit round sphere $\mr{\S}$, is the volume form on the spherical shell $\S_r$. The area element of the spherical shell $\S_r$ is $\sqrt{q} = r^{d-2} \sqrt{\mr{q}}$ and its area is given by
\begin{align}
\A  = \int_{\S_r} \volS = \int_{\S_r} \sqrt{ q} \ \exd^{d-2} z = r^{d-2}\int_{\mr{\S}} \sqrt{\mr{q}} \ \exd^{d-2} z  =\mr{\A} r^{d-2}, \lb{area}
\end{align}
where $\mr{\A} = \int_{\mr{\S}} \volSR$ denotes the area of the unit round sphere\footnote{The area of a unit round sphere in $d-2$ dimension is $\mr{\A} = \frac{2 \pi^{\frac{d-1}{2}}}{\Gamma(\frac{d-1}{2})}$, where $\Gamma$ is the Gamma function.}. For instance, we have $\mr{\A} = 4\pi$ in four dimensions. In addition, the \emp{area expansion} is defined as 
\begin{align}
\vtheta  := \frac{1}{\A}\frac{\exd}{\exd r}\A = \frac{d-2}{r}. \lb{expansion}
\end{align}
It should be noted that, for a round sphere, the area expansion $\vtheta$ coincides with its total mean curvature, defined as the sum of principal curvatures, and corresponds precisely to the trace of the second fundamental form when the sphere is viewed as embedded in flat ambient space $\mathbb{R}^{d-1}$. This interpretation of $\vtheta$ as an extrinsic curvature-related quantity becomes crucial when we later reinterpret spacetime as a soap bubble governed by a gravitational analog of the Young-Laplace equation. \\

In this setting, any vector field $X = X^\mu \pa_\mu$ on the spacetime $\M$ can be decomposed into its normal and angular components as $X^\mu = (X^a, X^A)$. Similarly, any 1-form $\form{\omega} = \omega_\mu \exd x^\mu$ decomposes as $\omega_\mu = (\omega_a, \omega_A)$. Owing to spherical symmetry, we will, in most cases, work with normal tensors whose angular components vanish, so that they lie entirely within the normal plane $\NP$.

We should remark that, even though the geometry of $\mr{\S}$ is fixed to that of a round sphere, the geometry of the full spacetime $\M$ remains dynamical and is governed by the Einstein equation. Its evolution is fully encoded in the geometry of the normal plane $\NP$ and the dynamics of the areal radius $r$. In this sense, the study of spherically symmetric spacetimes effectively reduces to a two-dimensional problem:\footnote{Note that even in spherically symmetric spacetime, quantum fluctuations must be considered in $d$ dimensions. For example, the UV divergence structure differs between two dimensions and four dimensions, leading to entirely different dynamics \cite{Birrell:1982ix,Kawai:2020rmt}.} the geometry of the normal plane $\NP$, coupled non-trivially to the scalar field $r$. This dimensional reduction is also the foundation of models such as Callan-Giddings-Harvey-Strominger (CGHS) \cite{Callan:1992rs} and Jackiw-Teitelboim (JT) gravity \cite{TEITELBOIM198341,JACKIW1985343}.

More details of the spherically symmetric geometry, such as curvatures, are reported in Appendix \ref{app:geometry}.  

\subsection{Time translation à la Kodama}\label{sec:Kodama}

In static spacetimes, the time translation symmetry is tied to the existence of a global timelike Killing vector. However, such an isometry is absent in dynamical spacetimes, which obstructs a universal definition of energy in gravitational systems.

For spacetimes with spherical symmetry in general relativity, there exists a special vector field called the \emp{Kodama vector} \cite{Kodama:1979vn}, defined as
\begin{align}
\kv^a := (\epsilon_{\sss \NP})^{ba}\DN_b r, \qquad \text{satisfying} \qquad \kv^a \DN_a r = 0 , \qquad \text{and} \qquad \DN_a \kv^a =0, \lb{Kodama-vector}
\end{align}
where $(\epsilon_{\sss \NP})^{ba}$ is the antisymmetric Levi-Civita tensor\footnote{The convention is, provided an antisymmetric symbol $\tilde{\epsilon}_{ab}$ with the condition $\tilde{\epsilon}_{01} =1$, the covariant tensor $(\epsilon_{\sss \NP})_{ab}$ is defined as $(\epsilon_{\sss \NP})_{ab} := \sqrt{-\det h}\tilde{\epsilon}_{ab}$, while the contravariant form is $(\epsilon_{\sss \NP})^{ab} = -\frac{1}{\sqrt{-\det h}}\tilde{\epsilon}^{ab}$ with $\tilde{\epsilon}^{01} =1$. The contraction formula is $(\epsilon_{\sss \NP})^{ac}(\epsilon_{\sss \NP})_{cb} = \delta^a_b$.} on the normal plane, whose components coincide with the components of the volume form $\volN$ with indices raised. The property $\kv^a \DN_a r = 0$ indicates that the vector $\kv^a$ is tangent to a hypersurface of constant areal radius. The lift of the Kodama vector to the spacetime, $\kv^\mu = (\kv^a,0)$, is also divergence-free; that is, $\nabla_\mu \kv^\mu = \DN_a \kv^a = 0$ (see the relation between spacetime and normal covariant derivatives in Appendix \ref{app:geometry}). Notice that the norm of the Kodama vector\footnote{In this article, we adopt the notations $\DN^2 r = \DN_a \DN^a r$ and $(\DN r)^2 = \DN_a r \DN^a r$.}, $\kv_a \kv^a = - (\DN r)^2$, can be spacelike, timelike, or null.

What makes the Kodama vector special is that it provides a preferred time direction to the dynamical spacetime, which can be viewed as a generalization of a global time translation vector field for the static case \cite{Abreu:2010ru}. Furthermore, it defines the conserved \emp{Kodama current} for spherically symmetric spacetime (we review the derivation in Appendix \ref{app:Kodama-derive}),
\begin{align}
J^a := \frac{1}{8\pi\Newton}G^a{}_b \kv^b,\qquad \text{satisfying} \qquad \nabla_\mu J^\mu = 0, \lb{Kodama-current}
\end{align}
where $G_{ab}$ are the normal components of the spacetime Einstein tensor, and $\Newton$ is the Newtonian gravitational constant in $d$ dimensions. It is important to note that this conservation laws is an "off-shell" identity, stemming from the warped product structure \eqref{metric} of the spherically symmetric spacetime (this point has been emphasized in \cite{Abreu:2010ru}). 

The corresponding charge is given by the integral of the current on a codimension-1 surface $\Sigma$. This charge, denoted by $\m = - \int_{\Sigma} J^a \rd \Sigma_a$, is the \emp{Misner-Sharp energy} (or mass) defined as (see also Appendix \ref{app:Kodama-derive})\footnote{It is worth mentioning that one can consider $\mr{\S}$ to be any maximally symmetric $(d-2)$-dimensional Riemannian space whose constant curvature is given by $\lambda = -1,0,1$. The Misner-Sharp energy is then,
\begin{align*}
\m_{\sss (\lambda)} := \frac{d-2}{16\pi\Newton} \mr{\A} r^{d-3} \left(\lambda - (\DN r)^2\right). 
\end{align*}
Here, we only focus on the spherically symmetric case where $\lambda=1$. }
\begin{align}
\m(y) := \frac{d-2}{16\pi\Newton} \mr{\A} r^{d-3} \left(1 - (\DN r)^2\right) = \frac{\vtheta \A}{16\pi\Newton} \left(1 - (\DN r)^2\right), \lb{Misner-Sharp}
\end{align}
where the second equality followed from \eqref{area} and \eqref{expansion}. This quasi-local energy measures the energy contained within the spacetime region enclosed by the spherical shell $\S_r$ (see also \cite{Misner:1973prb}). Indeed, for the vacuum and static case, the Misner-Sharp energy reduces to the Schwarzschild mass. Furthermore, when evaluated at null and spatial infinity in an asymptotically-flat spacetime, $\m$ coincides with the Bondi mass $M_B$ and Arnowitt-Deser-Misner (ADM) mass $M$, respectively \cite{Hayward:1994bu,Misner:1973prb}. We should note that $\m$ does not only capture such mass of the system but can also depend on other physical quantities. For instance, in the case of a Reissner-Nordström black hole, the Misner-Sharp energy $\m$ depends on both the ADM mass $M$ and the electric charge $Q$.

In addition, let us define the \emp{gravitational potential} as
\begin{align}
\Phi(y) := - \frac{1}{(d-2)}\frac{8\pi\Newton}{\mr{\A}} \frac{\m}{r^{d-3}}, \qquad \text{or equivalently} \qquad (\DN r)^2 = 1 + 2\Phi. \label{Npotential}
\end{align}
It is the analog of gravitational potential in Newtonian mechanics. Indeed, in four dimensions, we recover $\Phi \stackrel{d=4}{=} - \frac{\Newton \m}{r}$. 

It is useful to define the \emp{Misner-Sharp energy density} as
\begin{align}
\E := \frac{\m}{\A} =  -\frac{1}{8\pi\Newton}\vtheta \Phi \lb{Edensity}, 
\end{align}
which will be often used below.

The Kodama vector becomes locally null when $(\DN r)^2 = 0$, which corresponds to the condition $\Phi = -\frac{1}{2}$. The locus of points satisfying this condition defines a codimension-1 hypersurface in the spacetime,
\begin{align}
\H = \bigg\{ (y^a_{\sss \H}, z^A) \in \M \ \bigg| \  r(y_{\sss \H})^{d-3} = \frac{1}{(d-2)}\frac{16\pi\Newton}{\mr{\A}} \m(y_{\sss \H})\bigg\}, \label{AH-def}
\end{align}
referred to as the \emp{apparent horizon}. This definition is consistent with the standard geometric characterization of the apparent horizon as the hypersurface on which the expansion of outgoing null rays is zero (see Sec.\ref{sec:AH}). We should also note that the apparent horizon can be timelike, spacelike, or null. We will elaborate more on its geometry and discuss its geometro-hydrodynamic interpretation in Sec.\ref{sec:AH}. 



\section{Spherical Slices: Geometry \& Dynamics} \lb{sec:slice}

In the previous section, we have discussed the geometry of spacetimes with spherical symmetry based on the natural splitting of the spacetime into the $(d-2)$-dimensional spherical shells and the normal 2-plane. In addition, we have introduced the Kodama vector that generates a preferred flow of time associated with the Misner-Sharp energy, playing the role of a time translation generator even when no timelike Killing vector exists. 

An alternative way to discuss the spacetime geometry is to adopt the $(1+(d-1))$-decomposition picture in which the whole spacetime (or its portion) is now viewed as stacks of $(d-1)$-dimensional hypersurfaces, or \emp{slices}. To this end, we consider a foliation of the spacetime $\M$ into a family of codimension-1 spherical slices. Thanks to spherical symmetry, such a surface can be specified by a leaf of a foliation function $\sigma(y)$ on the normal plane $\NP$. More concretely, the spherical slice $\Sigma_\sigma$ admits the product structure: $\Sigma_\sigma = \{ (y_{\star}^a,z^A) | \sigma(y_{\star}) = \mathrm{constant} \} = \Time_\sigma \times \S_{r(y_{\star})}$, where $\Time_\sigma \subseteq \mathbb{R}$ is a curve on $\NP$ at which $\sigma(y) = \mathrm{constant}$. In this construction, $\Time_\sigma$ provides the \emp{vertical direction} that complements the \emp{horizontal directions} (or angular directions) tangential to the spherical shell $\S_r$. Moreover, the \emp{transverse direction} to the slice is the direction in which the value of $\sigma(y)$ changes. The bulk spacetime $\M$ is a stack of these slices, $\bigcup_\sigma \Sigma_\sigma \subseteq \M$ and, similarly, the normal plane $\NP$ is a union of the vertical lines, $\bigcup_\sigma \Time_\sigma \subseteq \NP$. 

In principle, we can choose any general foliation function $\sigma(y)$. In practice, however, there is a more suitable $\sigma(y)$ depending on the physical situation at hand. For spherically symmetric spacetime, the areal radius $r$ serves as a natural foliation function. As we have already seen, the vertical space $\Time_r$ (for $\sigma (y) = r (y) = \mathrm{constant}$) is generated by the flow of the Kodama vector $\kv^a$ \eqref{Kodama-vector}, whose norm is expressible in terms of the Misner-Sharp energy \eqref{Misner-Sharp}. Here, we will refer to the constant-$r$ hypersurfaces as \emp{radial slices}. Another physically interesting surface is the apparent horizon, which belongs to a family of \emp{equipotential slices} where the gravitational potential \eqref{Npotential} $\Phi$ is constant. 

To construct the geometry of a slice, we utilize the \emp{rigging technique}, first developed by Mars and Senovilla \cite{Mars:1993mj, Mars:2013qaa} and recently explored by one of the authors in the context of Carrollian geometry of stretched horizons \cite{Freidel:2022vjq,Freidel:2024emv} (see also the more complete discussion in the review \cite{Ciambelli:2025unn}). An advantage of this technique is that it is universally applicable for hypersurfaces of different causal types (timelike, null, and spacelike). It also permits circumstances where the surface's causal character varies from point to point, a useful feature when considering the apparent horizon. We will proceed by, first, presenting the rigging framework for a general spherical surface $\Sigma_\sigma$ (Sec.\ref{sec:rigging}), and then specializing to the case of radial slices, where many geometric expressions simplify and a hydrodynamic description of spacetime dynamics naturally emerges (Sec.\ref{sec:slices}).

\subsection{Spherical slices \& Rigging technique}
\lb{sec:rigging}
We now explain the construction of the geometry of the spherical slice based on the rigging technique. As explained, we treat a $(d-1)$-dimensional surface $\Sigma_\sigma$ embedded in the spacetime $\M$ as a leaf of the foliation function $\sigma (y) = \mathrm{constant}$, and it factorizes into the vertical space $\Time_\sigma$ and the spherical shell $\S_r$ as $\Sigma_\sigma = \Time_\sigma \times \S_{r}$.

The rigging technique starts from supplementing $\Sigma_\sigma$ with a \emp{null rigging structure}, comprising a doublet $\left(\form{\n}, \vec{\k}\right)$. Here, $\form{\n} = \n_\mu \exd x^\mu$ is a normal 1-form to $\Sigma_\sigma$, and a null rigging vector $\vec{\k} = \k^\mu \pa_\mu$ is transverse to $\Sigma_\sigma$ and tangent-bundle dual
\footnote{Here, a vector $\vec{X}$ and a 1-form $\omega$ is tangent-bundle dual if their interior product is $\iota_{\vec{X}} \omega = X^\mu \omega_\mu = 1$. This does not mean that they are metric-dual, $\omega_\mu \neq g_{\mu\nu}X^\nu$.}
to $\form{\n}$, thereby satisfying $\n_\mu \k^\mu = 1$ and $g(\vec{\k},\vec{\k}) = \k_\mu \k^\mu =0$. The norm squared of the normal form is $2\rho :=  g^{-1}(\form{\n}, \form{\n})$, whose sign locally determines the causal property of $\Sigma_\sigma$. In the following, we will refer to the function $\rho$ as the normalization factor of the normal 1-form.

Because of spherical symmetry, both $\form{\n}$ and $\vec{\k}$ are "normal" tensors that lie on the normal plane $\NP$, meaning that $\n_\mu = (\n_a, 0)$ and $\k^\mu = (\k^a,0)$. The rigging structure provides a transverse direction to $\Sigma_\sigma$. For instance, a vector $\vec{X} =X^\mu \pa_\mu$ and a 1-form $\form{\omega} = \omega_\mu \exd x^\mu$ are tangent to $\Sigma_\sigma$ if they obey $X^\mu \n_\mu = 0$ and $\k^\mu \omega_\mu =0$, respectively. Furthermore, spacetime tensors are projected onto $\Sigma_\sigma$ using the projector, $\Pi_\nu{}^\mu :=  \delta_\nu^\mu - \n_\nu \k^\mu$, which satisfies $\k^\nu\Pi_\nu{}^\mu =0$ and $\Pi_\nu{}^\mu \n_\mu =0$. The projector also admits the block-diagonal splitting as $\Pi_\mu{}^\nu = (\Pi_a{}^b, \Pi_A{}^B)$, where the normal components are $\Pi_a{}^b= \delta_a^b - \n_a \k^b$, while the angular components are $\Pi_A{}^B = \delta_A^B$. \\

Following the construction in \cite{Freidel:2022vjq}, given the spacetime metric and its inverse, an intrinsic geometry of $\Sigma_\sigma$ can be induced from the rigging structure. More precisely, we define a \emp{vertical vector} $\vec{\u} = \u^\mu \pa_\mu$ and its dual 1-form $\form{\k} = \k_\mu \exd x^\mu$, whose components are given by 
\begin{align}
\u^\mu := \n^\nu\Pi_\nu{}^\mu =  g^{\mu\nu}\n_\nu - 2\rho \k^\mu, \qquad \text{and} \qquad \k_\mu := \Pi_\mu{}^\nu \k_\nu = g_{\mu\nu} \k^\nu. \lb{vector-u}
\end{align}
It then follows by definitions that $\u^\mu \n_\mu =0 = \k^\mu \k_\mu$ and $\u^\mu \k_\mu =1$, implying that both $\vec{\u}$ and $\form{\k}$ are tangent to $\Sigma_\sigma$ and are tangent-bundle dual. Again, due to spherical symmetry, $\u^\mu = (\u^a, 0)$ and $\k_\mu = (\k_a,0)$ only have components on the normal plane. The flow of the vertical vector $\vec{\u}$ generates the vertical space $\Time_\sigma$, hence it spans a 1-dimensional vertical subspace of the tangent space $T\Sigma_\sigma$, while the dual $\k$ spans its dual space. By construction, its norm squared is $g (\vec{\u}, \vec{\u}) = - 2\rho$. 

When the normalization factor $\rho$ vanishes, the slice $\Sigma_\sigma$ becomes null. However, this only holds locally, as $\rho$ can be non-zero at another point on the surface. The surface is null entirely if and only if $\rho$ vanishes everywhere on the surface, implying the conditions $\rho \stackrel{\sss \mathrm{null}}{=} 0$ and $\Lie_{\u}\rho \stackrel{\sss \mathrm{null}}{=} 0$. This is the case for black holes' event horizons in non-dynamical backgrounds. We will see in Sec.\ref{sec:AH} that, the apparent horizon, nonetheless, allows for the local variation of its casual nature, which is determined by $\mathrm{sign}(\rho)$, and is controlled by the Einstein equation. \\

The tangent space $T\Sigma_\sigma$ and the cotangent space $T^*\Sigma_\sigma$ split into the vertical component, spanned by $(\vec{\u},\form{\k})$, and the angular (or tangential) components along the spherical shells $\S_r$, spanned by $(\vec{\pa}_A, \rd z^A)$. This infers that the projector to the surface decomposes as $\Pi_\nu{}^\mu = q_\nu{}^\mu + \k_\nu \u^\mu$, where $q_\nu{}^\mu$ is a projector onto the angular components on $\S_r$, satisfying $q_\mu{}^\alpha q_\alpha{}^\nu = q_\mu{}^\nu$ and $q_A{}^\mu q_B{}^\nu g_{\mu \nu} = q_{AB}$ is the metric on $\S_r$. It then follows that the non-zero components are $q_A{}^B = \delta_A^B$. 
Then, by equating $\Pi_\nu{}^\mu = \delta_\nu^\mu - \n_\nu \k^\mu$ and $\Pi_\nu{}^\mu = q_\nu{}^\mu + \k_\nu \u^\mu$, we obtain the following decomposition,
\begin{align}
\delta_\nu^\mu = \k_\nu \u^\mu + \n_\nu \k^\mu+ q_\nu{}^\mu. \lb{identity}
\end{align}
In addition, it implies that the spacetime metric \eqref{metric} and its inverse decompose as 
\begin{equation}
\lb{metric-gen}
\begin{aligned}
g_{\mu\nu} &= \k_\mu \u_\nu + \n_\mu \k_\nu +q_{\mu\nu} = -2\rho\k_\mu \k_\nu + 2 \n_{(\mu} \k_{\nu)} + q_{\mu\nu}, \\
g^{\mu\nu} &= \k^\mu \u^\nu + \n^\mu \k^\nu + q^{\mu\nu} = 2\rho \k^\mu \k^\nu + 2 \u^{(\mu} \k^{\nu)} + q^{\mu\nu}. 
\end{aligned}
\end{equation}
Hence, the metric on the normal plane $\NP$ and its inverse decompose as
\begin{equation}
\label{h_h}
\begin{aligned}
h_{\mu\nu} = k_\mu \u_\nu + \n_\mu \k_\nu &= -2\rho\k_\mu \k_\nu + 2 \n_{(\mu} \k_{\nu)} \,,\nonumber\\
h^{\mu\nu} = \k^\mu \u^\nu + \n^\mu \k^\nu &= 2\rho \k^\mu \k^\nu + 2 \u^{(\mu} \k^{\nu)}.
\end{aligned}
\end{equation}

The vertical basis $(\vec{\u}, \form{\k})$ and the angular basis $(\vec{\pa}_A, \exd z^A)$ span the tangent and cotangent spaces to the spherical slice $\Sigma_\sigma$. Combined with the transverse basis $(\vec{\k}, \form{\n})$, they form a complete basis for the tangent and cotangent spaces for the spacetime $\M$. In this work we will refer to the set of the basis vectors and the basis 1-forms
\begin{align}
e_{\mathscr{A}} = (\vec{\u},\vec{\k})\, , \qquad \text{and} \qquad e^{\mathscr{A}} = (\k,\n)
\end{align}
on the normal plane $\NP$ as the \emp{foliation-adapted frame} (or just frame for short). In components, we have $e_{\mathscr{A}}{}^a = (\u^a,\k^a)$ and $e_a{}^{\mathscr{A}} = (\k_a,\n_a)$ that obey\footnote{The contraction $e_\mu{}^{\mathscr{A}} e_{\mathscr{A}}{}^\nu = \delta_\mu^\nu - q_\mu{}^\nu = h_\mu{}^\nu$, following directly from \eqref{identity}, provides a projection from $T\M$ to $T\NP$. Certainly, $e_\mu{}^{\mathscr{A}} e_{\mathscr{A}}{}_\nu = h_{\mu\nu}$ is the lift of the normal-plane metric, $h_{\mu\nu} = (h_{ab},0)$.}
\begin{align}
e_{\mathscr{A}}{}^a e_a{}^{\mathscr{B}} = \delta_{\mathscr{A}}^{\mathscr{B}} \, , \qquad \text{and} \qquad e_a{}^{\mathscr{A}} e_{\mathscr{A}}{}^b = \n_a \k^b + \k_a \u^b = \delta_a^b \, ,
\end{align}
where the second condition follows from the completeness relation \eqref{identity}. 
\bbox
{\bf Foliation-adapted frame:} We summarize that the frame fields and the coframe fields satisfy the pairing conditions,
\begin{align}
\k^a \n_a = 1,  \ \ \ \ \k^a \k_a = 0, \ \ \ \ \u^a \k_a = 1,\ \ \ \ \text{and} \ \ \ \ \u^a \n_a =0,  \lb{pairing}
\end{align}
whereas the normalization function $\rho$ controls the norm squared,
\begin{align}
\n_a \n^a =  -\u^a \u_a = 2\rho. \lb{norm}
\end{align}
\ebox

A vector $\vec{X} = X^a \pa_a \in T\NP$ and a 1-form $\omega = \omega_a \exd y^a \in T^*\NP$ can be decomposed in the frame as 
\begin{align}
\vec{X} = X^a \delta_a^b \pa_b = (X^a e_a{}^{\mathscr{A}}) e_{\mathscr{A}} \, ,\qquad \text{and} \qquad \omega = \omega_a \delta_b^a \rd y^b = (e_{\mathscr{A}}{}^a\omega_a) e^{\mathscr{A}}.  
\end{align}
The components $X^a e_a{}^{\mathscr{A}}$ and $e_{\mathscr{A}}{}^a\omega_a$ are the frame-dressed vector and the 1-form, respectively. In terms of the (co)frame fields, we explicitly have 
\begin{align}
\vec{X} = (X^a \k_a) \vec{\u} + (X^a \n_a) \vec{\k}, \qquad \text{and} \qquad \omega = (\u^a \omega_a) \k + (\k^a \omega_a) \n, \lb{decomposition}
\end{align}
The differential of a function $F(y)$ on the normal-plane $\NP$ is given by 
\begin{align}
\exd F = (\e_{\mathscr{A}}{}^a \pa_a F)e^{\mathscr{A}}  = \left( \u^a\pa_a F \right) \form{\k} + \left( \k^a \pa_a F \right) \form{\n} = \left( \Lie_{\u} F \right) \form{\k} + \left( \Lie_{\k} F \right) \form{\n} \, , 
\label{df}
\end{align}
where we recall that $\Lie_{V}$ denotes the Lie derivative along a vector $\vec{V}$. \\

\noindent Let us provide some remarks: 

\begin{itemize}
\item Using the projector, we can define the induced rigged metric $g_{\sss \Sigma}$ on the slice $\Sigma_\sigma$, whose components are $(g_{\sss \Sigma})_{\mu\nu} := \Pi_\mu{}^\alpha \Pi_\nu{}^\beta g_{\alpha \beta} = -2\rho \k_\mu\k_\nu + q_{\mu\nu}$. The objects $(\vec{\u}, \k, g_{\sss \Sigma}, \rho)$ form what is called a stretched Carrollian structure of $\Sigma_\sigma$ \cite{Freidel:2022vjq, Freidel:2024emv}, which is considered a generalization of a Carrollian structure. The latter characterizes the geometry of a spacetime in the vanishing speed of light limit \cite{Leblond1965, SenGupta, Duval:2014uoa, Ciambelli:2019lap} and is an intrinsic geometry of a generic null hypersurface. The vertical vector $\vec{\u}$ is the Carrollian vector, the 1-form $\k$ is the Ehresmann connection, and the scalar $\rho$ is called the stretching of $\Sigma_\sigma$, which vanishes when $\Sigma_\sigma$ is null.

\item The frame $e_{\mathscr{A}} = \left( \vec{\u}, \vec{\k}\right)$ and its dual coframe $e^{\mathscr{A}} = \left(\n,\k\right)$, satisfying the conditions \eqref{pairing}, are defined up to internal local rescaling $\left(\n, \k, \vec{\u}, \vec{\k} \right) \to \left( \e^{\lambda}\n,  \e^{-\lambda}\k,  \e^{\lambda}\vec{\u}, \e^{-\lambda}\vec{\k} \right)$, for an arbitrary function $\lambda (y)$ on the plane $\NP$. Since the metric $g$ is invariant under the rescaling, the rescaling is considered a gauge transformatiion. Physical quantities constructed from the metric hence need to be scale-invariant. 

To define such scale-invariant quantities, we need to take care of the scaling ambiguity of the (co)frame fields. The simplest way to resolve the ambiguity, and the one we will use, is to demand that the normal form is closed, $\form{\exd \n} =0$, which also implies that $\form{\n} = \form{\exd} \sigma$. Goemetrically, this means the slice $\Sigma_\sigma$ is an integrable submanifold of the spacetime $\M$, whose normal form satisfying the Frobenius integrability condition, $\n \wedge \rd \n =0$. In general, we have $\rd n = a \wedge \n$ for a 1-form $a$ for an integral submanifold. Here, for simplicity, we fix $a =0$ and we have $\rd \n =0$.   

It is also worth mentioning that the other direction is to define the scale-invariant basis $\left(\e^{-\ba}\n, \e^{\ba} \k, \e^{-\ba}\vec{\u},\e^{\ba}\vec{\k} \right)$ by introducing an "edge mode" field $\ba(y)$ that transforms as $\ba \to \ba + \lambda$ under the rescaling \cite{Donnelly:2016auv}. In the context of the finite-distance null surfaces, the rescaling symmetry leads, from the Noether theorem, to the surface charge being the area of the sphere \cite{Adami:2021nnf, Freidel:2021fxf, Ciambelli:2023mir, Freidel:2024emv}. In this work, however, we will not consider the internal rescaling symmetry, and assume that the scales of the (co)frame fields are fixed by $\exd \form{\n} =0$.

\item While we primarily work in a coordinate-independent framework, it is sometimes useful to discuss coordinates on the normal plane $\NP$. For general coordinates $y^a = (y^0, y^1)$ on $\NP$, we can parameterize the frame and coframe fields as follows:
\begin{equation}
\lb{vector-form}
\begin{alignedat}{6}
&\text{\it Normal form}  \ \ \ &&\form{\n} &&= J\left( \exd y^1 - V \exd y^0 \right)  \qquad  &&\text{\it Vertical frame} \ \ \  && \vec{\u} &&= \e^{-\alpha}\left(\pa_0 + V\pa_1 \right) \\
&\text{\it Vertical coframe} \ \ \ && \form{\k} &&= \e^{\alpha}\left( \exd y^0 - \beta \form{\n} \right)  \qquad &&\text{\it Rigging vector}  \ \ \ && \vec{\k} &&= J^{-1}\pa_1 + \beta \e^{\alpha} \vec{\u}  
\end{alignedat}
\end{equation}
where $J, V, \alpha$, and $\beta$ are functions on $\NP$, subjected to the condition $\pa_0 J + \pa_1 (JV) = 0$ that stems from $\form{\exd \n} = 0$. The spacetime metric components \eqref{metric} given in this coordinate system are combinations of these variables. It is also possible to adopt a comoving coordinate system in which $J=1$ and $V=0$, and $\sigma$ is now an adapted coordinate. Denoting the comoving coordinates with $y^a = (t,\sigma)$, the spacetime line element is given from \eqref{metric-gen} by
\begin{align}
\rd s^2 = -2\rho \e^{2\alpha}\left( \rd t - \beta \rd \sigma \right)^2 + 2\e^\alpha \left( \rd t - \beta \rd \sigma \right) \rd \sigma + r^2 \rd \Omega_{d-2}^2 \ ,
\end{align}
where $\rd \Omega_{d-2}^2$ denotes the line element of the unit round sphere $\mr{\S}$. 

Furthermore, it is possible to remove $\beta$ through coordinate transformations, $(t,\sigma) \to (v, \sigma)$ such that $\rd v(t,\sigma) = \mu (\rd t - \beta \rd \sigma)$ for a certain integrating factor $\mu (t,\sigma)$. Overall, we can work with the adapted coordinates $y^a = (v,\sigma)$ such that the metric is given by
\begin{align}
\rd s^2 = -2\rho \e^{2\tilde{\alpha}}\rd v^2 + 2\e^{\tilde{\alpha}} \rd v \rd \sigma + r^2 \rd \Omega_{d-2}^2 \,, \lb{EFmetric}
\end{align}
where $\tilde{\alpha} = \alpha - \ln \mu$. 
Let us note that, interestingly, parts of the above discussion are reminiscent of Carrollian manifolds. For instance, the parameterization of the frame and coframe fields \eqref{vector-form} (which also stems from $\exd \n = 0$) coincides with the general parameterization of the ruled Carrollian structure \cite{Freidel:2022bai,Freidel:2022vjq} in two dimensions. Moreover, the coordinate transformations $t \mapsto v(t, \sigma)$ with $\sigma \mapsto \sigma$ are precisely Carrollian transformations. This naturally raises the question of whether one can find a correspondence between the geometry of the normal plane and that of a two-dimensional Carrollian manifold, and whether the full dynamics of spherically symmetric spacetimes can be reinterpreted in terms of Carrollian geometro-hydrodynamics. We leave this question for future work. \\
\end{itemize}

Having described the geometry of a spherically symmetric spacetime in terms of a foliation by spherical slices, we proceed to define the geometric objects that will play a role in the subsequent analysis.

\subsubsection{Expansions}
The expansion rate of the area $\A$ of the spherical shell $\S_r$ along a vector field $\vec{X} \in T\NP$ on the normal plane is defined as
\begin{align}
\theta_{ (X)} := \Lie_{X} \ln \A = \vtheta \Lie_{X} r, \lb{expansion-V}
\end{align}
where we recall the area expansion $\vtheta$ from \eqref{expansion}. The expansion rate is always proportional to the area expansion $\vtheta$ and is determined by the change of the radius along the vector. In addition, one can easily check that $\theta_{ (X)} = q_\mu{}^\nu \nabla_\nu X^\mu = \nabla_A X^A = \Gamma^A_{Aa}X^a$ by computing directly the covariant derivative with the Christoffel symbols (see Appendix \ref{app:geometry}). 

Since any vector field on $\NP$ can be decomposed in terms of the vertical vector $\u^a$ and the rigging vector $\k^a$, its expansion is thus the combination of the vertical expansion $\theta_{ (\u)}$ and the transverse expansion $\theta_{ (\k)}$, where 
\begin{align}
\theta_{ (\u)} := \Lie_{\u} \ln \A = \vtheta \Lie_{\u}  r , \qquad \text{and} \qquad \theta_{ (\k)} := \Lie_{\k} \ln \A = \vtheta \Lie_{\k} r  \label{expansion-basis}.
\end{align}
More precisely, for a vector $\vec{X} \in T\NP$ decomposing as in \eqref{decomposition}, its corresponding expansion is given by
\begin{align}
\theta_{(X)} = X^a e_a{}^{\mathscr{A}} \Lie_{e_{\mathscr{A}}} \ln \A = (X^a \k_a) \theta_{(\u)} + (X^a \n_a) \theta_{(\k)} \, . 
\end{align}

\subsubsection{Accelerations}

While we can impose that the exterior derivative of the normal form $\form{\n}$ is closed as stemming from the foliation, the exterior derivative (or the field strength) of the vertical coframe $\form{\k}$ is generally arbitrary. As the plane $\NP$ is 2-dimensional, the exterior derivative $\exd \form{\k}$, which is a 2-form on $\NP$, has one independent component. In total, we impose the following exterior derivatives of the coframe fields,
\begin{align}
\exd \form{\n} = 0, \qquad \text{and} \qquad \exd \form{\k} = \bkappa \form{\n} \wedge \form{\k}, \label{dk}
\end{align}
where $\bkappa$ is a function on $\NP$ and controls how the vertical coframe field is Lie-transported along the frame fields, that is, we have $\Lie_{\u} \form{\k} = - \bkappa \form{\n}$ and $\Lie_{\k} \form{\k} = \bkappa \form{\k}$.  

Many consequences follow from the exterior derivatives. First, since the rigging vector $\vec{\k}$ is null, it follows that $\Lie_\k \k_a = \k^b \DN_b \k_a - \k_b \DN_a \k^b = \DN_\k \k_a$, since $\k_b \DN_a \k^b = \frac{1}{2} \DN_a (\k_b \k^b) =0$. Then, $\Lie_{\k} \form{\k} = \bkappa \form{\k}$ implies that $\vec{\k}$ serves as the non-affine generator for a null geodesic, 
\begin{align}
\DN_{\vec{\k}} \k^a = \bkappa \k^a, \qquad \text{or equivalently} \qquad \nabla_{\vec{\k}} \k^\mu = \bkappa \k^\mu, 
\label{k_geodesic}
\end{align}
where $\bkappa$ plays the role of the transverse acceleration, quantifying the inaffinity of the null geodesic. 

In addition, the exterior derivatives of the coframe fields \eqref{dk} determine the following Lie bracket between the frames $(\vec{\u},\vec{\k})$ (to derive the Lie bracket from the exterior derivative, see the explanation in section 3.3.3 of \cite{Ciambelli:2025unn}):
\begin{align}
[\uline{\u},\uline{\k}] = \bkappa \vec{\u}. \lb{com-gen}
\end{align}
Here we see that $\bkappa$ plays a role of the anholonomicity of the basis vectors. 
We note that this Lie bracket closely resembles the Lie bracket of a Carrollian manifold in two dimensions, where $\bkappa$ plays a role of the Carrollian acceleration. Following the Koszul identity, this Lie bracket and the relations in \eqref{pairing} fully determine the spacetime covariant derivatives $\nabla_\mu \u^\nu$ and $\nabla_\mu \k^\nu$. Their expressions are given in Appendix \ref{app:geometry} (see also Appendix B of \cite{Freidel:2022vjq}).\\

In addition to the transverse acceleration $\bkappa$, we define the \emp{vertical acceleration} $\kappa$ of the vertical vector $\vec{\u}$ as 
\begin{align}
\kappa := \k_a \DN_{\vec{\u}} \u^a = \k_\mu \nabla_{\vec{\u}} \u^\mu.
\end{align}
By using the relation \eqref{vector-u}, $\u_a = \n_a - 2\rho \k_a$, and the null-ness of the rigging vector, the acceleration can also be written as $\kappa = \k^a \u^b \DN_b \n_a = \k^a \u^b \DN_a \n_b$, where the last equality followed from $\DN_{[a}\n_{b]} =0$. It can be expressed in terms of $\rho$, as one can show using the Leibniz rule and the norm $2\rho = \n^a \n_a$:
\begin{equation}
\lb{kappa-gen}
\begin{aligned}
\kappa  = \k^a \u^b \DN_a \n_b = (\n^a - 2\rho \k^a) \DN_{\vec{\k}} \n_a & = \frac{1}{2} \DN_\k (\n^a \n_a ) + 2\rho \n_a \DN_\k \k^a\\
& = \left(\Lie_{\k} + 2\bkappa\right)\rho.
\end{aligned}
\end{equation}
In general, $\DN_{\vec{\u}} \u$ also has a transverse component, that is $\n_a \DN_{\vec{\u}} \u^a = -\left( \Lie_{\u} - 2\kappa\right)\rho$. However, on the null surface where both $\rho \stackrel{\sss \mathrm{null}}{=} 0$ and $\Lie_{\u}\rho \stackrel{\sss \mathrm{null}}{=} 0$, we have $\DN_{\vec{\u}} \u^a \stackrel{\sss \mathrm{null}}{=}  \kappa \u^a$, and $\kappa \stackrel{\sss \mathrm{null}}{=} \Lie_{\k}\rho$ plays a role of the inaffinity of null geodesics generated by $\vec{\u}$. 

\subsubsection{Volume forms}
Having defined the foliation-adapted coframes, we write the volume form on the spherical surface $\Sigma_\sigma$, on the normal plane $\NP$, and on the spacetime $\M$ as follows: 
\begin{align}
\bvol_{\sss \Sigma} = \form{\k} \wedge \volS, \qquad \volN = \form{\k} \wedge \form{\n},  \qquad \text{and} \qquad \volM = \form{\k} \wedge \form{\n} \wedge \volS, \lb{volume}
\end{align}
where we recall that $\volS$ is the volume form on the spherical shell $\S_r$. The components of the normal plane's volume form are the 2-dimensional Levi-Civita tensor $(\epsilon_{\sss \NP})_{ab}$, which can now be expressed in terms of the coframe fields as
\begin{align}
(\epsilon_{\sss \NP})_{ab} = \k_a \n_b - \k_b \n_a, \lb{epsilon}
\end{align}

Using the formula \eqref{df}, the definition of expansions \eqref{expansion-basis} and the exterior derivatives \eqref{dk}, we can derive the following formulas that will be useful in our subsequent considerations:
\begin{align}
\exd \left( -F_1 \form{\k} + F_2 \form{\n} \right) & = \big(  (\Lie_{\k}+ \bkappa) F_1  + \Lie_{\u} F_2\big)\volN, \\
\exd \left( -F_1 \form{\k} \wedge \volS + F_2 \form{\n}\wedge \volS \right) & = \big(  (\Lie_{\k}+ \bkappa +\theta_{(\k)})F_1  + (\Lie_{\u}  +\theta_{(\u)})F_2\big)\volM,
\end{align}
for arbitrary functions $F_1$ and $F_2$ on $\NP$. We can use these formulas to perform integration by parts. For instance, the Stokes theorem dictates that 
\begin{align}
\int_\NP  \big(  (\Lie_{\k}+ \bkappa) F_1  + \Lie_{\u} F_2\big)\volN =  \int_{\pa \NP} \left( F_2 \form{\n} - F_1 \form{\k}\right). \lb{Stokes}
\end{align}

\subsubsection{Einstein equations} \lb{sec:Ein-gen}

We have laid down the geometric construction of spherically symmetric spacetimes based on the foliation by spherical slices $\Sigma_\sigma$, and introduced geometrical objects such as the accelerations $\kappa$ and $\bkappa$, as well as the expansions $\theta_{(\u)}$ and $\theta_{(\k)}$. These quantities are subjected to the Einstein equation,
$G_{\mu\nu} = 8\pi\Newton T_{\mu \nu}$.

In fact, the components, in the adapted frame, of the Einstein equation on the normal plane, $G_{ab} = 8\pi\Newton T_{ab}$, are represented as (see Appendix \ref{app:Ein-ten} for the derivation) 
\begin{subequations}
\lb{Ein-Ray}
\begin{alignat}{2}
\Lie_{\u}\theta_{(\n)} - \left(\kappa - \frac{1}{d-2}\theta_{(\n)} \right)\theta_{(\u)} - \theta_{(\k)}\Lie_{\u}\rho \ & = -&&8\pi\Newton T_{\u \n}  \lb{T-un}\\
\Lie_{\u}\theta_{(\k)} + \left( \kappa + \theta_{(\n)} \right)\theta_{(\k)} - \left(\frac{d-1}{d-2}\right) \rho \theta_{(k)}^2 - \frac{1}{2}\stackrel{\sss (\S)}{R} \ & = &&8\pi\Newton T_{\k \n} \\
\Lie_{\k}\theta_{(\n)}+ \left(\bkappa + \theta_{(k)} \right)\theta_{(\n)} -\kappa \theta_{(\k)} - \left(\frac{d-3}{d-2}\right) \rho \theta_{(k)}^2 - \frac{1}{2}\stackrel{\sss (\S)}{R} \ & = &&8\pi\Newton T_{\u \k} \\
\Lie_{\k}\theta_{(\k)} - \left(\bkappa - \frac{1}{d-2}\theta_{(\k)} \right)\theta_{(\k)} \ & = -&&8\pi\Newton T_{\k \k} \,.
\end{alignat}
\end{subequations}
Here, we use the notation $T_{\u \n} = \u^a T_a{}^b \n_b$ and so on.
The Ricci scalar of the spherical shell $\S_r$ is given by
\begin{align}
\stackrel{\sss (\S)}{R} = \frac{(d-3)(d-2)}{r^2} \,. \lb{RicciS}
\end{align}

These equations take the form of the \emp{Raychaudhuri equation}, governing the evolution of the expansions along the vertical and transverse directions. 
These components are not entirely independent, as one can verify $T_{\k \n} - T_{\u \k} = 2\rho T_{\k\k}$ from the relation \eqref{vector-u}, $\n^a = \u^a + 2\rho \k^a$. 

The remaining components of the Einstein equation are the angular ones, $G_{AB} = 8\pi\Newton T_{AB}$. Due to spherical symmetry, they contain only one independent component, which corresponds to their trace $q^{AB}G_{AB} = 8\pi\Newton q^{AB}T_{AB}$. We can show that it is given by
\begin{equation}
\begin{aligned}
& \Lie_{\u} \left(\bkappa + \frac{d-3}{d-2} \theta_{(\k)}\right) + \left( \Lie_{\k} + \bkappa \right) \left( \kappa + \frac{d-3}{d-2} \theta_{(\n)}  \right) \ = \ \frac{8\pi\Newton}{d-2}q^{AB}T_{AB} \\
& + \frac{1}{2} \left( \frac{d-3}{d-2} \right) \theta_{(\k)} \left( \theta_{(\n)} + \theta_{(\u)} \right)  - \frac{1}{2} \left( \frac{d-4}{d-2} \right) \stackrel{\sss (\S)}{R}.
\end{aligned}
\end{equation}

Note that the Einstein equations presented here can be viewed as the spherically symmetric version of those in \cite{Freidel:2024emv}, but 
our formulation applies to general spacetime dimensions and allows for nonzero $\bkappa$.\\

In the case of null surfaces, 
where $\rho \stackrel{\sss \mathrm{null}}{=} 0$ and $ \Lie_{\u}\rho \stackrel{\sss \mathrm{null}}{=} 0$, these equations simplify further. 
In particular, the first equation \eqref{T-un} reduces to the spherically symmetric version of the celebrated null Raychaudhuri equation, which can be expressed as  
\begin{align}
\Lie_{\u} \varepsilon_{\sss \mathrm{mem}} + \left( \varepsilon_{\sss \mathrm{mem}} + p_{\sss \mathrm{mem}} \right)\theta_{(u)}
\stackrel{\sss \mathrm{null}}{=}
T_{\u \n}, 
\end{align}
where 
\begin{align}
\varepsilon_{\sss \mathrm{mem}} := -\frac{\theta_{(\n)}}{8\pi\Newton} \qquad \text{and} \qquad p_{\sss \mathrm{mem}} := \frac{1}{8\pi\Newton} \left( \kappa + \frac{d-3}{d-2}\theta_{(n)} \right). 
\lb{membrane}
\end{align}
This can be interpreted as the gravitational analog of the Euler equation for energy in hydrodynamics, with $\varepsilon_{\sss \mathrm{mem}}$ and $p_{\sss \mathrm{mem}}$ interpreted as the energy density and pressure, respectively. 
This perspective is known as the \emp{membrane paradigm} of black holes \cite{Damour:1978cg,thorne1986black,Price:1986yy}.

However, this gravity-fluid dictionary relies on the presence of a null surface, such as a black hole horizon. Although the work \cite{Freidel:2022vjq} extends the identification into the spacetime region near the null surface, known as the stretched horizon, it still presupposes the existence of a null surface. Furthermore, in cases where the null surface is non-expanding (i.e., $\theta_{(n)} = 0$), as with isolated horizons, the associated energy density vanishes, rendering the fluid analogy less natural. This raises the question of whether an alternative geometro-hydrodynamic correspondence exists, especially one that does not hinge on the presence of a null surface. This is precisely what we aim to demonstrate in this work.\\

To summarize this subsection, we have described the geometry of a general spherically symmetric spacetime by considering a foliation of spherical hypersurfaces (or slices) $\Sigma_\sigma= \Time_\sigma \times \S_r$. The normal plane $\NP$ is sliced by the constant-$\sigma$ curves $\Time_\sigma$, and the rigging technique provides a systematic way to assign the frame fields $(\vec{\u}, \vec{\k})$ and the coframe fields $(\form{\k}, \form{\n})$ adapted to the foliation, from which the geometry can be reconstructed. The geometry is characterized by the function $\rho$, which computes the norm of $\n$ (and also $\vec{\u}$) and whose sign encodes the local causal character of $\Sigma_\sigma$, the expansions $\left( \theta_{ (\u)}, \theta_{ (\k)} \right)$, and the accelerations $\left( \kappa, \bkappa \right)$. All of them are variables that appear in the Einstein equation. See Tab.\ref{tab:frames}.

Then, which foliation function (or slice) $\sigma(y)$ is best suited for a given physical problem? In this work, we focus on the radial slice $\sigma(y) = r(y)$ and the equipotential slice $\sigma(y) = \vrho(y) := \frac{1}{2} + \Phi(y)$. The former, which we will discuss next, allows the connection between the rigging technique and the Kodama formalism, while the latter characterizes the dynamical apparent horizon and will be discussed in Sec.\ref{sec:AH}.

\subsection{Radial slices: From rigging to Kodama} \label{sec:slices}

\begin{table}[t!]
\renewcommand{\arraystretch}{1.3}
\centering
\begin{tabular}{| l | c | c | c | }
\hline 
\rowcolor{palegray}
\textbf{Geometrical objects} & \textbf{General ($\Sigma_\sigma$)} & \textbf{Radial ($\Sigma_r$)} & \textbf{Equipotential ($\Sigma_\vrho$)}    \\ \hline 
Foliation-adapted Frame & $e_{\mathscr{A}}$ & $\bdx{e}_{\mathscr{A}}$ & $\mathsf{e}_{\mathscr{A}}$ 
\\ \hline
Transverse coframe (normal form) & $\n$ & $\nn$ & $\nAH$ 
\\ \hline
Transverse frame (rigging vector) & $\k$ & $\kk$ & $\kAH$ 
\\ \hline
Vertical frame & $\vec{\u}$ & $\vec{\uu}$ & $\vec{\uAH}$ 
\\ \hline
Vertical coframe & $\vec{\k}$ & $\vec{\kk}$ & $\vec{\kAH}$ 
\\ \hline
Normalization factor & $\rho$ & $\vrho$ & $\rhoAH$ 
\\ \hline
Vertical acceleration & $\kappa$ & $\vkappa$ & $\kappaAH$ 
\\ \hline
Transverse acceleration & $\bkappa$ & $\bvkappa$ & $\bkappaAH$ 
\\ \hline
Vertical expansion & $\theta_{(\u)}$ & $\theta_{(\uu)} = 0$ & $\theta_{(\uAH)}$ 
\\ \hline
Transverse expansion & $\theta_{(\k)}$ & $\theta_{(\kk)} = \vtheta $ & $\theta_{(\kAH)}$ 
\\ \hline
\end{tabular}
\caption{The notations for the foliation-adapted frames and coframes, as well as various geometrical objects, corresponding to different slices defined by distinct foliation functions $\sigma(y)$. }
\label{tab:frames}
\end{table}

As explained in the Introduction, a notion of energy is central to thermodynamics. In spherically symmetric spacetimes, the Misner-Sharp energy $\m$, defined in \eqref{Misner-Sharp}, provides a natural and useful notion of energy, irrespective of the presence of a horizon. However, in our current formulation, $\m$ does not appear transparently in the equations of motion. The goal of this section, and indeed of the remainder of the paper, is to render $\m$ explicit within our formalism. We will show that, in doing so, the gravitational dynamics acquires a natural hydrodynamic interpretation. Our starting point are radial slices, where $\m$ naturally appears.

In this section, we focus on the radial slices, $ \Sigma_r = \{ (y^a_{\star}, z^A) | r(y_{\star}) = \mathrm{constant} \} = \Time_r \times \S_r$. Choosing the areal radius as the foliation function, $\sigma(y) = r(y)$, is the most convenient and straightforward choice. For spherically symmetric spacetimes, the areal radius is the only scalar present in the metric and can be defined everywhere in spacetime. Furthermore, as we will show below, working with the constant area slices reveals a connection between the rigging technique and the Kodama formalism, and some geometrical objects, such as $\rho$ and $\kappa$, can now be expressed in terms of physical quantities like the area $\A$, the Misner-Sharp energy $\m$, as well as their derivatives.  

\begin{figure}[t]
    \centering
    \begin{subfigure}[b]{0.5\textwidth} 
        \centering
        \adjustbox{valign=c}{\includegraphics[width=\textwidth]{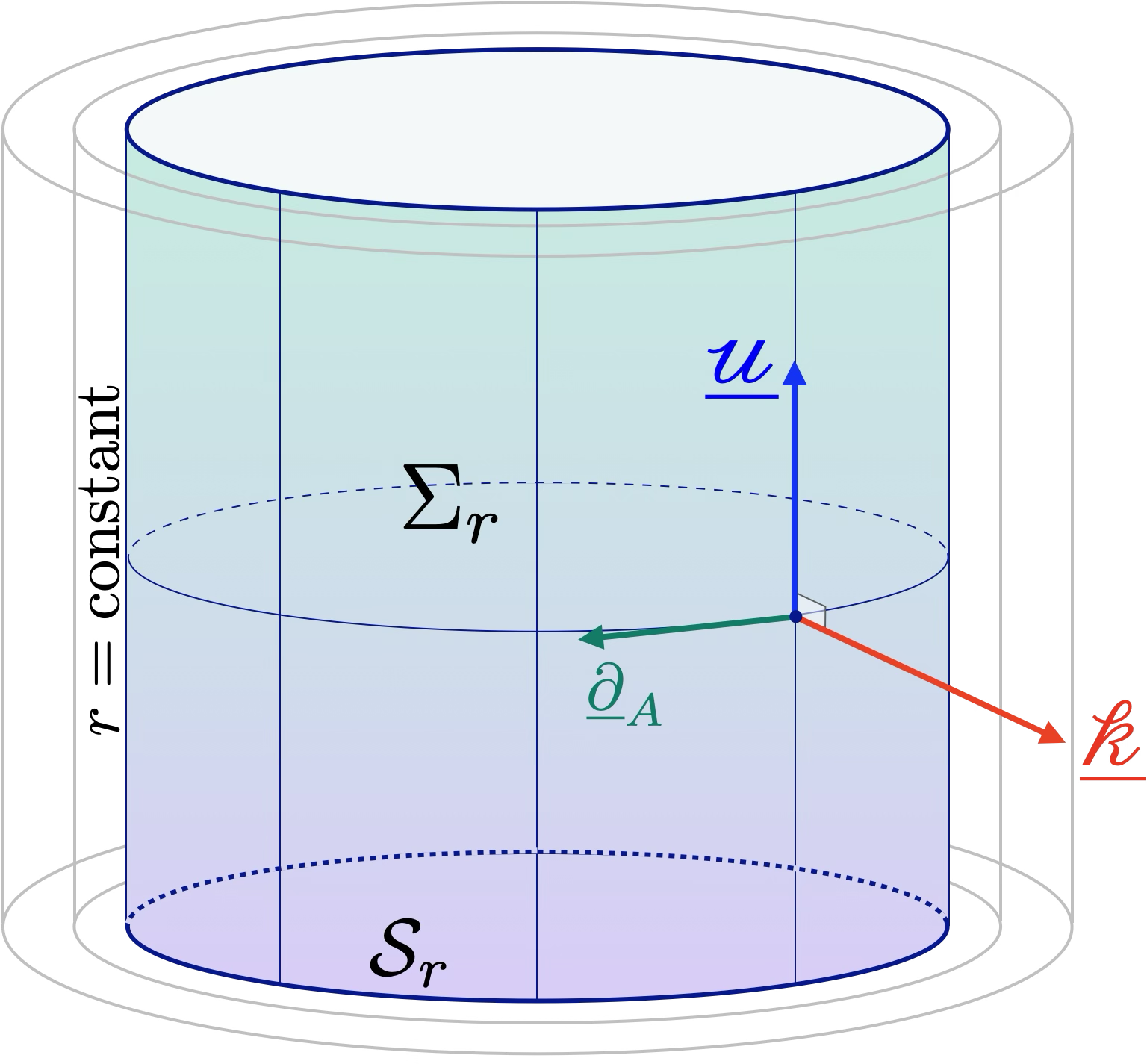}} 
    \end{subfigure}
    \hfill
    \begin{subfigure}[b]{0.45\textwidth} 
        \centering
        \adjustbox{valign=c}{\includegraphics[width=\textwidth]{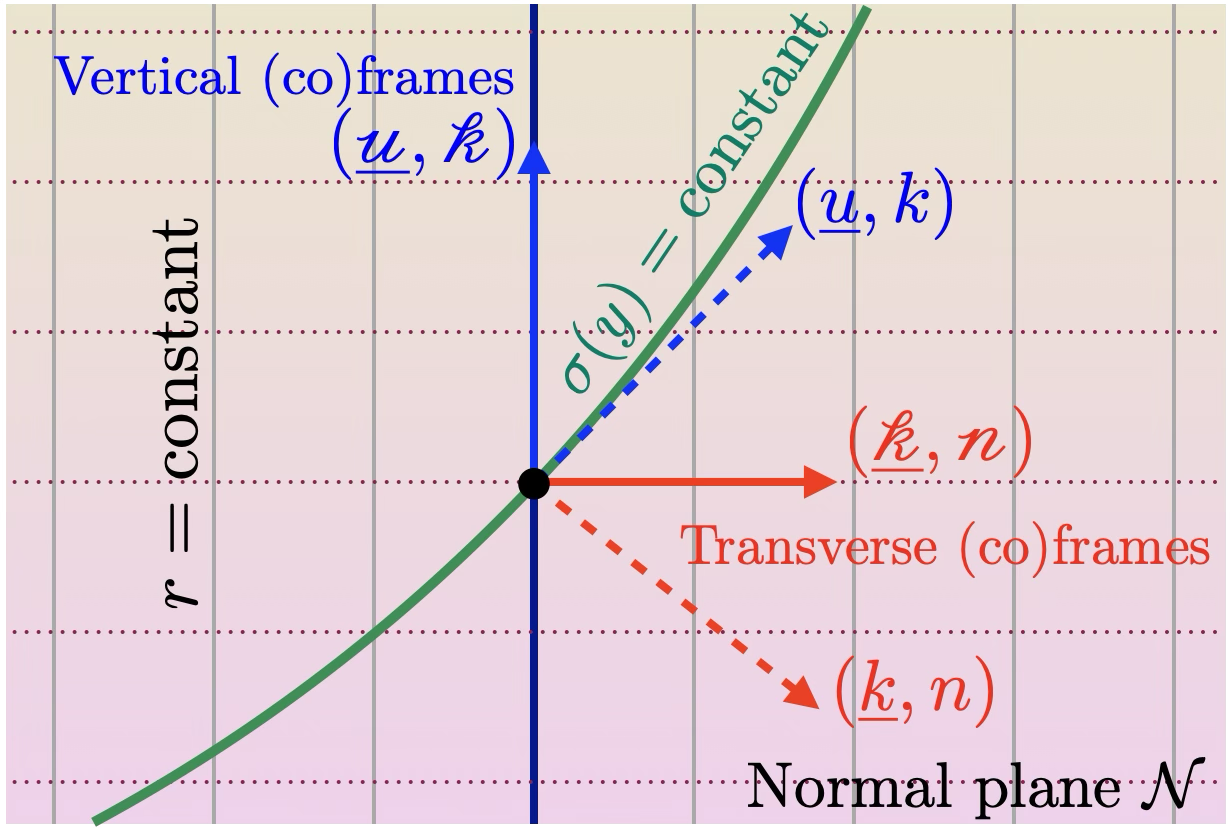}} 
    \end{subfigure}
    \caption{\textbf{(Left)} Spherically symmetric spacetime foliated by a series of radial slices $\Sigma_r = \Time_r \times \S_r$, which are non-expanding surfaces (i.e., $\theta_{(\uu)} =0$). \textbf{(Right)} The 2-dimensional normal plane $\NP$, whose points are attached to the round sphere $\S_r$. The radial slice $\Sigma_r$ is represented as a vertical line. The tangent space and cotangent space to $\NP$ split into the vertical (temporal) part spanned by $(\vec{\uu}, \kk)$ and the transverse (radial) part spanned by the null rigging structure $(\vec{\kk}, \nn)$. Locally, the bases $(\vec{\u}, \form{\k})$ and $(\vec{\k},\form{\n})$, adapted to an arbitrary foliation $\sigma(y) = \text{constant}$ (the green line), are linear combinations of those adapted to the radial slices.}
    \label{figure}
\end{figure}

The construction and the equations present in Sec.\ref{sec:rigging} can be readily applied to the special case of radial slices. To distinguish this particular case from the general one, we use "curly" symbols for the frame $\bdx{e}_{\mathscr{A}} = (\vec{\uu}, \vec{\kk})$ and the dual coframe $\bdx{e}^{\mathscr{A}} = (\kk, \nn)$, and also for the geometrical objects derived from them (see Tab.\ref{tab:frames}). Indeed, they respect the same pairings \eqref{pairing}, and we denote the norm squared with $2\vrho = \nn^a \nn_a = - \uu^a \uu_a$. In this case, the null rigging structure comprises
\begin{align}
\nn := \form{\exd} r = (\DN_a r) \exd y^a \qquad \text{and} \qquad \vec{\kk} := \frac{\exd }{\exd r} = \frac{\exd y^a}{\exd r} \pa_a.
\end{align} 
 Since the rigging vector $\vec{\kk}$ generates null geodesics, it is clear that $r$ also serves as the non-affine parameter for the geodesics. The inaffinity is characterized by the acceleration $\bvkappa := \nn_a \DN_{\vec{\kk}} \kk^a$. As in the general case, we can use the (co)frame fields adapted to the radial slice to span the tangent and cotangent spaces to the normal plane $\NP$. The frame fields $(\vec{\uu},\vec{\kk})$ define preferred directions across $\NP$, and by following their integral curves, we generate a grid-like structure on $\NP$ (see Fig.\ref{figure}). This process is akin to "Cartesianization", where the plane $\NP$ is endowed with anholonomic frame fields, instead of coordinates. We will refer to this frame as the \emp{radial frame}. 

In addition, the function $2\vrho = \nn^a \nn_a = (\DN r)^2$ is given in terms of the gravitational potential \eqref{Npotential} or the Misner-Sharp energy \eqref{Misner-Sharp} as
\begin{align}
2\vrho  = 1+2\Phi = 1 - \frac{16\pi\Newton}{(d-2)\mr{\A}} \frac{\m}{r^{d-3}}. \lb{rho}
\end{align}
From this relation, the normal plane metric $h_{ab}$ and its inverse $h^{ab}$ \eqref{h_h} are expressed as
\begin{equation}
\lb{metric-1}
\begin{aligned}
h_{ab} &= \kk_a \uu_b + \nn_a \kk_b = -\left(1 - \frac{16\pi\Newton}{(d-2)\mr{\A}} \frac{\m}{r^{d-3}} \right)\kk_a \kk_b + 2\nn_{(a} \kk_{b)}  \\
h^{ab} &= \kk^a \uu^b + \nn^a \kk^b =  \left(1 - \frac{16\pi\Newton}{(d-2)\mr{\A}} \frac{\m}{r^{d-3}} \right) \kk^a \kk^b +  2\uu^{(a} \kk^{b)} .
\end{aligned}
\end{equation}

Now, let us prove one key result here: Choosing the normal form to be $\nn = \rd r$ 
infers that the vertical vector $\vec{\uu}$ coincides with the Kodama vector defined in \eqref{Kodama-vector},
\bbox
\vspace{-13pt}
\begin{align}
\text{\bf Vertical vector for radial slices} \qquad \uu^a = \kv^a  \qquad \text{\bf Kodama vector}.
\end{align}
\ebox
\noindent To derive this result, we first consider the normal components of the projection tensor with the indices raised, which can be written as
\begin{align}
\Pi^{ba} = h^{bc}\Pi_c{}^a = h^{ba} - \nn^b \kk^a = \kk^a \uu^b - 2\kk^{[a} \nn^{b]} = \kk^a \uu^b + (\epsilon_{\sss \NP})^{ba}
\end{align}
where we recalled the inverse metric on $\NP$ \eqref{metric-1} and the expression \eqref{epsilon} for the Levi-Civita tensor. Then, the definition \eqref{vector-u} of the vertical vector for radial foliation becomes 
\begin{align}
\uu^a = \nn^b\Pi_b{}^a = \Pi^{ba}\nn_b = (\epsilon_{\sss \NP})^{ba}\nn_b = (\epsilon_{\sss \NP})^{ba}\DN_b r
\end{align}
where we used $\uu^a \nn_a = 0$ from the condition \eqref{pairing}. Hence, the frame $\vec{\uu}$ agrees with the original definition \eqref{Kodama-vector} of the Kodama vector\footnote{Throughout this work, we refer to the vector $\vec{\uu}$ interchangeably as the vertical frame field and the Kodama vector.}. \\

We conclude this subsection by providing relations between physical quantities.

The radial slice $\Sigma_r$ becomes a null hypersurface when $\vrho \stackrel{\sss \mathrm{null}}{=} 0$ and $\Lie_{\uu}\vrho \stackrel{\sss \mathrm{null}}{=} 0$, inferring (from \eqref{rho} and $\uu^a \DN_a r = 0 $ in \eqref{Kodama-vector}) the conditions 
\begin{align}
r^{d-3} \stackrel{\sss \mathrm{null}}{=} \frac{1}{(d-2)}\frac{16\pi\Newton}{\mr{\A}} \m, \qquad \text{and} \qquad \Lie_{\uu}\m \stackrel{\sss \mathrm{null}}{=} 0.
\end{align}
This is consistent with the event horizon of the Schwarzschild black hole; it is null and located at $r = 2\Newton \m$ for $d=4$, and the mass is constant.
Note that, as we will discuss in Sec.\ref{sec:AH}, a generic apparent horizon, where $\vrho \stackrel{\sss \mathrm{AH}}{=} 0$ but $\Lie_{\uu}\m \neq 0$, does not belong to the family of radial slices $\Sigma_r$, but the geometric framework for the radial frame is useful to describe the apparent horizon via the changes of frames. \\

The expansions \eqref{expansion-basis} of the Kodama vector $\uu^a$ and the null rigging vector $\kk^a$ are 
\begin{align}
\theta_{(\uu)} = \vtheta \Lie_{\uu}r=0 , \qquad \text{and} \qquad \theta_{ (\kk)} = \vtheta \Lie_{\kk}r = \vtheta,  \lb{expansion-slices}
\end{align}
following directly from the conditions $\Lie_{\uu}r = \uu^a \nn_a =0$ and $\Lie_{\kk}r = \kk^a \nn_a =1$. The expansion of the area in the direction tangential to the surface vanishes $\theta_{(\uu)} = 0$, as one would expect from the non-expanding, constant-radius, surface. \\

The acceleration $\vkappa$ (see equation \eqref{kappa-gen}) of the vertical field $\vec{\uu}$ adapted to the radial slices can be expressed (from the formula \eqref{rho} and the property $\Lie_{\kk}r =1$) as
\begin{equation}
 \label{kappa}
\begin{aligned}
\vkappa &= \left(\Lie_{\kk} + 2\bvkappa\right)\vrho \\
&= \Lie_{\kk}\Phi + (1+2\Phi) \bvkappa\\
& = \left(\frac{d-3}{d-2}\right)\frac{8\pi\Newton}{\mr{\A}} \frac{\m}{r^{d-2}} - \frac{8\pi\Newton}{(d-2)\mr{\A}} \frac{\Lie_{\kk}\m}{r^{d-3}} + \left(1 - \frac{16\pi\Newton}{(d-2)\mr{\A}} \frac{\m}{r^{d-3}} \right) \bvkappa.
\end{aligned}
\end{equation} 
In the context of black hole physics, $\vkappa$ evaluated at the horizon is also called the \emp{surface gravity} and obtained by the radial derivative of the potential, $\varkappa|_\H=(\Lie_{\kk}\Phi)|_\H$, which is consistent with the standard formula \cite{Poisson:2009pwt}. It indeed reproduces the expression $\vkappa =  1/(4\Newton \m)$ for the black hole horizon ($r = 2\Newton\m$) in the Schwarzschild spacetime in four dimensions. In general, $\vkappa$ also receives contributions from both the $\bvkappa$ term (which vanishes on the horizons from the definition \eqref{AH-def}) and the radial distribution of the Misner-Sharp energy $\Lie_{\kk}\m$.

\subsubsection{Gravitational dynamics on the radial slice} \label{sec:Ein}

We have connected the geometric construction of spherically symmetric spacetimes based on the rigging technique to the Kodama formalism. The purely geometric objects, such as the expansions and accelerations, now have a meaningful interpretation in terms of the physical quantities unique to spherical symmetry, such as the Kodama vector and the Misner-Sharp energy and their derivatives. 

We next express the components of the Einstein equation in the radial frame by using physical quantities like $\m$, $\A$, and $\P$. This allows for a more transparent interpretation of the gravitational dynamics from a hydrodynamics perspective.

\begin{itemize}
\item \textbf{Temporal evolution of energy:} The component $\uu^a \left( G_a{}^b - 8\pi\Newton T_a{}^b \right) \nn_b = 0$ of the Einstein equation dictates the temporal evolution of the Misner-Sharp energy $\M$:
\begin{align}
\Lie_{\uu}\m = \A T_{\uu \nn}, \qquad \text{or equivalently} \qquad  \Lie_{\uu} \E = T_{\uu \nn}. 
\label{G-rad-un}
\end{align} 
See Appendix \ref{app:Ein-ten} for the derivation. Here, we recalled again the Misner-Sharp energy density \eqref{Edensity}, $\E = \frac{\m}{\A}$, and we used the non-expanding condition, $\theta_{(\uu)} =\Lie_{\uu}\ln \A = 0$.  

\item \textbf{Radial distribution of energy:} The equation $\uu^a \left( G_a{}^b - 8\pi\Newton T_a{}^b \right) \kk_b = 0$ controls how the energy (or mass) distribute radially:
\begin{align}
-\Lie_{\kk}\m = \A T_{\uu \kk}, \qquad \text{or equivalently} \qquad - \left( \Lie_{\kk}  + \vtheta \right)\E = T_{\uu \kk}, 
\label{G-rad-uk}
\end{align}
where we used $\theta_{(\kk)} = \Lie_{\kk}\ln\A = \vtheta$ for radial slices \eqref{expansion-slices}. 

\item \textbf{Gravitational Young-Laplace equation:} The equation $\kk^a \left( G_a{}^b - 8\pi\Newton T_a{}^b \right) \nn_b = 0$ can be expressed as 
\begin{align}
\P \vtheta = T_{\kk \nn}\,, 
\label{G-rad-kn}
\end{align}
where we have introduced 
\begin{align}
\P := \frac{1}{8\pi\Newton}\vkappa -  \left(\frac{d-3}{d-2}\right) \E. \label{pressure}
\end{align}
We note that $\P$ has the dimension of force per unit length, $\vartheta=\frac{d-2}{r}$ is the extrinsic curvature of the sphere, and $T_{\kk \nn}$ represents the pressure in the radial direction. On the other hand, the Young-Laplace equation for a soap bubble with surface tension $\gamma$, $\gamma \vartheta=\Delta p$, describes the mechanical balance between the force of that tension and the difference $\Delta p$ in internal and external pressure \cite{landau1987fluid}. Therefore, comparing these two equations reveals that \eqref{G-rad-kn} corresponds to the gravitational analog of the Young-Laplace equation, where $- \P$ plays the role of surface tension. In this sense, we call $\P$ \emp{gravitational pressure}. Interestingly, this pressure $\P$ \eqref{pressure} has a similar form to that in the membrane paradigm dictionary \eqref{membrane}, although the energy density is given by the Misner-Sharp energy density $\E$ rather than the expansion.

\item \textbf{Inaffinity equation:} The last normal component, $\kk^a \left( G_a{}^b - 8\pi\Newton T_a{}^b \right) \kk_b = 0$, can be expressed as  
\begin{align}
\bP \vtheta = T_{\kk\kk}, \qquad \text{where we define for convenience} \qquad \bP := \frac{ \bvkappa }{8\pi\Newton}.  \label{bP}
\end{align}
On one hand, this dynamical equation determines the inaffinity (or acceleration) $\bvkappa$ of the null geodesics generated by the null rigging vector $\kk^a$, sourced by the matter content. In the absence of energy flux along the null direction, $T_{\kk \kk} = 0$, the inaffinity vanishes, $\bvkappa = 0$. On the other hand, the same equation also resembles the Young-Laplace equation, where $\overline{\gamma} = - \bP$ can be interpreted as a surface tension term, and $T_{\kk \kk}$ represents a pressure difference.
\end{itemize}

\begin{itemize}
\item \textbf{Tangential equation:} The last components of the Einstein equation are those that are tangent to the sphere, $G_{AB} = 8\pi\Newton T_{AB}$. As we have explained, due to spherical symmetry, there is only one independent component given by the trace,
\begin{align}
 \Lie_{\uu}\bP + \left( \Lie_{\kk} + \bvkappa \right)\P  + \left(\frac{d-3}{d-2}\right)\left(\vtheta \P+\bvkappa \E\right)  = \frac{1}{d-2}q^{AB} T_{AB}. 
 \label{G_AB}
\end{align}
\end{itemize}

We have therefore shown that the normal components of the Einstein equation, when expressed in the radial frame $\bdx{e}_{\mathscr{A}} = (\vec{\uu}, \vec{\kk})$ and coframe $\bdx{e}^{\mathscr{A}} = (\kk, \nn)$, take on much simpler forms with clearer physical interpretations: they describe the temporal and radial evolution of the Misner-Sharp energy, supplemented by the Young-Laplace constraint. This serves as a prelude to the hydrodynamic description of spacetime dynamics. In the next section, we will extend this geometro-hydrodynamic picture to general slices and see the physical picture and mathematical structure more clearly.\\

Let us end this section with some comments. First, not all the normal components of the Einstein equation (4 equations: \eqref{G-rad-un}, \eqref{G-rad-uk}, \eqref{G-rad-kn}, and \eqref{bP}) are independent because of the relation $G_{\nn \kk} - G_{\uu \kk} = 2\rho G_{\kk\kk}$, which originates from $\nn^a = \uu^a + 2\vrho \kk^a$ (see \eqref{vector-u}).
This gives the following identity,
\begin{align}
\frac{\rd \m}{\rd r} + \P \frac{\rd \A}{\rd r}= \left(\vtheta - 16\pi\Newton \E \right) \bP \A, 
\label{relationP-M}
\end{align}
where we used the equations \eqref{G-rad-uk}, \eqref{G-rad-kn} and \eqref{bP}, the definition \eqref{expansion} and the formula \eqref{rho}. This identity also follows from the expression of \eqref{kappa} of the acceleration $\vkappa$ and the definition \eqref{pressure} of the pressure. 

Here, an interesting relation holds when the radial slice is null (e.g. the event horizon). This is characterized by the special value of the potential, $\Phi = -\frac{1}{2}$, which leads (from \eqref{Edensity}) to
\begin{align}
\E \stackrel{\sss \mathrm{null}}{=} \frac{\vtheta}{16\pi\Newton} \,,
\end{align}
where $\vtheta$ is evaluated on the null surface. Then, the right-hand side of \eqref{relationP-M} vanishes, and the gravitational pressure is completely determined by the areal derivative of the Misner-Sharp energy:
\begin{align}
\P \stackrel{\sss \mathrm{null}}{=} - \frac{\rd \m}{\rd \A}.
\end{align}

Finally, as a consistency check, let us confirm Birkhoff's theorem \cite{Poisson:2009pwt} in the present framework. Suppose that the matter fields are absent, i.e., $T_{\mu\nu} =0$. Then, the normal components of the Einstein equations reduce to 
\begin{align}
\Lie_{\uu} \m =0, \qquad \Lie_{\kk} \m = 0, \qquad \P = 0, \qquad \text{and} \qquad \bP = 0. \lb{EH-vacuum}
\end{align}
The first two equations imply that the Misner-Sharp energy is constant, while the last equation, together with the definition \eqref{bP} of $\bP$, implies that the inaffinity of the transverse null geodesics vanishes, $\bvkappa = 0$. The third equation, $\P = 0$, then follows trivially from these conditions and from the expressions \eqref{pressure} and \eqref{kappa} for the pressure $\P$ and the acceleration $\vkappa$, respectively. This again reflects the fact that only three normal components of the Einstein equations are independent. Note that the tangential equation \eqref{G_AB} also follows automatically. 

To complete the proof, we next set up the comoving coordinates $y^a = (\tilde{v},r)$ such that the metric takes the form \eqref{EFmetric},
\begin{align}
\rd s^2 = - 2\vrho \e^{2\tilde{\alpha}} \rd \tilde{v}^2 + 2 \e^{\tilde{\alpha}} \rd \tilde{v} \rd r + r^2 \rd \Omega^2_{d-2}. \lb{EFmetric-1}
\end{align}
Here, we have $\nn=\rd r,~\kk=e^{\tilde \alpha}\rd v,~\vec{\uu}=e^{-\tilde{\alpha}}\partial_v,~\vec{\kk}=\partial_r$, and the definition \eqref{dk} provides $\bvkappa = \pa_r \tilde{\alpha}$. Then, the condition $\bvkappa = 0$ implies that the scale factor can only be a function of the time coordinate, that is, $\tilde{\alpha} = \tilde{\alpha} (\tilde{v})$. Upon applying the time reparameterization, $\e^{\tilde{\alpha}(\tilde{v})} \rd \tilde{v} = \rd v$, and recalling the expression \eqref{rho} for $\vrho$, we arrive at the solution: 
\begin{align}
\rd s^2 = - \left( 1 - \frac{16\pi\Newton}{(d-2)\mr{\A}} \frac{\m_0}{r^{d-3}} \right) \rd v^2 + 2  \rd v \rd r + r^2 \rd \Omega^2_{d-2} \,,
\end{align}
where $\m_0$ is constant. This is the $d$-dimensional Schwarzschild metric in the Eddington-Finkelstein coordinates $(v,r)$. Thus, we have shown Birkhoff's theorem in our formulation. This fact highlights the essential role of matter fields in giving rise to non-trivial and interesting dynamics in spherically symmetric spacetime.

\section{Geometro-hydrodynamics}\lb{sec:Hydro}
As we have seen in the previous section, considering the radial frame, the spacetime geometry and its dynamics are expressed in terms of physically meaningful quantities such as the area $\A$, the Misner-Sharp energy $\m$, and their derivatives, which allows a hydrodynamic-like interpretation of the Einstein equations, in particular, leading to the notion of gravitational pressure $\P$. Then, does this result hold only for this particular frame? Or does it hold for a general frame? In this section, we will show that the Einstein equations take the form of hydrodynamic equations in any foliation-adapted frame, providing an interpretation of the geometro-hydrodynamics, where an object called "gravitational bubble" appears. This can be achieved by using the local change of frames.\\

Choosing different functions $\sigma(y)$ simply leads to different choices of frame and coframe, $(e_{\mathscr{A}}, e^{\mathscr{B}})$. Since these frame fields span the tangent space to the normal plane and its dual, different choices of frames are related by local linear transformations. Such changes of frames, or frame reorientations, are entirely standard in classical physics. A familiar example is the use of different coordinate bases for vectors in the Euclidean plane: one may choose the Cartesian basis $(\hat{x}, \hat{y})$ or the polar basis $(\hat{r}, \hat{\theta})$, related through the linear transformations $\hat{r} = \cos \theta \, \hat{x} + \sin \theta \, \hat{y}$ and $\hat{\theta} = -\sin \theta \, \hat{x} + \cos \theta \, \hat{y}$. Similarly, in special relativity, different Lorentz frames $e_{\mathscr{A}}$ and $e'_{\mathscr{A}}$ are connected by the Lorentz transformation $e'_{\mathscr{A}} = \Lambda_{\mathscr{A}}{}^{\mathscr{B}} e_{\mathscr{B}}$. An analogous situation holds for the adapted frames on $\NP$, with the important distinction that the frame fields are not arbitrary but are instead constructed to be compatible with the chosen foliation. As in the case of the radial slice, we will distinguish between different sets of frame fields and their associated geometrical quantities using distinct notations, as summarized in Tab.\ref{tab:frames}.

A well-known example illustrating the usefulness of frame transformations is the derivation of the potential of a moving electric point charge \cite{Landau:1975pou}. The potential in a given laboratory frame can be obtained by first going to the comoving frame of the charge, where the solution is given simply by the Coulomb field, and then applying a Lorentz transformation to recover the potential in the laboratory frame. In our case, the radial frame plays a role analogous to the comoving frame of the moving charge, in which the Einstein equations take a simple form, while the frame transformation we will discuss is the analog of the Lorentz transformation. \\

More explicitly, any arbitrary frame $e_{\mathscr{A}} = (\vec{\u},\vec{\k})$ is related to the radial frame $\bdx{e}_{\mathscr{A}} = (\vec{\uu},\vec{\kk})$ via the local linear transformation,
\begin{align}
e_{\mathscr{A}} = \mathbb{M}_{\mathscr{A}}{}^{\mathscr{B}}\bdx{e}_{\mathscr{B}} \, , \qquad \text{for the transformation matrix} \qquad \mathbb{M}_{\mathscr{A}}{}^{\mathscr{B}} = e_{\mathscr{A}}{}^a \bdx{e}_a{}^{\mathscr{B}} \,. 
\end{align}
Correspondingly, the transformation of the dual coframe is 
\begin{align}
e^{\mathscr{A}} = \bdx{e}^{\mathscr{B}}(\mathbb{M}^{-1})_{\mathscr{B}}{}^{\mathscr{A}} \,,
\end{align}
for the inverse $(\mathbb{M}^{-1})_{\mathscr{B}}{}^{\mathscr{A}}$ satisfying $(\mathbb{M}^{-1})_{\mathscr{B}}{}^{\mathscr{C}} \mathbb{M}_{\mathscr{C}}{}^{\mathscr{A}}= \mathbb{M}_{\mathscr{B}}{}^{\mathscr{C}} (\mathbb{M}^{-1})_{\mathscr{C}}{}^{\mathscr{A}} = \delta^{\mathscr{A}}_{\mathscr{B}}$.

Then, the components of the Einstein tensor in the general frame, represented with $G^{(e)}_{\mathscr{A} \mathscr{B}} = e_{\mathscr{A}}{}^a e_{\mathscr{B}}{}^bG_{ab}$, are related to those in the radial frame, $G^{(\bdx{e})}_{\mathscr{C} \mathscr{D}} = \bdx{e}_{\mathscr{C}}{}^a \bdx{e}_{\mathscr{D}}{}^bG_{ab}$, by 
\begin{align}
G^{(e)}_{\mathscr{A} \mathscr{B}} = \mathbb{M}_{\mathscr{A}}{}^{\mathscr{C}}\mathbb{M}_{\mathscr{B}}{}^{\mathscr{D}}G^{(\bdx{e})}_{\mathscr{C} \mathscr{D}} \,.
\end{align}
For example, the component $G_{\u \n}$, which is originally expressed in terms of expansions and accelerations (see \eqref{Ein-Ray}), can be written as a linear combination of the radial-frame components, $G_{\u \n} = c_1 G_{\uu \nn} + c_2 G_{\uu \kk} + c_3 G_{\kk \nn} + c_4 G_{\kk \kk}$, for certain functions $c_1,c_2,c_3$ and $c_4$. Here, the right-hand side is expressed in terms of the Misner-Sharp energy $\m$, the area $\A$, and the pressure $\P$. Our task is to determine the matrix $\mathbb{M}_{\mathscr{A}}{}^{\mathscr{B}}$, which in turn yields the coefficients $c_1$, $c_2$, $c_3$, and $c_4$, and show that $G^{(e)}_{\mathscr{A} \mathscr{B}}$ also exhibit hydrodynamic interpretation.

\subsection{Change of frames} \lb{sec:frame-change}

Let us find the transformation (or reorientation) rule between two foliation-adapted frames. In what follows, we seek the relation between a general frame $e_{\mathscr{A}} = (\vec{\u},\vec{\k})$, with the corresponding dual coframe $e^{\mathscr{A}} = (\k,\n)$, adapted to the general spherical slice $\Sigma_\sigma$ and the radial frame $\bdx{e}_{\mathscr{A}} = (\vec{\uu},\vec{\kk})$, with the coframe $\bdx{e}^{\mathscr{A}} = (\kk,\nn)$, adapted to the radial slice $\Sigma_r$. Note that the procedure and the result presented in this section can be applied to the transformation between any two frames. 

First of all, for a general frame that is not adapted to any underlying structure, one may freely perform arbitrary local linear transformations. This implies that the transformation matrices\footnote{The superscript in $\mathrm{GL}(2,\mathbb{R})^\NP$ means that each point of the plane $\NP$ is assigned its own element of $\mathrm{GL}(2,\mathbb{R})$. More precisely, $\mathrm{GL}(2,\mathbb{R})^\NP = \mathrm{Map}(\NP, \mathrm{GL}(2,\mathbb{R}))$ is the group of $\mathrm{GL}(2,\mathbb{R})$-valued function on $\NP$. } 
belong to the group of spacetime-dependent $2\times 2$ invertible matrices,  $\mathbb{M} \in \mathrm{GL}(2,\mathbb{R})^\NP$. However, the frames used in this work are adapted to the foliation by spherical slices and therefore satisfy certain conditions. As a result, the group of foliation-adapted frame reorientations forms a subgroup of $\mathrm{GL}(2,\mathbb{R})^{\NP}$. In what follows, we determine the explicit form of these frame reorientations. \\

Let us begin by recalling that, for the hypersurface $\Sigma_\sigma$, the transverse coframe field is given by the exact normal form $\n = \rd \sigma$. This also infers that it is closed, $\rd \n =0$. Using the decomposition \eqref{df} of the differential in any frame, we can write $\n$ in terms of the radial frame fields as
\begin{align}
\n = \rd \sigma = \left(\Lie_{\kk}\sigma\right) \nn + \left(\Lie_{\uu}\sigma\right) \kk = \sigma' \nn + \dot{\sigma} \kk,
\end{align}
where we define for convenience $\dot{\sigma} := \Lie_{\uu}\sigma$ and  $\sigma' := \Lie_{\kk}\sigma$ as the temporal and radial derivatives of $\sigma$, respectively. The normalization factor of $\n$ is related to the those of $\nn$ by
\begin{align}
\rho := \frac{1}{2}\n_a\n^a = \left( \dot{\sigma} + \vrho \sigma' \right) \sigma' \, , \lb{norm-rela} 
\end{align}
where we recall that $2\vrho = \nn_a \nn^a$ is related to the gravitational potential or the Misner-Sharp energy by equation \eqref{rho}. 

We next consider the transformations of the remaining bases. 
Stemming from the construction in Sec.\ref{sec:rigging}, the (co)frame fields need to satisfy the pairing conditions \eqref{pairing}, the norm squared condition \eqref{norm}, and the relations between the vertical fields and the transverse fields \eqref{vector-u}. These conditions are
\begin{equation}
\lb{condition}
\begin{aligned}
\k^a \n_a = 1, \qquad \k^a \k_a = 0, \qquad \u^a \k_a = 1, \qquad \u^a \n_a = 0, \\
\u^a\u_a = - \n^a \n_a, \qquad  h_{ab}\k^b = \k_a, \qquad \u^a = h^{ab}\n_b + 2\rho \k^a.  
\end{aligned}
\end{equation}
Solving these equations yields the transformations of the basis fields. Interestingly, there are two sets of transformations preserving the above conditions. The first set of transformations is given by 
\begin{equation}
\lb{frame-decom}
\begin{aligned}
\n = \sigma' \nn + \dot{\sigma} \kk, 
\qquad \k = \frac{1}{\sigma'} \kk,
\qquad \vec{\u} = \sigma' \vec{\uu} - \dot{\sigma} \vec{\kk},
\qquad \text{and}
\qquad \vec{\k} = \frac{1}{\sigma'} \vec{\kk},
\end{aligned}
\end{equation}
and the second one by 
\begin{equation}
\lb{frame-decom-2}
\begin{aligned}
\check{\n} = \sigma' \nn + \dot{\sigma} \kk, 
\quad \check{\k} = \frac{1}{\dot{\sigma}+\vrho\sigma'} \left(\nn - \vrho\kk \right),
\quad \check{\vec{\u}} =  \dot{\sigma} \vec{\kk} -\sigma' \vec{\uu},
\quad \check{\vec{\k}} = \frac{1}{\dot{\sigma}+\vrho\sigma'} \left(\vec{\uu} + \vrho \vec{\kk} \right).
\end{aligned}
\end{equation}
The derivation 
is provided in Appendix \ref{app:frame-change}. We here give an intuitive explanation for the two sets of frame reorientations. 

There are many, but equivalent, ways one can understand why, by solving \eqref{condition}, we arrive at the two distinct transformations between the two frames. First, let us notice that the transformation of the transverse coframe field is completely fixed, and we have $\n = \check{\n}$. This is because, by construction, $\n$ is a normal form to the surface $\Sigma_\sigma$ and is thus fixed (up to sign) by $\n = \rd \sigma$ and obeys $\rd \n = 0$. Then, the form of $\n$ determines the changes of other fields. Let us look at the vertical frame field $\vec{\u}$ satisfying $\u^a \n_a = 0$ and $\u^a \u_a = - \n^a \n_a$. The first condition implies $\vec{\u} = \Omega (\sigma' \vec{\uu} - \dot{\sigma} \vec{\kk})$ for a function $\Omega$, and the second one fixes $\Omega^2 = 1$. This therefore means there are two choices of the vertical vector: $\vec{\u}$ for $\Omega = 1$ and $\check{\vec{\u}}$ for $\Omega = -1$, and $\check{\vec{\u}} = - \vec{\u}$. Different decompositions of the vertical frame, that is either $\u$ or $\check{\u}$, hence lead to different decompositions of $(\vec{\k},\k)$ in the radial frame. 

Another, more geometric, perspective has to do with null vectors. It is a fact that in two dimensions, there exist just two null rays passing through each point of the plane $\NP$. These null rays are generated by null vectors $\vec{\ell}$ and $\vec{\bar{\ell}}$ that are unique up to some scales and can be expanded in any frame. In the radial frame, we can choose $\vec{\bar{\ell}} = - \vec{\kk}$ and $\vec{\ell} = \vec{\uu} + \vrho \vec{\kk}$. Since any null vector has to be proportional to either $\vec{\ell}$ and $\vec{\bar{\ell}}$, it then follows that the null vector $\vec{k}$ has to be proportional to either $\vec{\kk}$ or $\vec{\uu}+ \vrho \vec{\kk}$, leading to the two sets of transformations given hereabove. Accordingly, different decompositions of $\vec{\k}$ result in different decompositions of $\vec{\u}$.\\

Now, let us discuss the properties of these transformation rules. First, there exists a notable distinction between the two sets of transformations: The ones in \eqref{frame-decom} are connected to the identity, while those in \eqref{frame-decom-2} are not. This means that under a continuous deformation of the foliation function, $\sigma(y) \to r(y)$, and consequently $\dot{\sigma} \to 0$ and $\sigma' \to 1$, the transformations \eqref{frame-decom} reduce to $(\vec{\u}, \vec{\k}, \k, \n) \to (\vec{\uu}, \vec{\kk}, \kk, \nn)$, whereas $(\vec{\check{\u}}, \vec{\check{\k}}, \check{\k}, \check{\n}) \to (-\vec{\uu}, \vrho^{-1}\vec{\ell}, \vrho^{-1}\ell, \nn)$ for the transformations \eqref{frame-decom-2}. This property implies that the frame transformations in \eqref{frame-decom-2} are related to those in \eqref{frame-decom} by the action of a discrete transformation $\mathbb{T}$, which reverses the vertical direction, $\vec{\u} \leftrightarrow -\vec{\u}$, and simultaneously swaps and rescales the null directions, $\vec{\bar{\ell}} \leftrightarrow \vrho^{-1}\vec{\ell}$.

This feature, in a sense, is analogous to the Lorentz group $\mathrm{O}(1,3)$, which possesses four disconnected components. The proper orthochronous Lorentz group $\mathrm{SO}(1,3)$ consists of all Lorentz transformations that are connected to the identity element. The remaining three components are disconnected from the identity element and can be reached via discrete transformations such as parity and time reversal. In our case, the discrete transformation $\mathbb{T}$ is analogous to the time reversal transformation.

To make the explanation clearer, let us examine the property of the frame reorientation in matrix form. The frame reorientation $e_{\mathscr{A}} = \mathbb{M}_{\mathscr{A}}{}^{\mathscr{B}} \bdx{e}_{\mathscr{B}}$ given by \eqref{frame-decom} can be written as 
\begin{align}
\label{matrix_M}
e = \mathbb{M} \bdx{e}, \qquad \text{where} \qquad 
e := 
\begin{pmatrix}
\vec{\u} \\[5pt]
\vec{\k}
\end{pmatrix}, 
\qquad
\bdx{e} := 
\begin{pmatrix}
\vec{\uu} \\[5pt]
\vec{\kk}
\end{pmatrix}, 
\qquad \text{and} \qquad
\mathbb{M} :=
\begin{pmatrix}
A & - B\\[5pt]
0 & \dfrac{1}{A}
\end{pmatrix}
\end{align}
where, in our case, $A = \sigma'$ and $B = \dot{\sigma}$. We can easily check that $\det \mathbb{M} =1$. In a similar manner, the transformations \eqref{frame-decom-2} can be written as
\begin{align}
\check{e} = \check{\mathbb{M}} \bdx{e}, \qquad \text{where} \qquad 
\check{e} := 
\begin{pmatrix}
\vec{\check{\u}} \\[5pt]
\vec{\check{\k}}
\end{pmatrix}, 
\qquad \text{and} \qquad
\check{\mathbb{M}} :=
\begin{pmatrix}
-A & B\\[5pt]
\dfrac{1}{B+\vrho A} & \dfrac{\vrho}{B+\vrho A}
\end{pmatrix}
\,,
\end{align}
where $\det \check{\mathbb{M}} = -1$ holds.

This suggests that the frame transformations are contained in the subgroup of $\mathrm{SL}^\pm(2,\mathbb{R})^{\NP}$, consisting of $\mathrm{SL}(2,\mathbb{R})$-valued functions on $\NP$ with determinant $\pm1$. 
Those that have unit determinant are connected to the identity, while those with determinant -1 are obtained by the action of a discrete transformation $\mathbb{T}$ with $\det \mathbb{T} = -1$ (similar to parity and time reversal in the Lorentz group). Indeed, defining the matrix $\mathbb{T}$ as
\begin{align}
\mathbb{T} := 
\begin{pmatrix}
 -1 & 0 \\[5pt]
\rho^{-1} & 1
\end{pmatrix},
\qquad \text{where } \qquad
\rho = A(A\vrho + B),
\end{align}
one can verify that $\mathbb{T}^2 = \mathbb{I}$, and $\check{\mathbb{M}} = \mathbb{T} \mathbb{M}$ and $\mathbb{M} = \mathbb{T} \check{\mathbb{M}}$. Therefore, $\mathbb{T}$ is the analog of time reversal in our case, which flips the vertical (temporal) direction and exchanges the null direction. For the trivial case where $A =1$ and $B = 0$, the discrete transformation gives
\begin{align}
\begin{pmatrix}
 -1 & 0 \\[5pt]
\vrho^{-1} & 1
\end{pmatrix}
\begin{pmatrix}
\vec{\uu} \\[5pt]
\vec{\kk}
\end{pmatrix}
= 
\begin{pmatrix}
- \vec{\uu} \\[5pt]
\vrho^{-1}\vec{\ell}
\end{pmatrix}
\ .
\end{align}

Also, we can consider the analog of parity transformation. Both $\n = \rd \sigma$ and $\n = - \rd \sigma$ can equally serve as the normal form to the slice $\Sigma_\sigma$, and the frames $(-\vec{\u}, -\vec{\k}, -\k,-\n)$ and $(-\vec{\check{\u}}, -\vec{\check{\k}}, -\check{\k},-\check{\n})$, both obeying the conditions \eqref{pairing}, can also serve as the basis for $T\NP$ and its dual. They are related to the original ones by the parity-like transformation, $\sigma \to - \sigma$. \\

In this work, we will only consider the frame transformations that are connected to the identity, i.e., those in the form of the matrix $\mathbb{M}$ \eqref{matrix_M}. Note that, in the group-theoretic language, the transformation matrices $\mathbb{M}$ form a Borel subgroup of $\mathrm{SL}(2,\mathbb{R})^{ \NP}$. We summarize the transformation laws below.

\bbox
\noindent \textbf{\sffamily Frame transformation laws:} Below lists the transformation between the frame $(\vec{\u}, \vec{\k}, \k, \n)$ adapted to a general, $\sigma(y) = \text{constant}$, slice $\Sigma_\sigma$ and the radial frame $(\vec{\uu}, \vec{\kk}, \kk, \nn)$ adapted to the radial, $r(y) = \text{constant}$ slice $\Sigma_r$:
\begin{equation}
\label{f-trans}
\begin{alignedat}{5}
&\n = \sigma' \nn + \dot{\sigma} \kk 
\qquad &&\k = \frac{1}{\sigma'} \kk
\qquad &&\vec{\u} = \sigma' \vec{\uu} - \dot{\sigma} \vec{\kk}
\qquad &&\text{and}
\qquad &&\vec{\k} = \frac{1}{\sigma'} \vec{\kk} 
\end{alignedat}
\end{equation}
and conversely
\begin{equation}
\label{f-trans-con}
\begin{alignedat}{5}
&\nn = \frac{1}{\sigma'} \n - \dot{\sigma} \k
\qquad &&\kk = \sigma' \k
\qquad &&\vec{\uu} = \frac{1}{\sigma'} \vec{\u} + \dot{\sigma} \vec{\k}
\qquad &&\text{and}
\qquad &&\vec{\kk} = \sigma' \vec{\k} 
\end{alignedat}
\end{equation}
\ebox

We have, so far, presented the transformation laws in terms of $\dot{\sigma} = \Lie_{\uu}\sigma$ and $\sigma' = \Lie_{\kk} \sigma$. We can alternatively write them in terms of $\Lie_{\u} r$ and $\Lie_{\k} r$ by using the relations
\begin{align}
\dot{\sigma} = \Lie_{\uu} \sigma = - \Lie_{\u} r \qquad \text{and}\qquad \sigma' = \Lie_{\kk} \sigma = \frac{1}{\Lie_{\k} r}. \label{f-rela}
\end{align}
These relations can be verified using the above frame decompositions and the fact that $\Lie_{\uu} r = 0$ and $\Lie_{\kk} r = 1$.

\subsection{Einstein equation as geometro-hydrodynamics} \label{sec:geohydro}
Now, armed with the Einstein equation in the radial frame (Sec.\ref{sec:Ein}) and the frame transformation laws (Sec.\ref{sec:frame-change}), we are ready to derive the Einstein equation in a general frame and construct the geometro-hydrodynamic picture. 

We start by defining the following quantities:
\begin{align}
K := \vkappa + \frac{\dot{\sigma}}{\sigma'}\bvkappa, \qquad P := \frac{1}{8\pi\Newton}K -  \left(\frac{d-3}{d-2}\right) \E, \qquad \text{and} \qquad  \overline{P} := \frac{1}{\left( \sigma'\right)^2} \frac{\bvkappa}{8\pi\Newton}. \lb{P-gen}
\end{align}
It is straightforward to derive the following relation between $P$ and $\P$:  
\begin{align}
P = \P + \frac{\dot{\sigma}}{\sigma'} \bP,
\label{P-rela}
\end{align}
where we used the definitions \eqref{pressure} and \eqref{bP}. In the case when $\sigma = r$, we have $P = \P$.\\

Let us now consider the components of the Einstein equation with respect to the general frame $(\vec{\u}, \vec{\k}, \k, \n)$. The first component is $G_{\u \n} = 8\pi\Newton T_{\u \n}$. We first call for the change of frames given in \eqref{f-trans} and write $T_{\u \n} = \u^a T_a{}^b \n_b$ as
\begin{align}
T_{\u \n}  = \left( \sigma' \uu^a - \dot{\sigma} \kk^a \right) T_a{}^b \left( \sigma' \nn_b + \dot{\sigma} \kk_b \right)
= \left(\sigma'\right)^2 T_{\uu \nn} + \sigma' \dot{\sigma} \left( T_{\uu \kk} - T_{\kk \nn} \right) - \left( \dot{\sigma} \right)^2 T_{\kk \kk}. 
\end{align}
We next recall the components of the Einstein equation in the radial frame given in Sec.\ref{sec:Ein}. Then, we can show that
\begin{equation}
\begin{aligned}
T_{\u \n} &=  \left(\sigma'\right)^2 \Lie_{\uu} \E - \sigma' \dot{\sigma} \left( \Lie_{\kk} \E + (\E+\P)\vtheta\right) - \left( \dot{\sigma} \right)^2 \bP \vtheta \\
\step{[\text{rearranging and factoring} ]} \qquad 
& = \sigma' \left[ \left( \sigma' \Lie_{\uu} - \dot{\sigma} \Lie_{\kk}\right)\E - \left(\E+\P + \frac{\dot{\sigma}}{\sigma'}\bP \right) \dot{\sigma} \vtheta \right] \\
\step{[\text{use \eqref{f-trans}, \eqref{f-rela}, and \eqref{P-rela}} ]} \qquad  
& = \sigma' \left[ \Lie_{\u}\E + \left(\E+P \right) \vtheta (\Lie_{\u}r) \right] \\
\step{[\text{use \eqref{expansion-basis}}]} \qquad 
& = \sigma' \left[ \Lie_{\u}\E + \left(\E+P \right) \theta_{(\u)} \right],
\end{aligned}
\end{equation}
where we explain each calculation step in \step{[.....]}. We therefore arrive at the Misner-Sharp energy evolution equation 
\bbox
\vspace{-13pt}
\begin{flalign}
&\text{\sffamily \textbf{Energy evolution equation:}} &&\Lie_{\u}\E + \left(\E+P \right) \theta_{(\u)} = (\sigma')^{-1} T_{\u\n}&& \label{Ein-un}
\end{flalign}
\ebox

\noindent This equation governs the evolution of the Misner-Sharp energy density $\E$ along the vertical direction $\vec{\u}$ on the surface $\Sigma_\sigma$ sourced by the matter energy flow $T_{\u \n}$ (see the left of Fig.\ref{fig:summary}). This can also be expressed in another form:  
\begin{align}
\nabla \dd \left( \E \vec{\u} \right) + P \left(\nabla \dd \vec{\u}\right) = Q^{\sss \mathrm{ext}} \qquad \text{with} \qquad Q^{\sss \mathrm{ext}} := (\sigma')^{-1} T_{\u\n} \,. \lb{Euler}
\end{align}
where we used $\theta_{(\u)} = \nabla_\mu \u^\mu = \nabla \dd \vec{\u}$ from \eqref{div}. This has exactly the same form as the energy sector of the Euler equations of a perfect fluid with internal energy density $\E$ and velocity field $\vec{\u}$ under an external energy source $Q^{\sss \mathrm{ext}}$ \cite{landau1987fluid} (where the coefficient $\sigma'$ can be viewed as arising from the Jacobian of the frame transformation). 
Taking $\sigma=r$, we have $ \theta_{(\u)}\to  \theta_{(\uu)}=0$ and $\sigma'\to r'=1$, and the energy conservation law \eqref{Ein-un} for a general frame reduces to the one \eqref{G-rad-un} for the radial frame. Note that, considering a general frame, $P$ \eqref{P-rela} appears naturally as the pressure in \eqref{Ein-un}.

The next component is $G_{\k \n} = 8\pi\Newton T_{\k \n}$. Using again the frame transformation \eqref{f-trans}, we can write $T_{\k \n} = T_{\kk \nn} + \frac{\dot{\sigma}}{\sigma'} T_{\kk \kk}$. Then, by recalling the Einstein equation \eqref{G-rad-kn} and \eqref{bP}, as well as the definition of the pressure \eqref{P-rela}, we can obtain
\bbox
\vspace{-13pt}
\begin{flalign}
&\text{\sffamily \textbf{Gravitational Young-Laplace equation:}} &&P\vtheta = T_{\k \n}.&& \label{Ein-kn}
\end{flalign}
\ebox
\noindent This is exactly the same form as the Young-Laplace equation for the surface pressure $P$ (see the right of Fig.\ref{fig:summary}). The difference from the radial-frame one \eqref{G-rad-kn} is the replacement of $\P$ with $P$ (except for $T_{\k \n}$), and $P$ plays the role of the gravitational pressure in a general frame. 

Similarly, we can employ the frame decomposition \eqref{f-trans} and represent the component $G_{\u \k} = 8\pi\Newton T_{\u \k}$ as 
\begin{equation}
\begin{aligned}
-T_{\u \k} &= -T_{\uu \kk} + \frac{\dot{\sigma}}{\sigma'} T_{\kk \kk} \\
\step{\text{[use \eqref{G-rad-uk} and \eqref{bP}]}} \qquad \quad 
& = \Lie_{\kk}\E + \E \vtheta +  \frac{\dot{\sigma}}{\sigma'} \bP \vtheta\\
\step{\text{[use \eqref{f-trans-con} and factoring]}} \qquad \quad
& = \sigma'\left[ \Lie_{\k} \E + \frac{1}{\sigma'}\E \vtheta +   \frac{\dot{\sigma}}{(\sigma')^2} \bP \vtheta  \right]\\
\step{\text{[use \eqref{f-rela} and $\overline{P}$ from \eqref{P-gen}]}} \qquad \quad
& = \sigma'\left[ \Lie_{\k} \E + \E \vtheta \Lie_{\k} r -  \overline{P} \vtheta \Lie_{\u} r  \right] \\
\step{\text{[recall the expansions \eqref{expansion-basis}]}} \qquad \quad
& = \sigma'\left[ \Lie_{\k} \E + \E \theta_{(\k)} -  \overline{P} \theta_{(\u)}  \right]. 
\end{aligned}
\end{equation}
Therefore, we derive the energy distribution equation:
\bbox
\vspace{-13pt}
\begin{flalign}
&\text{\sffamily \textbf{Energy distribution equation:}} &&\Lie_{\k}\E + \E \theta_{(\k)} -  \overline{P} \theta_{(\u)} = S^{\sss \mathrm{ext}}. &&  \label{Ein-uk}
\end{flalign}
\ebox
\noindent where we define the matter source
\begin{align}
S^{\sss \mathrm{ext}} := -(\sigma')^{-1} T_{\u\k}.
\end{align}
Given the form of $\overline{P}$ and the matter contribution $S^{\sss \mathrm{ext}}$, this equation governs how the Misner-Sharp energy density $\E$ is distributed along the transverse direction generated by $\vec{\k}$. Note that, since $\vec{\k} \sim \vec{\kk}$ as follows from \eqref{f-trans}, the transverse null direction also corresponds to the radial direction. (Again, $\sigma'$ arises as the Jacobian of the frame transformation.) 

We are left with the component $G_{\k \k} = 8\pi\Newton T_{\k \k}$. Again, by simply using the change of frame \eqref{f-trans} and the definition \eqref{P-gen} of $\overline{P}$, we derive the equation that determines $\overline{P}$:
\bbox
\vspace{-13pt}
\begin{flalign}
&\text{\sffamily \textbf{Inaffinity equation:}} &&\overline{P}\vtheta = T_{\k \k}&& \label{Ein-kk}
\end{flalign}
\ebox
\noindent Note again that this equation is not an independent component; it is determined through the relation $G_{\n \k} - G_{\u \k} = 2\rho G_{\k\k}$ (from $\n^a = \u^a + 2\rho \k^a$) by the other normal components. 

Furthermore, the equation for the tangent direction, $G^A{}_A=8\pi\Newton T^A{}_A $, exists (as in \eqref{G_AB}), but it automatically holds through the Bianchi identity $\nabla_\mu G^{\mu\nu}=0$ and the energy-momentum conservation $\nabla_\mu T^{\mu\nu}=0$ if the other components are satisfied. Therefore, the three independent equations \eqref{Ein-un}, \eqref{Ein-kn} and \eqref{Ein-uk}, together with the matter-field equations, determine the spherically symmetric spacetime. Indeed, we have seen an example of this procedure when deriving Birkhoff's theorem at the end of Sec.\ref{sec:slices}.\\

Now, we discuss the geometro-hydrodynamic interpretation of this mathematical framework. The Misner-Sharp energy density $\E$ and the gravitational pressure $P$ appear consistently in \textit{both} the gravitational Euler equation for energy \eqref{Ein-un} and the gravitational Young-Laplace equation \eqref{Ein-kn}. Noting from Birkhoff's theorem that the non-trivial dynamics in spherically symmetric spacetime stems from the coupling to matter degrees of freedom, this fact suggests that a mixture of gravity and matter fields on $\S_r$ corresponds to a gravitational fluid on $\S_r$ with the energy density $\E$ and pressure $P$. Here, the fluid does not flow in the tangential direction $\partial_A$ because of spherical symmetry, but rather just contracts or expands. Therefore, the gravitational fluid on $\S_r$ can be considered as a spherical collective mode of its (still unknown) microscopic constituents that compose the spacetime and also encode the effect of the matter fields. We call it a \emp{"gravitational bubble"}. The slice $\Sigma_\sigma$ is the worldvolume of a gravitational bubble on $\S_r$.\footnote{A similar picture, called a "gravitational screen", was proposed in \cite{Freidel:2014qya}, but the dictionary is different from ours and applicable to a general spacetime.} Furthermore, considering many gravitational bubbles, a picture of the bulk space emerges: the spherically symmetric spacetime is the worldvolume of a concentric stacking of many gravitational bubbles. See Fig.\ref{fig:summary}.

Here, the energy distribution equation \eqref{Ein-uk} determines how to connect gravitational bubbles. Note that this gravitational equation does not have a direct counterpart in the standard hydrodynamics, where the evolution equation is given by the conservation law $\nabla_\mu T^{\mu\nu} = 0$. For a usual perfect fluid, the temporal component of this conservation law yields the Euler equation for energy density $e$ as in \eqref{Euler}, while the radial component provides an equation governing the radial evolution of pressure $p$. To obtain the radial evolution of energy $e$, one must invoke an equation of state, $p=p(e)$, which is not provided a priori in the case of the geometro-hydrodynamics. Instead, in gravity, the radial distribution of $\E$ is determined directly from the Einstein equation \eqref{Ein-uk}. This distinction highlights a key feature that sets the geometro-hydrodynamics apart from the standard hydrodynamics.\\

We conclude this section by giving alternative (and simpler) expressions for the two evolution equations \eqref{Ein-un} and \eqref{Ein-uk}. We first note that the two Young-Laplace-like equations \eqref{Ein-kn} and \eqref{Ein-kk}, which do not have Lie derivatives, can be viewed as the constraint equations for $P$ and $\overline{P}$. In the absence of the matter contribution where $T_{\k \n} =0$ and $T_{\k \k} =0$, the two equations make $P$ and $\overline{P}$ vanish. In generic cases, we can substitute the Young-Laplace equation \eqref{Ein-kn} into the Euler equation \eqref{Ein-un}, to obtain
\begin{align}
\Lie_{\u}\E + \E \theta_{(\u)} = (\sigma')^{-1} T_{\u\n} +\dot{\sigma} T_{\k \n} = T_{\uu \n},
\end{align}
where we used the relation $\theta_{(\u)}=-\vartheta \dot \sigma$ (from \eqref{expansion-basis} and \eqref{f-rela}) and the frame transformation \eqref{f-trans-con}. This is the component $G_{\uu \n} = 8\pi \Newton T_{\uu \n}$ of the Einstein equation in the mixed frames between the general one and the radial one. Using $\m = \E \A$, we can write this as the temporal development of the Misner-Sharp energy $\m$ as 
\begin{align}
\Lie_{\u} \m  = \A T_{\uu \n}. 
\label{u-M}
\end{align}
Similarly, applying \eqref{Ein-kk} to \eqref{Ein-uk} yields the third equation in \eqref{summary_eq}:
\begin{align}
\Lie_{\k}\E + \E \theta_{(\k)} = -(\sigma')^{-1} T_{\u\k} -\dot{\sigma} T_{\k \k} = -T_{\uu \k},
\end{align}
which provides the equation that controls the transverse profile of $\m$:
\begin{align}
\Lie_{\k} \m  = -\A T_{\uu \k}. 
\label{k-M}
\end{align}
Therefore, these two equations \eqref{u-M} and \eqref{k-M} together with one constraint equation, say, \eqref{Ein-kn}, determine the spacetime. Note that \eqref{u-M} and \eqref{k-M} generalize the evolution equations \eqref{G-rad-un} and \eqref{G-rad-uk} along the Kodama vector and its transverse vector to arbitrary directions.

\section{Equipotential Slices \& Apparent Horizon} \lb{sec:AH}

We have discussed in detail the geometry of a general spherical hypersurface $\Sigma_\sigma$ and how the gravitational dynamics can be reinterpreted through the perspective of geometro-hydrodynamics. As an application, 
we now turn to the apparent horizon, which is defined as the hypersurface where the outgoing null expansion vanishes \cite{Poisson:2009pwt}.

Let us begin with the conventional treatment of the apparent horizon that lies in the uses of the double-null frames $\left(\vec{\ell}, \vec{\bar{\ell}}\right)$, where both null vectors satisfy $g(\vec{\ell},\vec{\ell})= g(\vec{\bar{\ell}}, \vec{\bar{\ell}}) =0$ and $g( \vec{\ell}, \vec{\bar{\ell}}) = \ell^a \bar{\ell}_a =-1$. They can be expressed in terms of the radial frames as follows\footnote{Both $\vec{\u}+\rho\vec{\k}$ and $-\vec{\k}$ are also null vectors. They are proportional to $\vec{\ell}$ and $\vec{\bar{\ell}}$ as one can check using the frame transformations \eqref{f-trans} that $\vec{\u}+\rho\vec{\k} = \sigma' \vec{\ell}$ and $-\vec{\k} = \frac{1}{\sigma'} \vec{\bar{\ell}}$.  }
\begin{align}
\vec{\ell} = \vec{\uu} + \vrho \vec{\kk}, \qquad \text{and} \qquad \vec{\bar{\ell}} = - \vec{\kk}. 
\end{align}

In general, the outgoing null vector $\vec{\ell}$ is not tangent to the radial slice $\Sigma_r$. It generates outgoing null geodesics, obeying 
\begin{align}
\nabla_{\vec{\ell}} \vec{\ell} = \kappa_{(\ell)} \vec{\ell}, \qquad \text{where} \qquad \kappa_{ (\ell)} = \vkappa + \vrho \bvkappa,
\end{align}
which can be derived using the covariant derivatives given in \eqref{nabla-inv}. In addition, it trivially follows from \eqref{k_geodesic} that $\vec{\bar{\ell}}$ is tangent to the ingoing null geodesics, obeying $\nabla_{\vec{\bar{\ell}}} \vec{\bar{\ell}} = \kappa_{ (\bar{\ell})} \vec{\bar{\ell}}$ where $\kappa_ { (\mathrm{\bar{\ell}})} = -\bvkappa$. The expansions of the null geodesics (see equations \eqref{expansion-V} and \eqref{expansion-basis}) are given by 
\begin{align}
\theta_{(\ell)} = \vrho \vtheta, \qquad \text{and} \qquad  \theta_{(\bar{\ell})} = - \vtheta. 
\end{align}
Since the value of the area expansion $\vtheta$ is always positive, the sign of the outgoing expansion $\theta_{(\ell)}$ depends on the sign of the function $\vrho$. The apparent horizon, denoted $\H$, is defined as the surface on which the outgoing expansion vanishes:
\begin{align}
\theta_{(\ell)} \stackrel{\AH}{=} 0 \qquad \text{implying} \qquad \vrho \stackrel{\AH}{=}0. 
\end{align}

This condition is consistent with the definition \eqref{AH-def}. Indeed, recalling the relation $\vrho = \frac{1}{2} + \Phi$ from \eqref{rho}, the condition $\vrho \stackrel{\AH}{=}0$ means $\Phi \stackrel{\AH}{=}-\frac{1}{2}$, leading to 
\begin{align}
r^{d-3} \stackrel{\sss \AH}{=} \frac{1}{(d-2)}\frac{16\pi\Newton}{\mr{\A}} \m 
\label{AH-condition} \,,
\end{align}
which is \eqref{AH-def}. It should be noted that from the perspective of spherical slices, the apparent horizon $\H$ belongs to a family of \emp{equipotential slices} $\Sigma_\vrho$ and is specified by $\H = \Sigma_{\vrho =0}$. \\

We now apply the rigging technique in Sec.\ref{sec:rigging}, to apparent horizons. We set $\sigma(y) = \vrho(y)$ and consider equipotential slices $\Sigma_\vrho$. We denote the adapted frames and coframes with $(\vec{\uAH}, \vec{\kAH})$ and $(\kAH, \nAH)$, respectively (see also Tab.\ref{tab:frames}). As in Sec.\ref{sec:Hydro}, to connect the geometry and dynamics of $\Sigma_\vrho$ with physical quantities, we apply the transformation rule \eqref{f-trans} and express these equipotential (co)frame fields in terms of radial (co)frame fields:
\begin{equation}
\label{basis-AH}
\begin{aligned}
\nAH = \dot{\vrho} \kk + \vrho' \nn \, ,
\qquad \kAH = \frac{1}{\vrho'} \kk \, ,
\qquad \vec{\uAH} = \vrho' \vec{\uu} - \dot{\vrho} \vec{\kk} \, ,
\qquad \text{and} 
\qquad \vec{\kAH} = \frac{1}{\vrho'} \vec{\kk} \, ,
\end{aligned}
\end{equation}
where, again, we adopt the notation $\dot{\vrho} = \Lie_{\uu} \vrho$ and $\vrho' =\Lie_{\kk} \vrho$. It follows from the construction that $\uAH^a \DN_a \vrho =0 $, inferring that the vertical frame $\vec{\uAH}$ is tangent to the constant-$\vrho$ surface $\Sigma_\vrho$. 

The apparent horizon $\H$ can be locally timelike, spacelike, or null. 
Using our formalism, we can construct a simple formula that determines locally the causal nature of the apparent horizon.
The causal nature of the equipotential surface $\Sigma_\vrho$ is determined by the norm squared $2\rhoAH = \nAH_a \nAH^a$ of the normal form. 
Setting $\sigma=\varrho$ in \eqref{norm-rela} leads to
\begin{align}
\rhoAH = \dot{\vrho}\vrho' + \vrho\left(\vrho'\right)^2.
\end{align}
Here, each derivative of $\vrho$ can be evaluated on the horizon, through \eqref{rho}, as follows. 
\begin{align}
\dot{\vrho} 
= - \frac{8\pi\Newton}{(d-2)\mr{\A}r^{d-3}}\Lie_{\uu}\m 
\stackrel{\AH}{=} - \frac{1}{2\m} \Lie_{\uu}\m 
=-\frac{1}{2\E} T_{\uu \nn},
\label{Lie-u-rho} 
\end{align}
where the horizon condition \eqref{AH-condition} and the Einstein equation \eqref{G-rad-un} are applied, and 
\begin{align}
\vrho' &= - \frac{8\pi\Newton}{(d-2)\mr{\A}r^{d-3}}\left(\Lie_{\kk}\m - \frac{d-3}{r}\m \right)  \stackrel{\AH}{=} \vkappa\nonumber\\
&= \frac{8\pi\Newton}{(d-2)\mr{\A}r^{d-3}}\left(\A T_{\uu \kk}+  \frac{d-3}{r}\m \right)\nonumber\\
&\stackrel{\AH}{=} \frac{1}{2\E}T_{\uu \kk}+\frac{d-3}{2r},
\label{Lie-k-rho}
\end{align}
where the horizon condition \eqref{AH-condition}, the relation \eqref{kappa} of $\vkappa$, and the Einstein equation \eqref{G-rad-uk} are used.\footnote{We see from \eqref{basis-AH} that on the slices $\Sigma_\vrho$, the components of the equipotential (co)frame fields are controlled by the Einstein equation. Observe that on the apparent horizon, the radial component $\vrho'$ is given by the acceleration (see \eqref{kappa}) $\vkappa$ of the Kodama vector.} \eqref{Lie-k-rho} provides a formula for the surface gravity on the apparent horizon, $\varkappa|_{\AH}$, including the effect of the matter fields. Combining these, we thus obtain the formula:
\begin{align}
\rhoAH \stackrel{\AH}{=} -\frac{\vkappa}{2\E} T_{\uu \nn}=-\frac{1}{4\E^2}T_{\uu \nn}\left(T_{\uu \kk}+(d-3)\frac{\E}{r}\right).
\label{rho_formula}
\end{align}
This determines locally the causal character of $\H$ depending on the energy-momentum tensor on the apparent horizon. For $T_{\uu \nn} = 0$, it is null locally, and if $T_{\mu\nu}$ vanishes entirely on $\H$, it is the event horizon. When classical matter enters the horizon from outside, $T_{\uu \nn}$ and $T_{\uu \kk}$ are positive, and thus $\rhoAH$ is negative, meaning that the apparent horizon expands in a spacelike manner \cite{Poisson:2009pwt}. Finally, to discuss an example of $T_{\uu \nn}<0$, let us consider the effect of vacuum fluctuations around the apparent horizon for $d=4$. In this case, the apparent horizon is located at $r=r_{\sss \H}=2\Newton \m$, and the Misner-Sharp energy density becomes $\E_{\sss \H}=\frac{1}{8\pi \Newton r_{\sss \H}}$. In the conventional scenario of evaporating black holes, we have $T_{\uu \nn}\sim T_{\uu \kk}\sim -\frac{\hbar}{r_{\sss \H}^4}$ around the apparent horizon \cite{Birrell:1982ix}. Therefore, we can estimate \eqref{rho_formula} as $\rhoAH\sim \Newton^2 r_{\sss \H}^2 \times \frac{\hbar}{r_{\sss \H}^4} \times \left(-\frac{\hbar}{r_{\sss \H}^4}+\frac{1}{\Newton r_{\sss \H}^2}\right)\sim \frac{\Newton\hbar}{r_{\sss \H}^4}>0$, and the apparent horizon is timelike there. \\

Next, we discuss the geometro-hydrodynamic equations in the equipotential frame. The components of the Einstein equation in this frame automatically follow from the ones provided in Sec.\ref{sec:geohydro} by setting $\sigma(y) = \vrho(y)$ and $(\vec{\u}, \vec{\k}, \k, \n) = (\vec{\uAH}, \vec{\kAH}, \kAH, \nAH)$:
\begin{subequations}
\lb{Ein-fluid}
\begin{align}
\left(\Lie_{\uAH} + \theta_{(\uAH)}\right)\E + \PAH \theta_{(\uAH)} &= \left( \vrho' \right)^{-1}T_{\uAH \nAH},  \lb{EF-1} \\
\left(\Lie_{\kAH} + \theta_{(\kAH)}\right)\E - \bPAH \theta_{(\uAH)} &=  -\left( \vrho' \right)^{-1}T_{\uAH \kAH}, \lb{EF-2} \\
\PAH \vtheta &= T_{\kAH \nAH}, \lb{EF-3} \\
\bPAH \vtheta & = T_{\kAH \kAH}, \lb{EF-4}
\end{align}
\end{subequations}
where the definition \eqref{P-gen} provides straightforwardly
\begin{align}
\KAH := \vkappa + \frac{\dot{\vrho} }{\vrho'}\bvkappa, \qquad \PAH := \frac{1}{8\pi\Newton}\KAH -  \left(\frac{d-3}{d-2}\right) \E, \qquad \text{and} \qquad  \bPAH := \frac{1}{\left( \vrho' \right)^2} \frac{\bvkappa}{8\pi\Newton}. \lb{P-AH}
\end{align}
This describes the apparent horizon as the special gravitational bubble that satisfies the condition \eqref{AH-condition}, and replaces the conventional black-hole membrane paradigm. \\

We conclude this section with a few mathematical results. The first one is a dual picture of dynamics in the surface area $\A$ and the energy $\m$. Using the relation \eqref{f-rela}, we can express the expansion $\theta_{(\uAH)}$ along the vertical vector $\vec{\uAH}$ tangent to the equipotential surface $\Sigma_\vrho$ as $\theta_{(\uAH)} = \vtheta \Lie_{\uAH} r = - \vtheta \Lie_{\uu} \vrho$. Then, recalling the definition of the expansion $\theta_{(\uAH)} = \Lie_{\uAH} \ln \A$ and the expression \eqref{Lie-u-rho} for $\dot{\vrho} = \Lie_{\uu}\vrho$ in terms of $\Lie_{\uu} \m$, 
we obtain 
\begin{align}
\Lie_{\uAH} \A = 8\pi\Newton \Lie_{\uu} \m = 8\pi\Newton \A T_{\uu \nn} \, .
\end{align}
This relation offers us a dual interpretation of the component $G_{\uu \nn} = 8\pi\Newton T_{\uu \nn}$ of the Einstein equation. On one hand, it dictates the temporal evolution, i.e., the flow along the Kodama vector $\vec{\uu}$, of the Misner-Sharp energy $\m$. On the other hand, it tells how the area $\A$ of the spherical shell $\S_r$ evolves along $\vec{\uAH}$, the direction of the equipotential slice $\Sigma_\vrho$. Note that if one follows a suggestion that the entropy of the apparent horizon is given by its area \cite{Hayward:1998ee,Hollands:2024vbe,Visser:2024pwz}, $\Ent = \frac{1}{4\Newton \hbar} \A$, the latter interpretation would provide the balance law for the entropy,
\begin{align}
\Lie_{\uAH} \Ent \stackrel{\AH}{=} \frac{2\pi}{\hbar} \A T_{\uu \nn} \,. 
\end{align}

The next result is the relation between $\KAH$ and the vertical acceleration $\kappaAH$. By definition, the vertical accelerations in the radial frame and the equipotential frame are $\vkappa = \kk_a \DN_{\vec{\uu}} \uu^a = - \uu^a \DN_{\vec{\uu}} \kk_a$ and $\kappaAH = \kAH_a \DN_{\vec{\uAH}}\uAH^a = - \uAH^a \DN_{\vec{\uAH}}\kAH_a$, respectively. They are related by the frame transformation \eqref{basis-AH}. We show that 
\begin{equation}
\begin{aligned}
\kappaAH &= - \vrho' \left( \uu^a - \dot{\vrho} \kAH^a \right)\DN_{\vec{\uAH}} \kAH_a  \\
\step{[\text{using $\kAH_a\kAH^a =0$} ]} \qquad 
& = -\vrho' \uu^a \DN_{\vec{\uAH}}\left(\frac{1}{\vrho'} \kk_a\right) \\
\step{[\text{Leibniz rule and $\kk_a\uu^a =1$} ]} \qquad 
& = - \uu^a \DN_{\vec{\uAH}} \kk_a + \Lie_{\uAH} \ln \vrho' \\
\step{[\text{using \eqref{basis-AH}} ]} \qquad 
& =  - \vrho' \uu^a \DN_{\vec{\uu}} \kk_a + \dot{\vrho} \uu^a \DN_{\vec{\kk}} \kk_a + \Lie_{\uAH} \ln \vrho' \\
&= \vrho' \vkappa + \dot{\vrho}\bvkappa +  \Lie_{\uAH} \ln \vrho'
\end{aligned}
\end{equation}
where to obtain the last equality, we recalled the definition of the transverse acceleration $\bvkappa := \nn_a \DN_{\vec{\kk}}\kk^a = \uu^a\DN_{\vec{\kk}}\kk_a$. We therefore obtain the relation
\begin{align}
\kappaAH = \vrho' \KAH  +  \Lie_{\uAH} \ln \vrho'. 
\end{align}

Here, in the context of black hole thermodynamics, the surface gravity $\vkappa$, when evaluated on the event horizon, plays the role of the temperature for a stationary black hole. In dynamical situations, it would be natural to expect that the temperature of a dynamical apparent horizon is associated with the acceleration $\kappaAH$ of the vertical vector $\vec{\uAH}$ tangential to the equipotential slice $\Sigma_\vrho$. However, we will show in our follow-up work \cite{JYreport} that the temperature of the apparent horizon should instead be related to $\KAH$.

\section{Geometro-hydrodynamics of Lovelock Gravity} \label{sec:LL} 

In the final section, we shift our attention to a theory beyond general relativity, focusing on Lovelock gravity. Up to this point, we have worked within Einstein's gravity 
and developed the hydrodynamic description of spacetime geometry and dynamics. 
Natural questions are whether this correspondence stems from the specific simplicity of the Einstein equations, and whether a similar correspondence holds in other theories of gravity, including higher-curvature, higher-derivative, and non-minimally coupled theories. Some progress in this direction has been made in the context of the black hole membrane paradigm \cite{Chatterjee:2010gp, Jacobson:2011dz, Kolekar:2011gg}. In the following, we will perform (almost) the same analysis for Lovelock gravity as we did on Einstein gravity, to show that a generalized geometro-hydrodynamics holds.

\emp{Lovelock gravity} (or Lanczos-Lovelock gravity) \cite{Lovelock:1971yv, Kastor:2006vw} (see also a comprehensive review \cite{Padmanabhan:2013xyr}) is a natural higher-dimensional generalization of the standard four-dimensional theory of general relativity. It includes higher-curvature terms whose equations of motion remain at most second-order in derivatives of the metric tensor, much like the Einstein equation. The gravity sector of the Lovelock Lagrangian density is constructed from antisymmetric contractions of the Riemann tensor, 
\begin{align}
\mathscr{L}_{\sss \mathrm{L}} := \frac{\sqrt{-g}}{16\pi\Newton}  \sum_{q=0}^{\lfloor d/2 \rfloor} c_{(q)} \mathscr{L}_{(q)} , \qquad \text{with} \qquad \mathscr{L}_{(q)} := \frac{1}{2^q}\delta^{\mu_1 ...\mu_q \nu_1 ... \nu_q}_{\rho_1 ...\rho_q \sigma_1 ... \sigma_q} R_{\mu_1\nu_1}{}^{\rho_1\sigma_1} ... R_{\mu_q\nu_q}{}^{\rho_q\sigma_q}
\end{align}
where $c_{(q)}$ is a coupling constant, the floor function $\lfloor d/2 \rfloor$ is the largest integer not exceeding $d/2$, and the delta symbol 
denotes the normalized, totally antisymmetric product of $2q$ Kronecker deltas. For instance, for $d=4$, the Lovelock Lagrangian contains the Einstein-Hilbert term with a cosmological constant, as well as the Gauss-Bonnet term, which is topological and does not contribute to the equations of motion. This feature persists in higher dimensions: the $q^{\mathrm{th}}$-order Lovelock Lagrangian density is non-zero only when $d \geq 2q$ and becomes topological when $d = 2q$. When coupling to matter fields, the corresponding Lovelock equations of motion are
\begin{align}
\esstix{G}_\alpha{}^\beta = 8\pi\Newton T_\alpha{}^\beta, \lb{LL-eom}
\end{align}
where $T_\alpha{}^\beta$ is the matter energy-momentum tensor and the Lovelock tensor $\esstix{G}_\alpha{}^\beta$ is given by 
\begin{align}
\esstix{G}_\alpha{}^\beta := \sum_{q=0}^{\lfloor d/2 \rfloor} c_{(q)} \esstix{G}^{(q)}_\alpha{}^\beta, \qquad \text{where} \qquad  \esstix{G}^{(q)}_\alpha{}^\beta := -\frac{1}{2^{q+1}}\delta^{\beta\mu_1 ...\mu_q \nu_1 ... \nu_q}_{\alpha\rho_1 ...\rho_q \sigma_1 ... \sigma_q} R_{\mu_1\nu_1}{}^{\rho_1\sigma_1} ... R_{\mu_q\nu_q}{}^{\rho_q\sigma_q}. 
\end{align}
For $d=4$, this reduces to the Einstein equation with a cosmological constant.  

In spherically symmetric spacetimes, the Lovelock tensor simplifies, and its expression has been given by Maeda, Willison, and Ray \cite{Maeda:2011ii}. As in general relativity, the Lovelock tensor factorizes into normal components, $\esstix{G}_a{}^b$, and angular components, $\esstix{G}_A{}^B = \frac{1}{d-2} (\esstix{G}_C{}^C) \delta_A{}^B$. In what follows, we focus only on the normal components of the Lovelock equation, which can be expressed in our variables as
\begin{align}
\frac{\esstix{G}_a{}^b}{8\pi\Newton}  = \sum_{q=0}^{\lfloor d/2 \rfloor} c_{(q,d)}\left( q \frac{\E^{q-1}}{r^{q-1}} \frac{G_a{}^b}{8\pi\Newton} + (q-1)(d-1) \frac{\E^{q}}{r^{q}}\delta_a^b \right)
\end{align}
where we define 
\begin{align}
c_{(q,d)} := c_{(q)} \frac{(d-1)!}{(d-2q-1)!} \left(\frac{16\pi\Newton}{d-2}\right)^{q-1}.
\end{align}

For the purposes of the discussion that follows, let us define the following field-dependent, dimensionless quantities:\footnote{In the natural unit where $c=1$, the length-dimensions of various variables include $[r] = 1$, $[\m] = -1$,  $[\E] =[\P] = -3$, $[\Newton] = 2$, $[T_{\alpha\beta}] = -4$, $[\esstix{G}_\alpha{}^\beta] = -2$, $[c_{(q)}] = 2(q-1)$, and $[c_{(q,d)}] = 4(q-1)$.}
\begin{align}
\esstix{a}_{(d)}\left(r,\E\right) := \sum_{q=0}^{\lfloor d/2 \rfloor} c_{(q,d)} \frac{\E^{q-1}}{r^{q-1}},\qquad \text{and} \qquad  \esstix{b}_{(d)}\left(r,\E\right) := \sum_{q=0}^{\lfloor d/2 \rfloor} c_{(q,d)} \frac{\E^{q-1}}{r^{q-1}}q. 
\end{align}
The Lovelock tensor $\esstix{G}_a{}^b$ is therefore related to the Einstein tensor $G_a{}^b$ via an affine transformation\footnote{For a vector space $V$, whose typical examples include functions, vectors, matrices, or tensors when viewed as elements of suitable vector spaces, an affine transformation is a map of the form $X' = L(X) + T$, where $X, X' \in V$, $L: V \rightarrow V$ is a linear map on $V$, and $T$ is a fixed element of $V$. In our case, the linear map is just a multiplication by a scalar. However, the transformations constructed from $\esstix{a}_{(d)}$ and $\esstix{b}_{(d)}$ are local and field-dependent.},
\begin{align}
\frac{\esstix{G}_a{}^b}{8\pi\Newton} = \esstix{b}_{(d)} \frac{G_a{}^b}{8\pi\Newton} + \left(\esstix{b}_{(d)} - \esstix{a}_{(d)}\right) \left( \frac{(d-1)\E}{r} \delta_a^b \right).
\end{align}
It can be expected from this structure that it is possible to express the components of the Lovelock equation in terms of the frame and physical quantities used in the geometro-hydrodynamics for the Einstein gravity; indeed, this is the case.\\

In the radial frame, the normal components of the Lovelock equation read
\begin{align} \label{LL-eq}
\Lie_{\uu} \E_{\sss \mathrm{L}} = T_{\uu \nn}, \qquad -\left(\Lie_{\kk} +\vtheta \right) \E_{\sss \mathrm{L}} = T_{\uu \kk}, \qquad \P_{\sss \mathrm{L}} \vtheta = T_{\kk \nn}, \qquad \text{and} \qquad \bP_{\sss \mathrm{L}} \vtheta = T_{\kk \kk}. 
\end{align}
These equations take the same form as those in the Einstein gravity (in Sec.\ref{sec:Ein}),
with the geometro-hydrodynamic quantities redefined for Lovelock gravity as 
\bbox
{\sffamily \textbf{Lovelock geometro-hydrodynamic dictionary:}}
\begin{flalign}
&\text{Energy density:} &&\E_{\sss \mathrm{L}} := \sum_{q=0}^{\lfloor d/2 \rfloor} c_{(q,d)} \frac{\E^q}{r^{q-1}} 
\label{energy_L}
&&\\
&\text{Pressure:} &&\P_{\sss \mathrm{L}}  := \sum_{q=0}^{\lfloor d/2 \rfloor} c_{(q,d)} \frac{\E^{q-1}}{r^{q-1}} \left( q \P + \frac{(q-1)(d-1)}{d-2} \E \right) 
\label{pressure_L}
\\
&\text{Dual pressure:} &&\bP_{\sss \mathrm{L}} := \sum_{q=0}^{\lfloor d/2 \rfloor} c_{(q,d)} \frac{\E^{q-1}}{r^{q-1}} q \bP 
\label{barP_L}
\end{flalign}
\ebox
\noindent Alternatively, they can be written as field-dependent affine transformations as
\begin{align}
\E_{\sss \mathrm{L}} = \esstix{a}_{(d)} \E, \qquad \P_{\sss \mathrm{L}} = \esstix{b}_{(d)} \P + \left(\esstix{b}_{(d)} - \esstix{a}_{(d)}\right) \left( \frac{d-1}{d-2}\E\right), \qquad \text{and} \qquad \bP_{\sss \mathrm{L}} = \esstix{b}_{(d)} \bP. \lb{LL-dictionary}
\end{align}

Having checked the geometro-hydrodynamic picture of Lovelock dynamics in the radial frame, we now proceed to show that Lovelock gravity also 
admits a hydrodynamic interpretation in a general frame. 

Since the situation closely mimics the Einstein-gravity case, it is both instructive and convenient to define analogs of the accelerations $\vkappa$ and $\bvkappa$ in Lovelock theory. These quantities are required in order to define the analog of the gravitational pressure $P$ for Lovelock gravity in an arbitrary frame, as in \eqref{P-gen} for the Einstein-gravity case. Considering the forms of \eqref{pressure} and \eqref{bP} and using the new quantities \eqref{energy_L}, \eqref{pressure_L} and \eqref{barP_L},
we define $\vkappa_{\sss \mathrm{L}}$ and $\bvkappa_{\sss \mathrm{L}}$ as 
\begin{alignat}{2}
\frac{\vkappa_{\sss \mathrm{L}}}{8\pi\Newton} := \P_{\sss \mathrm{L}} + \frac{d-3}{d-2} \E_{\sss \mathrm{L}} &= \sum_{q=0}^{\lfloor d/2 \rfloor} c_{(q,d)} \frac{\E^{q-1}}{r^{q-1}} \left( q \frac{\vkappa}{8\pi\Newton} + \frac{2(q-1)}{d-2} \E \right)  \nonumber \\
& = \esstix{b}_{(d)} \left( \frac{\vkappa}{8\pi\Newton} \right) + \left(\esstix{b}_{(d)} - \esstix{a}_{(d)} \right) \left(\frac{2}{d-2}\E \right) \label{kappa-LL}\\
\frac{\bvkappa_{\sss \mathrm{L}}}{8\pi\Newton} := \bP_{\sss \mathrm{L}} &=  \sum_{q=0}^{\lfloor d/2 \rfloor} c_{(q,d)} \frac{\E^{q-1}}{r^{q-1}} \left( q \frac{\bvkappa}{8\pi\Newton} \right) \nonumber \\
& = \esstix{b}_{(d)} \left( \frac{\bvkappa}{8\pi\Newton} \right). \label{bkappa-LL}
\end{alignat}
With these, the geometro-hydrodynamic dictionary and equations in the general frame can be achieved easily by applying the frame transformation (Sec.\ref{sec:Hydro}) to those in the radial frame \eqref{LL-eq}, since the transformation rule \eqref{f-trans} is independent of the dynamical equation. The dictionary is 
\begin{align}
K_{\sss \mathrm{L}} := \vkappa_{\sss \mathrm{L}} + \frac{\dot{\sigma}}{\sigma'}\bvkappa_{\sss \mathrm{L}}, \quad P_{\sss \mathrm{L}} := \frac{1}{8\pi\Newton}K_{\sss \mathrm{L}} -  \left(\frac{d-3}{d-2}\right) \E_{\sss \mathrm{L}} = \P_{\sss \mathrm{L}} + \frac{\dot{\sigma}}{\sigma'}\bP_{\sss \mathrm{L}}, \quad \text{and} \quad  \overline{P}_{\sss \mathrm{L}} := \frac{1}{\left( \sigma'\right)^2} \bP_{\sss \mathrm{L}}. 
\end{align}
The geometro-hydrodynamic equations in the general frame are
\bbox
{\sffamily \textbf{Lovelock geometro-hydrodynamic equations}}
\begin{subequations}
\begin{align}
\left(\Lie_{\u} + \theta_{(\u)}\right)\E_{\sss \mathrm{L}} +P_{\sss \mathrm{L}}\theta_{(\u)}  &= (\sigma')^{-1}T_{\u \n} \\
\left(\Lie_{\k} + \theta_{(\k)} \right)\E_{\sss \mathrm{L}} - \overline{P}_{\sss \mathrm{L}}\theta_{(\u)} & = -(\sigma')^{-1}T_{\u \k} \\
P_{\sss \mathrm{L}} \vtheta &= T_{\k \n} \\
\overline{P}_{\sss \mathrm{L}} \vtheta &= T_{\k \k}. 
\end{align}
\end{subequations}
\ebox
\noindent These results have therefore demonstrated the robustness of the geometro-hydrodynamic interpretation of the dynamics of spherically symmetric spacetimes beyond general relativity. \\

We conclude with several remarks. First, the generalization of the Misner-Sharp energy to Lovelock gravity can be obtained from the Lovelock energy density as $\m_{\sss \mathrm{L}} := \E_{\sss \mathrm{L}} \A$. This implies (from \eqref{energy_L}) that $\m_{\sss \mathrm{L}}$ is a polynomial in the original Misner-Sharp energy $\m$:
\begin{align}
\m_{\sss \mathrm{L}} 
= \sum_{q=0}^{\lfloor d/2 \rfloor} c_{(q,d)} \frac{\E^{q-1}}{r^{q-1}} \m
= \sum_{q=0}^{\lfloor d/2 \rfloor} c_{(q,d)} \frac{\m^q} {\mr{\A}^{q-1}r^{(q-1)(d-1)}} = \esstix{a}_{(d)} \m.
\label{MS_L}
\end{align}
This agrees with the formula for the Lovelock generalization of the Misner-Sharp energy given in \cite{Maeda:2011ii}, and with \cite{Nozawa:2008rjk} for the special case of Einstein-Gauss-Bonnet gravity, which is obtained by truncating the summation at $q=2$. Let us also remark that, as explained in \cite{Maeda:2011ii}, the energy $\m_{\sss \mathrm{L}}$ can be understood as a charge associated with the conserved current $(J_{\sss \mathrm{L}})^a := \frac{1}{8\pi\Newton} \esstix{G}^a{}_b \uu^b$, where we remind the readers that $\uu^a$ is the Kodama vector. This is directly analogous to the situation in the Einstein gravity \eqref{Kodama-current}. \\

Second, we should note that only three of the normal components of the Lovelock equation, $\esstix{G}_{ab} = 8\pi\Newton T_{ab}$, are independent. More precisely, one uses $\esstix{G}_{\kk \nn} = \esstix{G}_{\uu \kk} + 2\vrho \esstix{G}_{\kk \kk}$ (from $\nn^a = \uu^a + 2\vrho \kk^a$) and applies the radial-frame equations \eqref{LL-eq}, to obtain
\begin{align}
\P_{\sss \mathrm{L}}  = -\frac{1}{\vtheta}\Lie_{\kk}\E_{\sss \mathrm{L}} - \E_{\sss \mathrm{L}} + 2\vrho \bP_{\sss \mathrm{L}} = - \frac{1}{\vtheta \A} \Lie_{\kk} \m_{\sss \mathrm{L}}+ 2\vrho \bP_{\sss \mathrm{L}}. 
\end{align}
This is a natural generalization of \eqref{relationP-M}, providing the consistency of the definitions \eqref{kappa-LL} and \eqref{bkappa-LL}.\\ 

Furthermore, we should note that both $\bvkappa_{\sss \mathrm{L}}$ defined in \eqref{bkappa-LL} and $\bP_{\sss \mathrm{L}}$, which are proportional to $\bvkappa$, can still be interpreted as the acceleration of the null rigging vector $\vec{\kk}$. However, the quantity $\vkappa_{\sss \mathrm{L}}$ no longer retains the interpretation as the acceleration of the Kodama vector $\vec{\uu}$ due to the contributions from higher-curvature terms. Instead, it is given by a field-dependent affine transformation of $\vkappa$ as displayed in \eqref{kappa-LL}. Nonetheless, $\vkappa_{\sss \mathrm{L}}|_\H$, the one evaluated on the apparent horizon, can still be regarded as the \emp{Lovelock surface gravity}, as it can be obtained as the radial derivative (or gradient) of the \emp{Lovelock gravitational potential} $\Phi_{\sss \mathrm{L}}$, defined analogously to \eqref{Npotential} as
\begin{align}
\Phi_{\sss \mathrm{L}} := - \frac{8\pi\Newton}{(d-2)\mr{\A}} \frac{\m_{\sss \mathrm{L}}}{r^{d-3}} 
= \esstix{a}_{(d)} \Phi, 
\label{L-potential}
\end{align}
where the definition \eqref{MS_L} is used. Then, we can check the following relation:
\begin{align}
\vkappa_{\sss \mathrm{L}} = \Lie_{\kk} \Phi_{\sss \mathrm{L}} + 2\vrho \bvkappa_{\sss \mathrm{L}},
\end{align}
which is similar to \eqref{kappa}. Therefore, we obtain the desired expression, $\vkappa_{\sss \mathrm{L}}|_\H = (\Lie_{\kk} \Phi_{\sss \mathrm{L}})|_\H$.\\

Finally, we explain schematically how to obtain the spacetime metric \eqref{EFmetric} by solving the Lovelock equations in the geometro-hydrodynamic form \eqref{LL-eq}, and verify Birkhoff's theorem in Lovelock gravity \cite{Zegers:2005vx}. We first note that the Lovelock gravitational potential $\Phi_{\sss \mathrm{L}}$, defined in \eqref{L-potential}, can be expressed as a polynomial in the Newtonian potential $\Phi$ \eqref{Npotential} as
\begin{align}
-\frac{2\Phi_{\sss \mathrm{L}}}{r^2} = \sum_{q=0}^{\lfloor d/2 \rfloor} c_{(q)}\frac{(d-1)!}{(d-2q-1)!} \left(-\frac{2\Phi}{r^2}\right)^q. \lb{Wheeler}
\end{align}
When working in the radial frame, the Newtonian potential $\Phi$ 
appears in the spacetime metric (see \eqref{EFmetric} and \eqref{metric-1}), 
\begin{align}
\rd s^2 = - (1+2\Phi) \e^{2\tilde{\alpha}} \rd \tilde{v}^2 + 2 \e^{\tilde{\alpha}} \rd \tilde{v} \rd r + r^2 \rd \Omega^2_{d-2} \,, \lb{LL-alpha}
\end{align}
where $(\tilde{v}, r)$ are coordinates on the plane $\NP$, and $\tilde{\alpha}$ is a function on $\NP$. We would like to determine $\Phi$ and $\tilde{\alpha}$. 

For the potential $\Phi$, we solve the polynomial equation \eqref{Wheeler} for $\Phi$, given that the Lovelock analog $\Phi_{\sss \mathrm{L}}$ is known. The latter can be obtained by solving the first two equations of \eqref{LL-eq} for the Lovelock energy density $\E_{\sss \mathrm{L}}$ (or the energy $\m_{\sss \mathrm{L}} = \E_{\sss \mathrm{L}} \A$), then using the definition \eqref{L-potential} to get $\Phi_{\sss \mathrm{L}}$. For the scale factor $\tilde{\alpha}$, we recall that $\bvkappa = \pa_r \tilde{\alpha}$ (see the discussion below \eqref{EFmetric-1}) and, from the definition \eqref{LL-dictionary} of $\bP_{\sss \mathrm{L}}$ and \eqref{bP} of $\bP$, we have that
\begin{align}
\bP_{\sss \mathrm{L}} = \esstix{b}_{(d)} \frac{\bvkappa}{8\pi\Newton} = \esstix{b}_{(d)} \frac{\pa_r \tilde{\alpha}}{8\pi\Newton}.
\end{align}
Therefore, solving the last equation of \eqref{LL-eq} fixes $\bP_{\sss \mathrm{L}}$, and in turn determines $\tilde{\alpha}$.  

To demonstrate this procedure and also verify the Lovelock analog of Birkhoff's theorem, let us consider the vacuum case 
in which $T_{\mu\nu} = 0$. Similar to the case of general relativity \eqref{EH-vacuum}, the Lovelock geometro-hydrodynamic equations \eqref{LL-eq} in the radial frame reduce to 
\begin{align}
\Lie_{\uu} \m_{\sss \mathrm{L}} =0\,, \qquad \Lie_{\kk} \m_{\sss \mathrm{L}} = 0\,, \qquad \P_{\sss \mathrm{L}} = 0\,, \qquad \text{and} \qquad \bP_{\sss \mathrm{L}} = 0 \,.
\end{align}
The first two equations dictate that the Misner-Sharp energy for Lovelock theory is constant, $\m_{\sss \mathrm{L}} = \m_{\sss \mathrm{L}}{}_{0}$. The last equation together with \eqref{LL-alpha} imposes that $\tilde{\alpha} = \tilde{\alpha}(\tilde{v})$, which can then be reabsorbed into the new time coordinate $\e^{\tilde{\alpha}} \rd \tilde{v} = \rd v$. Hence, the spacetime metric reads
\begin{align}
\rd s^2 = - (1+2\Phi_0(r)) \rd v^2 + 2  \rd v \rd r + r^2 \rd \Omega^2_{d-2} \,, \lb{LL-BH}
\end{align}
where the time-independent Newtonian potential $\Phi_0(r)$ is the solution to the polynomial equation,
\begin{align}
-\frac{2\Phi_{\sss \mathrm{L}}{}_0}{r^2} = \sum_{q=0}^{\lfloor d/2 \rfloor} c_{(q)}\frac{(d-1)!}{(d-2q-1)!} \left(-\frac{2\Phi_0}{r^2}\right)^q \,, \quad \text{where} \quad \Phi_{\sss \mathrm{L}}{}_0 (r) = - \frac{8\pi\Newton}{(d-2)\mr{\A}} \frac{\m_{\sss \mathrm{L}}{}_0}{r^{d-3}} \,. \lb{Wheeler-old}
\end{align}
Note that $\Phi_{\sss \mathrm{L}}{}_0 (r)$, which follows from \eqref{L-potential}, is the static Lovelock gravitational potential. With this result, we have proven Birkhoff's theorem for the Lovelock gravity \cite{Zegers:2005vx} --- the spherically symmetric solutions to the vacuum Lovelock equations are locally isometric to the static Lovelock black hole solutions \eqref{LL-BH}, where $\Phi_0(r)$ is a root of the polynomial \eqref{Wheeler-old}. 

This polynomial equation \eqref{Wheeler-old} for the vacuum case was first discovered by Wheeler \cite{Wheeler:1985qd} and is referred to as the \emp{Wheeler polynomial equation}. Let us appreciate that, in our geometro-hydrodynamic formulation, this solution-finding procedure centered around the Wheeler polynomial equation readily extends to non-vacuum cases as well. The key difference is that the Lovelock energy $\m_{\sss \mathrm{L}}$ is no longer constant and the gravitational potential $\Phi_{\sss \mathrm{L}}$ may depend on time and radius. Nevertheless, once $\Phi_{\sss \mathrm{L}}$ is known, the Newtonian potential $\Phi$ can still be obtained by solving \eqref{Wheeler}, just as in the vacuum case.

\section{Conclusion and Outlook} \lb{sec:conclusion}
This series of works revisits the gravitational physics of spherically symmetric spacetimes as a foundation for understanding the thermodynamics of spacetime. If spacetime consists of microscopic degrees of freedom, the Einstein equations, expected to appear in a limit with many of those, may exhibit some thermodynamic/hydrodynamic behavior. Here, a general spherically symmetric spacetime, a suitable arena for exploring the interaction between matter and gravity, has the Kodama vector as a preferred time vector, providing the Misner-Sharp energy $\m$---a notion of locally conserved energy that can play the role of energy in hydrodynamics and thermodynamics. In this first article, we described the geometry of general spherically symmetric spacetimes as a foliation by spherical slices $\Sigma_\sigma$ and found a hydrodynamic interpretation of the gravitational dynamics. 

Based on the rigging technique, we first reconstructed the geometry and dynamics fully in terms of the foliation-adapted frames. In particular, we proved that the evolution vector in the radial frame agrees with the Kodama vector, and described apparent horizons in the equipotential frame to derive several formulas, such as the dynamical surface gravity. These results provide a new mathematical toolbox for studying the dynamics of general spherically symmetric spacetimes. 

Within this framework, we then demonstrated that the Einstein equation in any adapted frame takes the same form as the hydrodynamic equation, leading to a hydrodynamic interpretation of the gravitational system---geometro-hydrodynamics. More precisely, the gravitational analogs of the Euler equation for energy and the Young-Laplace equation hold on the slice $\Sigma_\sigma = \Time_\sigma \times \S_r$, where the Misner-Sharp energy density $\E$ and the gravitational pressure $P$ appear consistently. Therefore, a mixture of gravity and matter fields on $\S_r$ behaves as a spherical collective mode that merely contracts or expands---a gravitational bubble---and its worldvolume corresponds to the slice $\Sigma_\sigma$. Then, the spherically symmetric spacetime can be regarded as the worldvolume of a concentric stacking of many gravitational bubbles, which is restricted by another component of the Einstein equation. We also extended the geometro-hydrodynamic picture beyond the Einstein gravity to Lovelock gravity, showing that similar structures persist even in the presence of higher-curvature terms. 
  
Thus, the geometro-hydrodynamics holds universally for any spherical slice in general spherically symmetric spacetimes (even without a horizon structure), any associated frame, and different types of gravitational dynamics. These results provide a fundamental framework for exploring the thermodynamic properties of spacetime from a perspective of dynamics. \\

In the second article \cite{JY2}, based on the geometric toolbox developed in this work, we will consider actions corresponding to various types of boundary conditions and investigate the covariant phase space of spherically symmetric spacetimes carefully. In particular, for a thermodynamic action, we will demonstrate that the symplectic geometry of the phase space on $\Sigma_\sigma$ exhibits a thermodynamic-like structure, incorporating geometro-hydrodynamic quantities (such as $\A$, $\m$, and $P$) as well as matter fields. We will then derive expressions for the entropy and temperature associated with the dynamical apparent horizon. \\

\noindent Let us also mention some other research directions worth investigating:

\begin{itemize}
\item \textbf{Beyond spherical symmetry:} While our analysis has focused on spherical symmetry, it would be interesting to explore more general configurations and examine whether a similar geometro-hydrodynamics works. In particular, extending the framework to axisymmetric spacetimes or to cases with either reduced or enhanced symmetry may offer deeper insights into the hydrodynamic interpretation of spacetime geometry and its dynamics.

\item \textbf{Carrollian perspective:} 
As touched in several places so far, there appear to be intriguing connections between the geometry of the normal plane $\NP$ and the 2-dimensional Carrollian manifold. In particular, the angular component of the Einstein equation \eqref{G_AB}, which we have not fully elaborated, resembles the momentum evolution equation of Carrollian hydrodynamics, where $\bvkappa$ (or $\bP$) plays the role of the Carrollian fluid's momentum. Further investigation may help reveal the Carrollian structure underlying the full dynamics of spherically symmetric spacetimes.

\item \textbf{Equation of state in the geometro-hydrodynamics:} Although we have expressed all the components of the Einstein equation in terms of $(\M,\A,P)$ and the frame, their hydrodynamic picture is not complete yet; it is not clear how to interpret the equation \eqref{Ein-uk} determining the radial distribution of $\E$, the one \eqref{Ein-kk} involving $\overline{P}$, and the tangential one \eqref{G_AB}, in terms of hydrodynamics. In particular, the equation of state is missing. As discussed in Sec.\ref{sec:geohydro}, it would be interesting to compare the radial equation for $\E$ \eqref{Ein-uk} and the one for pressure $P$ in the standard hydrodynamics and find a relation $P(\E)$. This may lead to a notion of entropy density $s$ and local temperature $T$ by assuming local thermodynamic relations, $\rd\E=T\rd s$ and $Ts=\E+P$ \cite{landau1987fluid}.

\item \textbf{Symmetry of general spherically symmetric spacetimes:} A spherically symmetric spacetime, a kind of mini-superspace, can be obtained by imposing spherical symmetry on a full superspace and making the number of configuration modes finite. This procedure could correspond to taking a "fluid limit" of the full superspace, and a universal symmetry may emerge independent of the details of quantum gravity models 
\cite{Oriti:2024elx,BenAchour:2024gir}. Indeed, Schrödinger symmetry holds as a physical (not gauge) symmetry in various mini-superspace models \cite{BenAchour:2022fif, BenAchour:2023dgj,Sano:2025xit}. Therefore, it is highly intriguing to utilize our formulation to investigate what symmetries are inherent in general spherically symmetric spacetime and whether Schrödinger symmetry generally holds.

\item \textbf{Killing-Yano tensor:} In this work, we have revealed the connection between the rigging technique and Kodama's formalism. Another perspective relates the Kodama vector to the conformal Killing-Yano tensor \cite{Nozawa:2008rjk, Kinoshita:2024wyr}. It would be interesting to further explore how our geometric framework and its geometro-hydrodynamic interpretation can be related to the Killing-Yano tensor.

\item \textbf{Higher-derivative theories:} An interesting question is whether the geometro-hydrodynamic interpretation presented here for Einstein-Hilbert and Lovelock gravity still persists in higher derivative theories, such as $f(R)$ gravity. A major obstacle is that, in such theories, even in the case of spherically symmetric spacetimes, there is no well-defined notion of Misner-Sharp energy unless certain additional conditions are imposed \cite{Zhang:2014goa}. To develop a geometro-hydrodynamic picture in these settings requires a better understanding of quasi-local energy in higher derivative theories.

\item \textbf{Toward gravity quantization:}  
Our geometric tools and hydrodynamic picture may support the quantization of spacetime, at least in a spherically symmetric setting.

A related direction has been explored for finite-distance null hypersurfaces \cite{Ciambelli:2023mir, Ciambelli:2024swv} (see also \cite{Wieland:2025qgx}). Their work relies on the ultra-local property of null hypersurfaces, which allows one to study each null ray independently, ultimately leading to a molecular description of spacetime geometry (see also \cite{Donnelly:2020xgu}). In our case, instead of null rays, we work with a congruence of fluid worldlines with velocity $\u^a$ distributed uniformly over the sphere, that is, the worldvolume of a gravitational bubble.  This point effectively reduces the analysis to a single representative fluid trajectory. We expect that the techniques developed in these works can be adapted to our simplified setting as well. Although limited in scope, our framework does not require the hypersurface to be null, thereby broadening its applicability. Note that this idea has also recently entered the program of quantum reference frames \cite{Freidel:2025ous}. 

Another approach is to consider the radial Wheeler-DeWitt equation \cite{Freidel:2008sh, Cianfrani:2013oja} within our framework. The spherical symmetry effectively reduces the system to a two-dimensional one, potentially leading to simplifications that allow for explicit treatment of the problem.

\end{itemize}

\section*{Acknowledgement}

We would like to thank Wei-Hsiang Shao and Sotaro Sugishita for their fruitful feedback during an informal presentation of this work in progress. P.J. thanks Luca Ciambelli for collaboration on related projects in which the understanding of the rigging technique was further refined. The presentation of this work, particularly the part concerning frames and their transformations, benefited greatly from the lectures of Philipp Höhn at the 8th Intensive Lectures by Quantum Gravity Gatherings in RIKEN-iTHEMS. Y.Y. was partially supported by Japan Society for the Promotion of Science (No.21K13929). This work was supported by the RIKEN Center for Interdisciplinary Theoretical and Mathematical Sciences (iTHEMS).


\appendix
\renewcommand{\theequation}{\thesection.\arabic{equation}}
\setcounter{equation}{0}
\makeatletter
\@addtoreset{equation}{section}
\makeatother

\section{Details of geometry} \lb{app:geometry}
In this appendix, we provide several formulas relevant to the geometry of spherically symmetric spacetimes.


\subsection{Covariant derivatives} 

We derive the relation between the spacetime covariant derivative $\nabla_\mu$ and the normal-plane covariant derivative $\DN_a$. First, with straightforward computation using the metric \eqref{metric}, the spacetime Christoffel symbols are given by
\begin{align*}
\Gamma^a_{bc} = \stackrel{\sss (\NP)}{\Gamma}\!{}^a_{bc}, \qquad  \Gamma^A_{BC} = \stackrel{\sss (\S)}{\Gamma}\!{}^A_{BC}, \qquad \Gamma^a_{AB} = - \frac{1}{r}(\DN^a r)  q_{AB}, \qquad \text{and} \qquad \Gamma^A_{aB} = \left(\DN_a \ln r \right) \delta^A_B,
\end{align*}
where $\stackrel{\sss (\NP)}{\Gamma}\!{}^a_{bc}$ and $\stackrel{\sss (\S)}{\Gamma}\!{}^A_{BC}$ denote, respectively, the Christoffel symbols on the normal plane $\NP$ and the unit round sphere $\mr{\S}$. 

For a vector $V^a$ on $T\NP$, the spacetime covariant derivative of its lift $V^\mu = (V^a, 0)$ can be expressed as the normal covariant derivative as,
\begin{align}
\nabla_\nu V^\mu = \left(\delta^b_\nu \delta_a^\mu \DN_b + \delta_A^\mu \delta_\nu^A \DN_a \ln r \right) V^a.
\end{align}
From this, one deduces the normal components of the spacetime derivative, $\nabla_b V^a = \DN_b V^a$. Furthermore, the spacetime divergence of the vector $V^\mu$ is
\begin{align}
\nabla_\mu V^\mu = \left(\DN_a +(d-2) \DN_a \ln r \right) V^a =  \frac{1}{r^{d-2}} \DN_a \left(r^{d-2} V^a\right). \lb{div-rela}
\end{align}
Note that, in terms of the expansion $\theta_{(V)} = \Lie_V \ln \A = \vtheta \Lie_V r$ along the vector $V$, we have the relation,
\begin{align}
\nabla_\mu V^\mu = \DN_a V^a + \theta_{(V)}. 
\end{align}
Similarly, for a rank-2 normal tensor $T^{ab}$ direct computation yields,
\begin{align}
\nabla_\mu T^{\mu a} = \left( \DN_b  + (d-2) \DN_b \ln r \right) T^{ba} = \frac{1}{r^{d-2}} \DN_b \left(r^{d-2} T^{ba}\right) .
\end{align}

\subsection{Curvature tensors} \lb{app:curvature}
The non-trivial components of the spacetime Riemann tensor are
\begin{subequations}
\lb{Riemann}
\begin{align}
R_{abcd} &= \frac{\stackrel{\sss (\NP)}{R}}{2} \left( h_{ac} h_{bd} - h_{ad}h_{bc} \right) \\
R_{aAbB} &= - \frac{1}{r} \left(\DN_a \DN_b r\right) q_{AB} \\
R_{ABCD} & = \frac{16\pi \Newton}{d-2} \frac{\m}{\mr{\A} r^{d-1}} \left(q_{AC} q_{BD} - q_{AD} q_{BC}\right) \, ,
\end{align}
\end{subequations}
where $\stackrel{\sss (\NP)}{R}$ denote a Ricci scalar of the normal plane $\NP$ and we recall the definition of the Misner-Sharp energy $\m$ in \eqref{Misner-Sharp}.

One can then derive the non-zero components of the spacetime Ricci tensor. They are given by
\begin{align}
R_{ab} = \frac{\stackrel{\sss (\NP)}{R}}{2} h_{ab} - \frac{(d-2)}{r}\DN_a \DN_b  r, \ \ \ \ \text{and} \ \ \ \ R_{AB} &= \left( \frac{16 \pi \Newton}{\mr{\A}}\frac{(d-3)}{(d-2)}  \frac{\m}{r^{d-1}} - \frac{1}{r}\DN^2 r \right)q_{AB}. \lb{Ricci-tensor}
\end{align}
The Ricci scalar is thus
\begin{align}
\frac{1}{2}R = \frac{1}{2}\stackrel{\sss (\NP)}{R} - (d-2) \frac{\DN^2 r}{r} + (d-3) \frac{8\pi\Newton}{\mr{\A}} \frac{\m}{r^{d-1}} \ .\lb{Ricci-scalar} 
\end{align}
Following from these results, the components of the spacetime Einstein tensor are 
\begin{subequations}
\lb{Einstein-tensor}
\begin{align}
G_{ab} &= - \frac{d-2}{r}\DN_a \DN_b  r + \left(\frac{(d-2)}{r}\DN^2 r- (d-3) \frac{8 \pi \Newton }{\mr{\A} } \frac{\m}{r^{d-1}} \right)h_{ab} \\
G_{AB} &= (d-3)\left(\frac{1}{r}\DN^2 r  -\frac{ 1}{2 (d-3)}\stackrel{\sss (\NP)}{R} - \frac{(d-4)}{(d-2) }  \frac{8\pi\Newton }{\mr{\A}} \frac{\m}{r^{d-1}} \right)q_{AB}. 
\end{align}
\end{subequations}
Taking the trace of the normal components, we can show the relation
\begin{align}
h^{ab}G_{ab} &= \frac{d-2}{r}\DN^2  r - 2(d-3)\frac{8\pi\Newton}{\mr{\A} } \frac{\m}{r^{d-1}} \, . \lb{trace-G}
\end{align}

\subsection{Kodama conservation laws} \lb{app:Kodama-derive}

We review the derivation of the Kodama conservation laws \eqref{Kodama-current}. Starting from the normal components of the Einstein tensor \eqref{Einstein-tensor}, we show that 
\begin{equation}
\begin{aligned}
G^a{}_b \kv^b &= - \frac{(d-2)}{r}\left(\DN_c r \right) \DN_b \left( (\epsilon_{\sss \NP})^{cb}\DN^a  r \right)  + (d-2)\left(\frac{ \DN^2 r}{r}- \frac{(d-3)}{(d-2)} \frac{8\pi \Newton}{\mr{\A}} \frac{\m}{r^{d-1}} \right)\kv^a \\
& = \frac{(d-2)}{r}\left(\DN_c r\right) \DN_b \left( (\epsilon_{\sss \NP})^{ba}\DN^c  r - (\epsilon_{\sss \NP})^{ca}\DN^b  r \right)  + (d-2)\left(\frac{ \DN^2 r}{r}- \frac{(d-3)}{(d-2)} \frac{8\pi\Newton}{\mr{\A}}\frac{\m}{r^{d-1}}  \right)\kv^a \\
& = (d-2) \left(\frac{1}{2r}(\epsilon_{\sss \NP})^{ba} \DN_b \left(\DN r  \right)^2 -  \frac{(d-3)}{(d-2)} \frac{8\pi\Newton}{\mr{\A}} \frac{\m}{r^{d-1}} \kv^a \right) \\
& = -\frac{1}{r^{d-2}}(\epsilon_{\sss \NP})^{ba} \DN_b \left( \frac{8\pi\Newton }{\mr{\A}} \m \right), 
\end{aligned}
\end{equation}
where in the second equality, we simply used the identity $(\epsilon_{\sss \NP})^{[ab} \DN^{c]} r =0$, and in the third equality, we recalled the definition \eqref{Misner-Sharp} of the Misner-Sharp energy. It simply follows from the torsion-free property of the derivative $\DN_a$ that
\begin{align}
\DN_a (r^{d-2} J^a) = \frac{1}{2}\epsilon^{ab} [\DN_a, \DN_b] \left( \frac{8\pi\Newton }{\mr{\A}} \m \right) = 0, 
\end{align}
where the Kodama current is $8\pi\Newton J^a = G^a{}_b \kv^b$. Using the relation between the spacetime divergence and the normal divergence \eqref{div-rela}, we can show that the Kodama current is locally conserved, 
\begin{align}
\nabla_\mu J^\mu = 0. 
\end{align}

Integrating the current over a spherically symmetric codimension-1 surface $\Sigma$ yields the Misner-Sharp energy,
\begin{equation}
\begin{aligned}
- \int_\Sigma J^a \exd \Sigma_a & = \int_\Sigma \frac{1}{\A}(\epsilon_{\sss \NP})^{ba} (\DN_b \m) (\epsilon_{\sss \NP})_{ac} \exd y^c \wedge \volS \\
& = \int (\pa_a \m)  \exd y^a \\
& = \m,
\end{aligned}
\end{equation}
where the covariant volume form of $\Sigma$ is given by the contraction of the spacetime volume form, $\exd \Sigma_a = \iota_{\pa_a} \volM = (\epsilon_{\sss \NP})_{ac} \exd y^c \wedge \volS$. We also used the contraction $(\epsilon_{\sss \NP})^{ba} (\epsilon_{\sss \NP})_{ac} = \delta^b_c$ and integrated over the spherical shell $\S_r$ to obtain the second equality. 

\subsection{Covariant derivative of the adapted frame}
The spacetime covariant derivatives of the frame fields and coframe fields adapted to the spherical slice $\Sigma_\sigma$ decompose as follows:
\begin{subequations}
\lb{nabla-inv}
\begin{align}
\nabla_\mu \u^\nu  & = \frac{1}{d-2} \theta_{ (\u)}q_\mu{}^\nu-\bkappa \n_\mu \u^\nu + \kappa  \k_\mu \u^\nu - \kappa \n_\mu \k^\nu - \left( \Lie_{\u}\rho - 2\rho \kappa \right) \k_\mu \k^\nu \\
\nabla_\mu \k^\nu  & = \frac{1}{d-2}  \theta_{ (\k)} q_\mu{}^\nu  + \bkappa \n_\mu \k^\nu - \kappa \k_\mu \k^\nu  \\
\nabla_\mu \n_\nu & = \frac{1}{d-2} (\theta_{ (\u)} + 2\rho \theta_{ (\k)}) q_{\mu \nu} - \bkappa \n_\mu \n_\nu + \kappa \n_\mu \k_\nu + \kappa \k_\mu \n_\nu  + \left( \Lie_{\u}\rho - 2\rho \kappa \right) \k_\mu \k_\nu. 
\end{align}
\end{subequations}
We recall the expansions $\theta_{ (\u)} = \vtheta \Lie_{\u} r$ and $\theta_{ (\k)} = \vtheta \Lie_{\k} r$. The spacetime divergences are 
\begin{align}
\nabla_\mu \u^\mu = \theta_{ (\u)}, \qquad \nabla_\mu \k^\mu = \theta_{ (\k)} + \bkappa, \qquad \text{and} \qquad \nabla_\mu \n^{\mu}  = \theta_{ (\n)} + \kappa + \Lie_{\k} \rho,\lb{div}
\end{align}
where we used that $\theta_{ (\n)} = \theta_{ (\u)} + 2\rho \theta_{ (\k)}$ which followed from the definition of the vertical frame \eqref{vector-u}. The normal covariant derivative of the (co)frame fields can be obtained by simply dropping the angular components, which are terms proportional to the sphere metric $q_{\mu\nu}$. The normal divergences are 
\begin{align}
\DN_a \u^a = 0, \qquad \DN_a \k^a =  \bkappa, \qquad \text{and} \qquad \DN_a \n^a  =  \kappa + \Lie_{\k} \rho = 2\kappa - 2\rho \bkappa. \lb{normal-div}
\end{align}

When working with the radial frame adapted to the radial slice $\Sigma_r$, where $\nn = \rd r$, the vertical expansion vanishes, $\theta_{ (\uu)} = 0$, and the transverse expansion agrees with the area expansion, $\theta_{ (\kk)} = \vtheta$. The normal-plane's Laplace-Beltrami operator acting on the areal radius is given by the normal divergence of the normal form, $\DN^2 r = \DN^a  \DN_a r = \DN^a \nn_a$, and can be expressed as
\begin{equation}
\begin{aligned}
\DN^2 r  =  2\vkappa - (1+2\Phi)\bvkappa = 2 \Lie_{\kk} \Phi  +(1+2\Phi) \bvkappa  = \vkappa + \Lie_{\kk}\Phi,
\end{aligned}
\end{equation}
where we recalled that $2\vrho = (1+ 2\Phi)$ for the gravitational potential $\Phi$. 

\section{Change of foliation-adapted frames} \lb{app:frame-change}

We provide in this section the detailed derivation of the change of foliation-adapted frames discussed in Sec.\ref{sec:frame-change}. Relevant to the present work, we will focus on the relation between an arbitrary frame $e_{\mathscr{A}} = (\vec{\u}, \vec{\k})$ and the dual coframe $e^{\mathscr{A}} = (\k,\n)$ adapted to an arbitrary $\sigma$-constant slice $\Sigma_\sigma$ and the radial counterparts, $\bdx{e}_{\mathscr{A}} = (\vec{\uu}, \vec{\kk})$ and $\bdx{e}^{\mathscr{A}} = (\kk,\nn)$, adapted to the radial slice $\Sigma_r$. We emphasize again that the result here is also applicable for the transformation between any two reference frames. 

Locally, the general (co)frames can be written as linear transformations of the radial ones, and vice versa, $e_{\mathscr{A}} = \mathbb{M}_{\mathscr{A}}{}^{\mathscr{B}} \bdx{e}_{\mathscr{B}}$. The elements $\mathbb{M}_{\mathscr{A}}{}^{\mathscr{B}}$ of the transformation matrices are determined by the conditions the frame fields need to obey. These conditions, stemming from the rigging construction presented in Sec.\ref{sec:rigging}, comprise the relations \eqref{vector-u} between the vertical (co)frames and transverse (co)frame fields, the pairings \eqref{pairing}, the norm squared \eqref{norm}, and the exterior derivatives of the coframe fields \eqref{dk}. In total, the (co)frame fields satisfy,  
\begin{equation}
\lb{condition-app}
\begin{alignedat}{7}
\k^a \n_a &= 1, \quad &&\k^a \k_a &&= 0, \quad &&\u^a \k_a &&= 1, \quad &&\u^a \n_a &&= 0, \\
\u^a\u_a &= - \n^a \n_a, \quad  &&h_{ab}\k^b &&= \k_a, \quad &&\u^a &&= \n^a + 2\rho \k^a, \quad &&\rd \n &&=0,  
\end{alignedat}
\end{equation}
and similarly for the radial frame. Among these bases, the transverse coframe $\n$ is most constrained. Let us recall that we choose $\n = \rd \sigma$, which satisfies $\rd \n =0$, and we can decompose the differential of a function in terms of the frame fields using \eqref{df}. This means we can express the components of $\n$ in the radial frame as
\begin{align}
\n = \sigma' \nn + \dot{\sigma} \kk, \qquad \text{where} \qquad \dot{\sigma} := \Lie_{\uu}\sigma \quad \text{and} \quad \sigma' := \Lie_{\kk}\sigma. 
\end{align}
Although it is not obvious at the first glance that the right-hand side of $\n = \sigma' \nn + \dot{\sigma} \kk$ is closed, one can verify that 
\begin{equation}
\begin{aligned}
\rd \left( \sigma' \nn + \dot{\sigma} \kk \right) & = \rd \sigma' \wedge \nn + \sigma'\rd \nn+ \rd \dot{\sigma} \wedge \kk + \dot{\sigma} \rd \kk \\
&= \left([\Lie_{\uu},\Lie_{\kk}]\sigma \right)\kk \wedge \nn + \dot{\sigma} (\bvkappa \nn \wedge \kk) \\
& = \left( \Lie_{[\vec{\uu},\vec{\kk}]} \sigma - \bvkappa \dot{\sigma} \right) \kk \wedge \nn \\
& = 0
\end{aligned}
\end{equation}
is indeed zero, where we used the exterior derivatives \eqref{dk}, $\rd \nn =0$ and  $\rd \kk = \bvkappa \nn \wedge \kk$, as well as the Lie bracket \eqref{com-gen}, $\big[\vec{\uu},\vec{\kk}\big] = \bvkappa \vec{\uu}$. 

To systematically find the decomposition of the remaining frame fields in the radial frame, let us parameterize them as follows:
\begin{align}
\vec{\u} = A \vec{\uu} + B\vec{\kk}, \qquad \vec{\k} = C \vec{\uu} + D\vec{\kk}, \qquad \text{and} \qquad \k = C \nn + (D-2\vrho C)\kk
\end{align}
where the unknowns $A,B,C$ and $D$ are to be determined from \eqref{condition-app}. The decomposition of $\k$ also follows from the decomposition of the transverse frame $\vec{\k}$ as $\k_a = h_{ab} \k^b = C\uu_a +D \kk_a = C \nn + (D-2\vrho)\kk$, following from the relation $\uu^a = \nn^a - 2\vrho \kk^a$. 

The coefficients $A$ and $B$ are determined from the conditions $\u^a \n_a = 0$ and $\u^a \u_a = - \n^a \n_a$, which impose 
\begin{align}
A \dot{\sigma} + B \sigma' = 0 \qquad \text{and} \qquad A\left(\vrho A - B\right)  =  \sigma' \left( \dot{\sigma} + \vrho \sigma' \right).
\end{align}
There are two solutions to the above equations, 
\begin{align}
B =  - \frac{\dot{\sigma}}{\sigma'} A \, , \qquad \text{and} \qquad A = \pm \ \sigma' \, .
\end{align}
The vertical frame field $\vec{\u}$ is therefore given by 
\begin{align}
\vec{\u} = \varepsilon \left( \sigma' \vec{\uu} - \dot{\sigma}\vec{\kk} \right), \qquad \text{where} \qquad \varepsilon = \pm 1. 
\end{align}

Finally, the values of $C$ and $D$ can obtained by simultaneously solving the conditions $\u^a \k_a = 1$, $\k^a\n_a =1$,  and $\k^a \k_a =0$, which impose
\begin{align}
\varepsilon\left( \sigma' (D-2\vrho C) - C \dot{\sigma} \right) =1, \quad C\dot{\sigma}+D \sigma' = 1, \quad \text{and} \quad C(D -2\vrho C) = 0. 
\end{align}
The solutions for the transverse (co)frame field are 
\begin{align}
\vec{\k}  &= \frac{1}{\dot{\sigma} + \vrho \sigma'} \left[ \left(\frac{1-\varepsilon}{2} \right)\vec{\uu} + \left( \frac{(1+\varepsilon) \dot{\sigma} + 2\vrho \sigma'}{2\sigma'} \right)\vec{\kk} \right] \\
\k  &= \frac{1}{\dot{\sigma} + \vrho \sigma'} \left[ \left(\frac{1-\varepsilon}{2} \right)\nn + \left( \frac{(1+\varepsilon) \dot{\sigma} + 2\vrho \varepsilon \sigma'}{2\sigma'} \right)\kk \right] . 
\end{align}


\section{Einstein tensor} \label{app:Ein-ten}
The expressions for the components of the Einstein tensor in an arbitrary frame can be obtained using the formulae given in Appendix \ref{app:curvature}. The normal components of the Einstein tensor $G_{ab}$, given in \eqref{Einstein-tensor}, can be expressed as
\begin{equation}
\begin{aligned}
G_{ab} &= - \vtheta \DN_a \DN_b r + \left( \vtheta \DN^2 r - \frac{1}{2} \left( \frac{d-3}{d-2} \right) \vtheta{}^2\left(\DN r\right)^2 - \frac{(d-3)(d-2)}{2r^2}  \right) h_{ab} \\
& = - \vtheta \DN_a \DN_b r + \left( \vtheta \DN^2 r - \frac{1}{2} \left( \frac{d-3}{d-2} \right) \vtheta{}^2\left(\DN r\right)^2 - \frac{1}{2}\stackrel{\sss (\S)}{R} \right) h_{ab}
\end{aligned}
\end{equation}
where in the first equality, we used the definition of the radial expansion \eqref{expansion} and the definition \eqref{Misner-Sharp} of the Misner-Sharp energy. We also used that the Ricci scalar of the spherical shell $\S_r$ is 
\begin{align}
\stackrel{\sss (\S)}{R} = \frac{(d-3)(d-2)}{r^2}
\end{align}
to obtain the second equality. Using the decompositions \eqref{df}, $\DN_a r = \left(\Lie_{\u}r\right) \k_a + \left(\Lie_{\k}r\right)\n_a$ and $\DN^a r = \left(\Lie_{\k}r\right)\u^a + \left(\Lie_{\n}r\right)\k^a$, we can show that
\begin{align}
\vtheta{}^2\left(\DN r\right)^2 = \theta_{(\k)} \left( \theta_{(\n)} + \theta_{(\u)} \right). \lb{formula-1}
\end{align}
In addition, we have 
\begin{equation}
\lb{formula-2}
\begin{aligned}
\vtheta \DN^2 r &= \vtheta \DN_a \left(\left(\Lie_{\k}r\right)\u^a + \left(\Lie_{\n}r\right)\k^a \right) \\
\step{[\text{Leibniz rule, \ } \DN_a \vtheta = \tfrac{1}{d-2}\vtheta{}^2 \DN_a r ]} \quad 
& = \DN_a \left( \theta_{(\k)}\u^a + \theta_{(\n)}\k^a  \right) + \frac{1}{d-2}\theta_{(\k)} \left( \theta_{(\n)} + \theta_{(\u)} \right) \\
\step{[\text{use \eqref{normal-div}}]} \quad
& = \Lie_{\u}\theta_{(\k)} + \Lie_{\k}\theta_{(\n)}+ \frac{1}{d-2}\theta_{(\k)} \left( \theta_{(\n)} + \theta_{(\u)} \right) + \bkappa  \theta_{(\n)} \\
& =  \Lie_{\k}\left(\theta_{(\n)} + \theta_{(\u)}\right)+ \frac{1}{d-2}\theta_{(\k)} \left( \theta_{(\n)} + \theta_{(\u)} \right) + \bkappa  \left( \theta_{(\n)} + \theta_{(\u)} \right).
\end{aligned} 
\end{equation}
In the last equality, we used the fact that $\Lie_{\u}\theta_{(\k)} = \Lie_{\k}\theta_{(\u)} + \bkappa \theta_{(\u)}$, which one can check using the identity $\Lie_{\u}\Lie_{\k}  = \Lie_{\k}\Lie_{\u} + \Lie_{[\u,\k]}$
and the Lie bracket \eqref{com-gen}, $\big[\vec{\u},\vec{\k}\big] = \bkappa \vec{\u}$.

\begin{enumerate}[left=0pt, labelsep=0em, labelindent=0pt, align=parleft, listparindent=0pt, label = \textbf{\roman*.)}]
\item For the component $G_{\u \n} := \u^a G_a{}^b \n_b$, we apply the Leibniz rule and the covariant derivative \eqref{nabla-inv} to derive
\begin{equation}
\begin{aligned}
G_{\u \n} &= - \vtheta \n^a \DN_\u \DN_a r  = - \Lie_{\u}\theta_{(\n)} + \left(\kappa - \frac{1}{d-2}\theta_{(\n)} \right)\theta_{(\u)} +  \theta_{(\k)}\Lie_{\u}\rho, 
\end{aligned}
\end{equation}
where we also used that $\DN_a \vtheta = \frac{1}{d-2}\vtheta{}^2 \DN_a r$ and $\theta_{(\u)} = \vtheta \Lie_{\u}r$ for a vector $\u$. 

The equation simplifies when working with the radial frame basis $(\vec{\uu}, \vec{\kk}, \nn, \kk)$ as the expansions become $\theta_{(\uu)} = 0$ and $\theta_{(\kk)} = \vtheta$, and that $\Lie_{\uu}\vtheta =0$. Expressing the function $\vrho$ in terms of the Misner-Sharp energy (see equation \eqref{rho}), we obtain
\begin{equation}
G_{\uu \nn} = -\vtheta  \Lie_{\uu}  \vrho =  \frac{8\pi\Newton}{\A} \Lie_{\uu} \m = 8\pi\Newton \Lie_{\uu} \E,
\end{equation}
where we recalled the definition \eqref{Edensity} for the Misner-Sharp energy density, $\E = \m/\A$, and that $\Lie_{\uu}\A =0$.  
\item Similarly, for the component $G_{\k \k} := \k^a G_a{}^b \k_b$, we show that
\begin{equation}
\begin{aligned}
G_{\k \k} &= - \vtheta \k^a \DN_\k \DN_a r  = - \Lie_{\k}\theta_{(\k)} + \left(\bkappa - \frac{1}{d-2}\theta_{(\k)} \right)\theta_{(\k)}. 
\end{aligned}
\end{equation}
When using the radial frame, we have $\theta_{(\kk)} = \vtheta$ and $\Lie_{\kk}\vtheta = -\frac{1}{d-2} \vtheta{}^2$. This component of the Einstein tensor simply becomes
\begin{equation}
\begin{aligned}
G_{\kk \kk} &= \bvkappa \vtheta = 8\pi\Newton \bP  \vtheta,
\end{aligned}
\end{equation}
where we defined $\bP := \bvkappa/{8\pi\Newton}$. 

\item The next component is $G_{\k \n} := \k^a G_a{}^b \n_b$, which can be expressed as
\begin{equation}
\begin{aligned}
G_{\k \n} 
& = - \vtheta \n^a \DN_\k \DN_a r + \vtheta \DN^2 r - \frac{1}{2} \left( \frac{d-3}{d-2} \right) \left(\vtheta\DN r\right)^2 - \frac{1}{2}\stackrel{\sss (\S)}{R}.
\end{aligned}
\end{equation}
The first term can be evaluated using the Leibniz rule and the covariant derivative \eqref{nabla-inv}, which yields
\begin{align}
- \vtheta \n^a \DN_\k \DN_a r = - \Lie_{\k}\theta_{(\n)} - \frac{1}{d-2}\theta_{(\n)}\theta_{(\k)} + \kappa \theta_{(\k)} - \bkappa \theta_{(\n)}.
\end{align}
Using the derived results \eqref{formula-1} and the third equality in \eqref{formula-2}, we obtain 
\begin{equation}
\begin{aligned}
G_{\k \n} 
& = \Lie_{\u}\theta_{(\k)} + \left( \kappa + \theta_{(\n)} \right)\theta_{(\k)} - \left(\frac{d-1}{d-2}\right) \rho \theta_{(k)}^2 - \frac{1}{2}\stackrel{\sss (\S)}{R}.
\end{aligned}
\end{equation}

Again, the equation greatly simplifies in the radial frame. We can show that
\begin{equation}
\begin{aligned}
G_{\kk \nn} = \vkappa \vtheta + \left(\frac{d-3}{d-2}\right)\vrho \vtheta{}^2- \frac{1}{2}\stackrel{\sss (\S)}{R}  &= 8\pi\Newton\left( \frac{\vkappa}{8\pi\Newton} - \frac{d-3}{d-2} \frac{\m}{\A} \right) \vtheta \\
& = 8\pi\Newton \P \vtheta,
\end{aligned}
\end{equation}
where we used \eqref{rho} to write $\vrho$ in terms of the Misner-Sharp energy, and we recalled the definition of the pressure \eqref{pressure}. 

\item The last normal component is $G_{\u \k} := \u^a G_a{}^b \k_b$, which can be written as
\begin{equation}
\begin{aligned}
G_{\u \k} 
& = - \vtheta \k^a \DN_\u \DN_a r + \vtheta \DN^2 r - \frac{1}{2} \left( \frac{d-3}{d-2} \right) \left(\vtheta\DN r\right)^2 - \frac{1}{2}\stackrel{\sss (\S)}{R}.
\end{aligned}
\end{equation}
Similar to the previous case, the first term can expressed as
\begin{align}
- \vtheta \k^a \DN_\u \DN_a r = - \Lie_{\u}\theta_{(\k)} - \frac{1}{d-2}\theta_{(\k)}\theta_{(\u)} - \kappa \theta_{(\k)}.
\end{align}
Again, by using the derived results \eqref{formula-1} and the third equality in \eqref{formula-2}, we show that
\begin{equation}
\begin{aligned}
G_{\u \k} 
& = \Lie_{\k}\theta_{(\n)}+ \theta_{(\n)}\theta_{(\k)} +\bkappa \theta_{(\n)} -\kappa \theta_{(\k)} - \left(\frac{d-3}{d-2}\right) \rho \theta_{(k)}^2 - \frac{1}{2}\stackrel{\sss (\S)}{R}.
\end{aligned}
\end{equation}

Finally, in the radial frame where the expansions are $\theta_{(\kk)} = \vtheta$ and $\theta_{(\nn)} = 2\vrho \vtheta$, we can write the Einstein tensor as
\begin{equation}
\begin{aligned}
G_{\uu \kk}  = \vtheta\Lie_{\kk}\vrho + \left(\frac{d-3}{d-2}\right) \vrho \vtheta{}^2 - \frac{1}{2}\stackrel{\sss (\S)}{R}  = -\frac{8\pi\Newton}{\A}\Lie_{\kk}\m \, , 
\end{aligned}
\end{equation}
where we expressed $\vrho$ in terms of $\m$ (see \eqref{rho}). 
\end{enumerate}
These components are not independent, as one can verify that $G_{\k \n} - G_{\u \k} = 2\rho G_{\k\k}$, coming from the relation $\n^a = \u^a + 2\rho \k^a$.\\

The remaining components are the angular components $G_{AB}$, which only have one independent component that is its trace, $q^{AB} G_{AB}$. First, we can show that the trace $q^{AB}G_{AB}$ from \eqref{Einstein-tensor} can be written as
\begin{align}
\frac{1}{d-2}q^{AB} G_{AB}= -\frac{ 1}{2 }\stackrel{\sss (\NP)}{R} + \left(\frac{d-3}{d-2}\right)\vtheta \DN^2 r + \frac{(d-4)(d-3)}{2 (d-2)^2} \vtheta (\DN r)^2  - \frac{(d-4)}{2(d-2)} \stackrel{\sss (\S)}{R},
\end{align}
where we used the definition of the Misner-Sharp energy \eqref{Misner-Sharp} and the Ricci scalar of the round sphere $\S_r$. We then need to express the Ricci scalar of the normal plane in terms of the geometric quantities in rigging framework. This can be achieved by considering the Ricci tensor $R_{\u \k}$, which can be written in as the commutator of covariant derivative as $R_{\u \k} = \u^\mu [\nabla_\nu, \nabla_\mu] \k^\nu$, and using the formula \eqref{Ricci-tensor} for the normal component of the spacetime Ricci tensor. We show that
\begin{equation}
\begin{aligned}
 -\frac{ 1}{2 }\stackrel{\sss (\NP)}{R} &=  - R_{\u \k} - \vtheta \u^b \k^a \DN_a \DN_b r \\
 & = \u^\mu [\nabla_\mu, \nabla_\nu] \k^\nu  - \vtheta \u^b \k^a \DN_a \DN_b r  \\
 & = \Lie_{\u}(\nabla \dd \vec{\k}) - \nabla_\mu (\nabla_{\u} \k^\mu) + (\nabla_\mu \u^\nu)(\nabla_\nu \k^\mu) - \left( \Lie_{\u}\theta_{(\k)} + \frac{1}{d-2}\theta_{(\u)} \theta_{(\k)} + \kappa \theta_{(\k)} \right) \\
 & = \Lie_{\u}\bkappa + \Lie_{\k}\kappa + \kappa \bkappa,
\end{aligned}
\end{equation}
where we employed the covariant derivatives \eqref{nabla-inv}. Using the formulae \eqref{formula-1} and \eqref{formula-2}, we obtain 
\begin{equation}
\begin{aligned}
\frac{1}{d-2}q^{AB} G_{AB} = \ & \Lie_{\u} \left(\bkappa + \frac{d-3}{d-2} \theta_{(\k)}\right) + \left( \Lie_{\k} + \bkappa \right) \left( \kappa + \frac{d-3}{d-2} \theta_{(\n)}  \right) + \frac{1}{2} \left( \frac{d-3}{d-2} \right) \theta_{(\k)} \left( \theta_{(\n)} + \theta_{(\u)} \right) \\
&  - \frac{(d-4)}{2(d-2)} \stackrel{\sss (\S)}{R}.
\end{aligned}
\end{equation}

In the radial frame, the above equation can be expressed in terms of the energy density \eqref{Edensity} and the pressure \eqref{pressure} as follows
\begin{align}
\frac{1}{d-2}q^{AB} G_{AB}= 8\pi \Newton \left( \Lie_{\uu}\bP + \left( \Lie_{\kk} + \bvkappa \right)\P  + \left(\frac{d-3}{d-2}\right)\left(\vtheta \P+\bvkappa \E\right) \right). 
\end{align}

\subsection{Ricci scalar}

With these expressions for the components of the Einstein tensor in an arbitrary frame, we can derive the expression for the spacetime Ricci scalar. Using the relation $-\frac{1}{2}R = \frac{1}{d-2} g^{\mu\nu} G_{\mu\nu}$, we derive the following result 
\begin{equation}
\begin{aligned}
- \frac{1}{2} R &= \frac{1}{d-2} \left( G_{\u \k} + G_{\k \n} \right) + \frac{1}{d-2} q^{AB} G_{AB} \\
& = \Lie_{\u}\left( \bkappa + \theta_{(\k)} \right) + \left(\Lie_{\k} + \bkappa \right) \left( \kappa + \theta_{(\n)} \right) + \frac{d-1}{2(d-2)} \theta_{(\k)} \left( \theta_{(\n)} + \theta_{(\u)} \right) -\frac{1}{2}\stackrel{\sss (\S)}{R}
\end{aligned} 
\end{equation}
where we employed the decomposition of the metric \eqref{metric-gen} and the previously derived expressions for components $G_{\u \k}$, $G_{\k \n}$, and $q^{AB}G_{AB}$ of the Einstein tensor. 

As usual, simplification occurs when adopting the radial slice basis, and the Einstein-Hilbert Lagrangian density becomes
\begin{align}
\frac{1}{16\pi\Newton} R = - \Lie_{\uu} \left( \frac{\bvkappa}{8\pi\Newton} \right) -  \left(\Lie_{\kk} + \bvkappa +\vtheta \right) \left( \frac{\vkappa}{8\pi\Newton} \right) + \frac{1}{\A}\Lie_{\k}\m.
\end{align}
Note that, the first two terms can be written as the total derivative terms (see the Stokes formula \eqref{Stokes}). By integrating out the angular directions, we finally obtain the dimensionally reduced gravitational Lagrangian
\begin{align}
\frac{1}{16\pi\Newton} R \A \volN = \Lie_{\kk}\m \volN+ \exd \left[ \frac{\vkappa \A}{8\pi\Newton} \kk -\frac{\bvkappa \A}{8\pi\Newton} \nn \right],
\end{align}
or alternatively
\begin{align}
\frac{1}{16\pi\Newton} R \A \volN = -\bvkappa \m \volN+ \exd \left[ \left(\frac{\vkappa \A}{8\pi\Newton} - \m \right) \kk -\frac{\bvkappa \A}{8\pi\Newton} \nn \right],
\end{align}
which also followed from the Stokes theorem \eqref{Stokes}.

\bibliography{Biblio.bib}
\bibliographystyle{Biblio}
\end{document}